\documentclass[12pt]{iopart}

\usepackage{graphicx}
\usepackage[british]{babel}
\usepackage{color}
\usepackage{mathrsfs}
\usepackage{setstack}
\usepackage{iopams}
\usepackage[T1]{fontenc}
\usepackage{array}
\usepackage{multirow}
\usepackage{cite}

\begin{document}

\title{First passage time moments of asymmetric L\'evy flights}

\author{Amin Padash$^{\dagger, \sharp}$, Aleksei V. Chechkin$^{\sharp,\ddagger}$,
Bart{\l}omiej Dybiec$^{\flat}$, Marcin Magdziarz$^{\S}$, Babak Shokri$^{\dagger,
\P}$, and Ralf Metzler$^{\sharp}$}
\address{$^\dagger$ Physics Department of Shahid Beheshti University, 19839-69411
Tehran, Iran}
\address{$^\sharp$ Institute for Physics \& Astronomy, University of Potsdam,
14476 Potsdam-Golm, Germany}
\address{$^\ddagger$ Akhiezer Institute for Theoretical Physics, 61108 Kharkov,
Ukraine}
\address{$^\flat$ Marian Smoluchowski Institute of Physics, and Mark Kac Center
for Complex Systems Research, Jagiellonian University, ul. St. Lojasiewicza 11,
30-348 Krakow, Poland}
\address{$\S$ Faculty of Pure and Applied Mathematics and Hugo Steinhaus Centre,
Wroc{\l}aw University of Science and Technology, Wyspianskiego 27, 50-370,
Wroc{\l}aw, Poland}
\address{$^\P$ Laser and Plasma Research Institute, Shahid Beheshti University,
19839-69411 Tehran, Iran}
\ead{rmetzler@uni-potsdam.de (Corresponding author)}

\begin{abstract}
We investigate the first-passage dynamics of symmetric and asymmetric L\'evy
flights in a semi-infinite and bounded intervals. By solving the space-fractional
diffusion equation, we analyse the fractional-order moments of the first-passage
time probability density function for different values of the index of stability
and the skewness parameter. A comparison with results using the Langevin approach
to L\'evy flights is presented. For the semi-infinite domain, in certain special
cases analytic results are derived explicitly, and in bounded intervals a general
analytical expression for the mean first-passage time of L\'evy flights with
arbitrary skewness is presented. These results are complemented with extensive
numerical analyses.
\end{abstract}

\section{Introduction}

L\'evy flights (LFs) correspond to a class of Markovian random walk processes
that are characterised by an asymptotic power-law form for the distribution of
jump lengths with a diverging variance \cite{BMandelbrot1997,BDHughes1995,
JPBouchaud1990,RMetzler2000,RMetzler2004}. The name "L{\'e}vy flight" was
coined by Beno{\^i}t Mandelbrot, in honour of his formative teacher, French
mathematician Paul Pierre L\'evy \cite{BMandelbrot1997,gordon}. The trajectories
of LFs are statistical fractals \cite{BMandelbrot1997}, characterised by local
clusters interspersed with occasional long jumps. Due to their self-similar
character, LFs display "clusters within clusters" on all scales. This emerging
fractality \cite{BMandelbrot1997,BDHughes1995,JPBouchaud1990,MVahabi2013} makes
LFs efficient search processes as they sample space more efficiently than normal
Brownian motion: in one and two dimensions\footnote{For most search processes of
animals for food or other resources these are the relevant dimensions: the case
of one dimension is relevant for animals whose food sources are found along
habitat borders such as the lines of shrubbery along streams or the boundaries
of forests. Two dimensional search within given habitats is natural for land
bound animals, but even airborne or seaborne animals typically forage within a
shallow depth layer compared to their horizontal motion.} Brownian motion is
recurrent and therefore oversamples the search space. LFs, in contrast, reduce
oversampling due to the occurrence of long jumps \cite{shlesinger,
gandhi,ganhdi1,watkins,sims,sims1,lomholt,lomholt1,vlad,vlad3,gandhibook}. As
search strategies LFs were argued to be additionally advantageous as, due to
their intrinsic lack of length scale they are less sensitive to time-changing
environments \cite{lomholt1}. Concurrently in an external bias LFs may lose their
lead over Brownian search processes \cite{vlad1,vlad2}. LFs were shown to underlie
human movement behaviour and thus lead to more efficient spreading of diseases as
compared to diffusive, Brownian spreading \cite{dirk,dirk1,barabasi}. LFs appear
as traces of light beams in disordered media \cite{wiersma}, and in optical
lattices the divergence of the kinetic energy of single ions under gradient
cooling are related to L{\'e}vy-type fluctuations \cite{walther}. Finally, we
mention that L{\'e}vy statistics were originally identified in stock market
price fluctuations by Mandelbrot and Fama \cite{mandelbrotfin,fama}, see also
\cite{mantegna}.

Mathematically, LFs are based on $\alpha$-stable distributions (or L{\'evy}
distributions) \cite{PPLevy1954,gnedenko} which emerge as limiting distributions
of sums of independent, identically distributed (i.i.d.) random variables according
to the generalised central limit theorem---that is, they have their own, well-defined
domains of attraction \cite{BDHughes1995,JPBouchaud1990,PPLevy1954,gnedenko}. The
characteristic function of an $\alpha$-stable process, which is a continuous-time
counterpart of an LF, is given as
\cite{Samorodnitsky-Taqqu,Gikhman-Skorokhod1975}
\begin{equation}
\fl\hat{\ell}_{\alpha,\beta}(k,t)=\int\limits_{-\infty}^{\infty}\ell_{\alpha,
\beta}(x,t)\e^{ikx}\mathrm{d}x=\exp\Big(-tK_{\alpha}|k|^{\alpha}[1-i\beta\mathrm{
sign}(k)\omega(k,\alpha)]+i\mu kt\Big),
\label{eq:charecA}
\end{equation}
with the stability index (L{\'e}vy index) $\alpha$ that is allowed to vary in the
interval $0<\alpha\leq2$. Moreover, equation (\ref{eq:charecA}) includes the
skewness parameter $\beta$ with $-1\leq\beta\leq1$, and $K_\alpha>0$ is a scale
parameter. The shift parameter $\mu$ can be any real number, and the phase factor
$\omega$ is defined as
\begin{equation}
\label{eq:charecA1}
\omega(k,\alpha)=\left\{\begin{array}{ll}\tan(\frac{\pi\alpha}{2}),&\alpha\neq 1\\
-\frac{2}{\pi}\ln|k|,&\alpha=1\end{array}\right..
\end{equation}
Physically, the parameter $\mu$ accounts for the constant drift in the
system. In this paper, we consider the first-passage time moments in the
absence of a drift, $\mu=0$. The stable index $\alpha$ is responsible for
the slow decay of the far asymptotics of the $\alpha$-stable probability
density function (PDF). Indeed, symmetric $\alpha$-stable distributions
in absence of a drift ($\beta=\mu=0$) have the characteristic function
$\exp(-K_{\alpha}|k|^{\alpha}t)$, whose asymptote in real space has the
power-law form $\simeq K_{\alpha}t|x|^{-1-\alpha}$ ("heavy tail" or "long
tail"), and thus absolute moments $\langle|x|^{\delta}\rangle$ of order
$\delta<\alpha$ exist \cite{BDHughes1995,JPBouchaud1990,Samorodnitsky-Taqqu,
Zolotarev1986}. The scale parameter $K_\alpha$ (along with the stable index
$\alpha$) physically sets the size of the LF-jumps. The skewness $\beta$ may
be related to an effective drift or counter-gradient effects \cite{para,para1}.
LFs have been applied to explain diverse complex dynamic processes, where
scale-invariant phenomena take place or can be suspected \cite{BMandelbrot1997,
PPLevy1954}. According to the generalised central limit theorem, each
$\alpha$-stable distribution with fixed $\alpha<2$ attracts distributions with
infinite variance which decay with the same law as the attracting stable
distribution. A particular case of a stable density is the Gaussian for $\alpha=2$,
for which moments of all orders exist. We note that the Gaussian law not only
attracts distributions with finite variance but also distributions decaying as
$\simeq|x|^{-3}$; that is, distributions, whose variance is marginally infinite
\cite{khin,gnedenko}. To fit real data, in particular, in finance, which
feature heavy-tailed distributions on intermediate scales, however, with
finite variance, the concept of the truncated LFs has been introduced 
according to which the truncation of the heavy tail at larger scales is
achieved either by an abrupt cutoff \cite{mantegna1}, an exponential cutoff
\cite{koponen}, or by a steeper power-law decay \cite{SokCheKla,chechkin_diss,
chechkin_pre}. 

The efficiency of the spatial exploration and search properties of a
stochastic process is quantified by the statistics of the "first-hitting"
or the "first-passage" times \cite{Feller1971,SRedner2001,CGardiner2009,
SidneyRedner2014}. For instance, the first-passage of a stock price crossing
a given threshold level serves as a trigger to sell the stock. The event of
first-hitting would correspond to the event when exactly a given stock price
is reached. Of course, when stock prices change continuously (as is the case
for a continuous Brownian motion) both first-passage and first-hitting are
equivalent \cite{CGardiner2009}. In contrast, for an LF with the propensity
of long, non-local jumps the two definitions lead to different results. In
general, the first-passage will be realised earlier: it is more likely that
an LF jumps across a point in space \cite{jpa2003} effecting so-called "leapovers"
\cite{Koren2007PhysicaA,TKoren2007}. For a foraging albatross, for instance,
the first-hitting would correspond to the moment when it locates a single,
almost point-like, forage fish. The first-passage would correspond to the
event when the albatross crosses the circumference of a large fish shoal.
We here focus on the first-passage time statistic of LFs, and our main
objective is the study of the moments of the first-passage time for
\emph{asymmetric\/} LFs in semi-infinite and bounded domains. Such moments
can be conveniently used to quantify search processes. The most commonly
used moment is the mean first-passage time (MFPT) $\langle\tau\rangle=\int_0^{
\infty}\wp(\tau)\tau\mathrm{d}\tau$ in terms of the first-passage time density $\wp
(\tau)$ (see below), when it exists. However, other definitions such as the mean
of the inverse first-passage time, $\langle1/\tau\rangle$ have also been studied
\cite{vlad1,vlad2}. More generally, the spectrum of fractional order
first-passage time moments $\langle\tau^q\rangle$ is important to characterise
the underlying stochastic process from measurements. The characteristic times
$\langle\tau\rangle$ and $\langle1/\tau\rangle$ thus correspond to $q=1$ and
$q=-1$, respectively. In what follows we study the behaviour of the spectrum
of $\langle\tau^q\rangle$ as function of the LF parameters.

A set of classical results exists for the first-passage time properties of LFs
in a semi-infinite domain. In particular, \cite{NHBingham1973-1,NHBingham1973-2}
used limit theorems of i.i.d. random variables to obtain the asymptotic
behaviour of the first-passage time distribution. Based on a continuous-time
storage model the first-passage time of a general class of L\'evy processes
was studied in \cite{NUPrabhu1981}. By applying the laws of ladder processes
the asymptotic of the first-passage time distribution of L\'evy stable processes
was investigated in \cite{Bertoin1996}. After becoming clear that LFs have
essential applications in different fields of science, several remarkable
results were established. Thus, in \cite{UFrisch1995} it was reported that
one-dimensional symmetric random walks with independent increments in half-space
have universal property. Also \cite{GZumofen1995} showed that the survival
probability of symmetric LFs in a one-dimensional half-space with an absorbing
boundary at the origin is independent of the stability index $\alpha$ and thus
displays universal behaviour. It is by now well-known that the mentioned results
are a consequence of the celebrated Sparre Andersen theorem \cite{ESparre1953,
ESparre1954}. Accordingly, the PDF of the
first-passage times of any symmetric and Markovian jump process originally
released at a fixed point $x_0$ from an absorbing boundary in semi-infinite
space has the universal asymptotic scaling $\wp(\tau)\simeq\tau^{-3/2}$
\cite{SRedner2001,jpa2003,Koren2007PhysicaA,TKoren2007}. This law has been
confirmed by extensive numerical simulations of the first-passage time PDF
\cite{Koren2007PhysicaA} and the associated survival probability
\cite{Dybiec2016} of symmetric LFs within a Langevin dynamic approach (see
below). Furthermore, the asymptotic of the survival probability of symmetric,
discrete-time LFs was studied in \cite{SNMajumdar2010, SNMajumdar2017}, and
based on the space-fractional diffusion equation the first-passage time
PDF and the survival probability was investigated in \cite{AminP2019}.
Starting from the Skorokhod theorem, the Sparre Andersen theorem could be
successfully reproduced analytically \cite{TKoren2007, AminP2019}. Other
analytical and numerical results that concern the first-passage properties
of asymmetric LFs in a semi-infinite domain are the following. For one-sided
$\alpha$-stable process ($0<\alpha<1$ with $\beta=1$) the first-passage time
PDF and the MFPT was studied in \cite{TKoren2007}. In \cite{Koren2007PhysicaA}
the authors used Langevin dynamic simulations to study the asymptotic behaviour
of the first-passage time PDF of extremal two-sided ($1<\alpha<2$ with $\beta=
-1$) $\alpha$-stable laws. Moreover, by employing the space-fractional
diffusion equation the first-passage time PDF and the survival probability of
extremal two-sided $\alpha$-stable laws ($1<\alpha<2$ with $\beta=1$) and the
asymptotic of the first-passage PDF of general, asymmetric LFs was investigated
in \cite{AminP2019}.

With respect to the first-passage from a \emph{finite\/} interval a number of
classical results for symmetric $\alpha$-stable process were reported in a
series of papers in the 1950s and 1960s. To name a few, the MFPT of
one-dimensional symmetric ($\beta=0$) Cauchy ($\alpha=1$) processes
\cite{MKac1950}, the MFPT of two-dimensional Brownian motion \cite{FSpitzer1958},
and the MFPT of one-dimensional symmetric $\alpha$-stable process with stability
index $0<\alpha<1$ were studied \cite{JElliott1959}. Moreover, for the case
$0<\alpha\leq2$ and $\beta=0$ the results of the first-passage probability in
one dimension \cite{RMBlumenthal1961}, the MFPT as well as the second moment of
the first-passage time PDF in $N$ dimensions were reported \cite{RKGetoor1961}.
One-sided $\alpha$-stable processes with $0<\alpha<1$ and $\beta=1$ in
a finite interval were studied with the help of arc-sine laws of renewal theory
in \cite{EBDynkin1961} and by using the harmonic measure of a Markov process in
\cite{NIkeda1962}. A closed form for the MFPT by potential theory method was
obtained in \cite{SCPort1970}. For completely asymmetric LFs the first-passage
time of the two-sided exit problem was addressed in \cite{SCPort1970,LTakacs1966,
NHBingham1975,JBertoin1996,ALambert2000,FAvram2004}. The residual MFPT of LFs in
a one-dimensional domain was investigated in \cite{vincent}. We also mention that
necessary and sufficient conditions for the finiteness of the moments of the
first-passage time PDF of a general class of L\'evy processes in terms of the
characteristics of the random process $X(t)$ were shown by \cite{RADoney2004}.
Additionally, harmonic functions in a Markovian setting were defined by the
mean value property concerning the distribution of the process being stopped
at the first exit time of a domain \cite{KBogdan2009}. Finally, the authors in
\cite{CProfeta2016}, by using the Green's function of a L\'evy stable process
\cite{AEKyprianou2014}, obtained the non-negative harmonic functions for the
stable process killed outside a finite interval, allowing the computation of
the MFPT.

We also mention that various problems of the first-passage for symmetric and
asymmetric $\alpha$-stable processes, as well as for two- and three-dimensional
motions, were considered by different approaches. These include Monte-Carlo
simulations and the Fredholm integral equation \cite{SVBuldyrev2001-1,
SVBuldyrev2001-2}, Langevin dynamics simulations \cite{BDybiec2006,BDybiec2017},
fractional Laplacian operators \cite{AZoia2007,ZQChen2010}, eigenvalues of the
fractional Laplacian \cite{EKatzav2008}, and the backward fractional Fokker-Plank
equation \cite{AADubkov2009}. Moreover, noteworthy are simulations of radial LFs
in two dimensions \cite{MVahabi2013}, the effect of L\'evy noise on a gene
transcriptional regulatory system \cite{YXu2013}, the study of the mean exit time
and the escape probability of one- and two-dimensional stochastic dynamical systems
with non-Gaussian noises \cite{JDuan2014, JDuan2015, JDuan2018}. The tail
distribution of the first-exit time of LFs from a closed $N$-ball of radius
$R$ in a recursive manner was constructed in \cite{YKim2015}. Very recently,
extensive simulations of the space-fractional diffusion equation and the
Langevin equation were used to investigate the first-passage properties of
asymmetric LFs in a semi-infinite domain in \cite{AminP2019}. In the
same work application of the Skorokhod theorem allowed to derive a closed form
for the first-passage time PDF of extremal two-sided $\alpha$-stable laws with
stability index $1<\alpha<2$ and skewness $\beta=\pm1$, as well as the
first-passage time PDF asymptotic for asymmetric L\'evy stable laws with
arbitrary skewness parameter $\beta$.

The first part of this paper, based on our previous results in \cite{AminP2019},
is devoted to the study of fractional order moments of the first-passage time PDF
of LFs in a semi-infinite domain for symmetric ($0<\alpha<2$ with $\beta=0$),
one-sided ($0<\alpha<1$ with $\beta=1$), extremal two-sided ($1<\alpha<2$ with
$\beta=\pm1$), and a general form ($\alpha\in(0,2]$ with $\beta\in[-1,1]$,
excluding $\alpha=1$ with $\beta\ne0$) $\alpha$-stable laws. Specifically we
obtain a closed-form solution for the fractional moments of the first-passage
time PDF for one-sided and extremal two-sided $\alpha$-stable processes, and we
report the conditions for the finiteness of the fractional moments of the
first-passage time PDF for the full class of $\alpha$-stable processes. We
also present comparisons with numerical solutions of the space-fractional
diffusion equation. In the second part we derive a closed form of the MFPT of
asymmetric LFs in a finite interval by solving the fractional differential
equation for the moments of the first-passage time PDF. In particular cases
we present a comparison between our analytical results with the numerical
solution of the space-fractional diffusion equation as well as simulations
of the Langevin equation. Moreover, we show that the MFPT of LFs in a finite
interval is representative for the first-passage time PDF by analysing the
associated coefficient of variation.

The structure of the paper is as follows. In section \ref{SFDE-Alpha} we
introduce the space-fractional diffusion equation in a finite interval. In
section \ref{NumSch}, the numerical schemes for the space-fractional diffusion
equation and the Langevin equation are presented. We set up the corresponding
formalism to study the moments of the first-passage time PDF in section
\ref{frac-FPT}. Section \ref{mean semi-infi} then presents the analytic
and numerical results of the fractional moments of the first-passage time
PDF for symmetric, one-sided, and extremal two-sided stable distributions
in semi-infinite domains. We derived a closed-form solution of the MFPT for
asymmetric LFs in a finite interval in section \ref{mean bound} and compare
with the numerical solution of the space-fractional diffusion equation and the
Langevin dynamics simulations.
We draw our conclusions in section \ref{concl},
and details of the mathematical derivations are presented in the appendices.

\section{Space-fractional diffusion equation in a finite domain}
\label{SFDE-Alpha}

Fractional derivatives have been shown to be convenient when formulating the
generalised continuum diffusion equations for continuous time random walk
processes with asymptotic power-law asymptotes for both the distributions of
sojourn times and jump lengths \cite{RMetzler2000,RMetzler2004,compte,mebakla1,
mebakla2}. We here use the space-fractional diffusion equation for infinite
domains and its extension to semi-infinite and finite domains to describe the
dynamics of LFs. From a probabilistic point of view, the basic Caputo and
Riemann-Liouville derivatives of order $\alpha\in(0,2)$ can be viewed as
generators of LFs interrupted on crossing a boundary \cite{jpa2003,TKoren2007,
Kolokoltsov2015}. The corresponding equation to describe LFs has the following
expression for the PDF $P_{\alpha,\beta}(x,t|x_0)$
\begin{equation}
\label{eq:ffpe}
\frac{\partial P_{\alpha,\beta}(x,t|x_0)}{\partial t}=K_{\alpha}\,D^{\alpha}_x
P_{\alpha,\beta}(x,t|x_0)
\end{equation}
with initial condition $P_{\alpha,\beta}(x,0|x_0)=\delta(x-x_0)$, where $D_x^{
\alpha}$ is the space-fractional operator for motion confined to the interval
$[-L, L]$,
\begin{equation}
\label{eq:fraccoef}
D^{\alpha}_xf(x)=L_{\alpha,\beta}\,\, _{-L}D_x^{\alpha}
f(x)+R_{\alpha,\beta}\,\, _xD_L^{\alpha}f(x).
 \end{equation}
Here $_{-L}D_x^{\alpha}$ and $_xD_L^{\alpha}$ are left and right
space-fractional derivatives, respectively. Let us first consider the case
$\alpha\neq1$ and $-1\leq\beta\leq1$. We use the Caputo form of the
fractional operators defined by $(n-1<\alpha <n)$ as \cite{Podlubny1999}
\begin{equation}
\label{eq:lcapu}
_{-L}D_x^{\alpha}f(x)=\frac{1}{\Gamma(n-\alpha)}\int\limits_{-L}^{x}\frac{f^{(n)}
(\zeta)}{(x-\zeta)^{\alpha-n+1}}\,\mathrm{d}\zeta,
\end{equation}
and
\begin{equation}
\label{eq:rcapu}
_xD_L^\alpha f(x)=\frac{(-1)^n}{\Gamma(n-\alpha)}\int\limits_{x}^{L}\frac{f^{(n)}
(\zeta)}{(\zeta-x)^{\alpha-n+1}}\,\mathrm{d}\zeta.
\end{equation}
$L_{\alpha,\beta}$ and $R_{\alpha,\beta}$ are the left and right weight
coefficients, defined as \cite{DCNegrete2006,DCNegrete2007}
\begin{equation}
L_{\alpha,\beta}=-\frac{1+\beta}{2\cos(\frac{\alpha\pi}{2})},\quad
R_{\alpha,\beta}=-\frac{1-\beta}{2\cos(\frac{\alpha\pi}{2})}.
\label{eq:weightcoeffRL}
\end{equation}
For the case $\alpha=1$ and $\beta=0$ we have $L_{1,0}=R_{1,0}=1/\pi$, and the
left and right space-fractional operators respectively read
\cite{SGSamko1993}
\begin{eqnarray}
\label{eq:alpaLdiffeq}
_{-L}D_x^1f(x)=-\int\limits_{-L}^x\frac{f^{(1)}(\zeta)}{x-\zeta}\mathrm{d}\zeta,\\
_xD_L^1f(x)=\int\limits_x^L\frac{f^{(1)}(\zeta)}{\zeta-x}\mathrm{d}\zeta.
\label{eq:alpaRdiffeq}
\end{eqnarray}
In the present paper, we do not consider the particular case $\alpha=1$, $\beta
\neq0$ since it cannot be described in terms of a space-fractional operator.

We end this section by adding a remark concerning our choice of the Caputo form
of the fractional derivatives (\ref{eq:lcapu}) and (\ref{eq:rcapu}): it is known
that there are different equivalent definitions of the fractional Laplacian operator
in unbounded domains \cite{kwasnicky}, which in general case loose their equivalence
in bounded domains, see, e.g., \cite{hilfer,song,cusimano}. Such ambiguity, however,
does not hold in case of the first passage problem when absorbing boundary
conditions are applied. In this case it is easy to verify that the Riemann-Liouville
derivatives are equivalent to the Caputo derivatives \cite{SGSamko1993,Podlubny1999}.
However, in the general case for bounded domains the use of the Caputo derivative is
preferrable in applied problems for the following reason: the Riemann-Liouville
approach leads to boundary conditions, which do not have known direct physical
interpretation \cite{Podlubny1999}, and thus the left and right Riemann-Liouville
derivatives might be singular at the lower and upper boundaries, respectively, as
discussed in \cite{DCNegrete2006} in detail---a problem circumvented by defining the
fractional derivative in the Caputo sense.

\section{Numerical schemes}
\label{NumSch}

Apart from analytical approaches to be specified below, to determine the moments
of the first-passage time PDF of $\alpha$-stable processes we will employ two
numerical schemes based on the space-fractional diffusion equation and the
Langevin equation for LFs. We here detail their specific implementation.

\subsection{Diffusion description}
\label{Diffdesc}

Numerical methods to solve space-fractional diffusion equations are relatively
sparse, and the majority of the publications are based on the finite-difference
scheme \cite{Jia2015, Shimin2018} and finite-element methods \cite{Deng-Weihua2008,
W. Melean2007,Fix2004} as well as the spectral approach \cite{Bhrawy2016,Li2009}.
In this paper, we use the finite-difference scheme to solve the space-fractional
diffusion equation introduced in the preceding section. Here we only outline the
essence of the method and refer to \cite{AminP2019} for further details. The
computationally most straightforward method arises from the forward-difference
scheme in time on the left hand side of equation (\ref{eq:ffpe}),
\begin{equation}
\label{eq:disctime}
\frac{\partial}{\partial t}f(x_i,t_j)=\frac{f^{j+1}_i-f^j_i}{\Delta t}+\mathcal{O}(
\Delta t),
\end{equation}
where $f^j_i=f(x_i,t_j)$, $x_i=(i-I/2)\Delta x$, and $t_j=j\Delta t$, where $\Delta
x$ and $\Delta t$ are step sizes in position and time, respectively. The $i$ and
$j$ are non-negative integers, $i=0,1,2,\ldots,I$, and $\Delta x=2L/I$. Similarly,
$j=0,1,2,\ldots,J-1$, $t_0=0$, $t_J=t$, and $\Delta t=t/J$. Absorbing boundary
conditions for the determination of the first-passage events imply $f_0^j=f_I^j=0$
for all $j$. The integrals on the right hand side of equation (\ref{eq:ffpe}) are
discretised as follows. For $0<\alpha<1$,
\begin{equation}
\label{eq:discleft1}
\int\limits_{-L}^{x_{i}}\frac{f^{(1)}(\zeta,t_{j})}{(x_{i}-\zeta)^{\alpha}}
\mathrm{d}\zeta=\displaystyle\sum_{k=1}^i\frac{f^j_k-f^j_{k-1}}{\Delta x}\int
\limits_{x_{k-1}}^{x_k}\frac{1}{(x_{i}-\zeta)^{\alpha}}\mathrm{d}\zeta+\mathcal{O}
(\Delta x^{2-\alpha})
\end{equation}
for the left derivative, and
\begin{equation}
\label{eq:discright1}
\int\limits_{x_i}^L\frac{f^{(1)}(\zeta,t_{j})}{(\zeta-x_i)^{\alpha}}\mathrm{d}
\zeta=\displaystyle\sum_{k=i}^{I-1}\frac{f^j_{k+1}-f^j_k}{ \Delta x}\int\limits_{
x_k}^{x_{k+1}}\frac{1}{(\zeta-x_i)^{\alpha}}\mathrm{d}\zeta+\mathcal{O}(\Delta
x^{2-\alpha})
\end{equation}
for the right derivative. This scheme is called L1 scheme and is an efficient
way to approximate the Caputo derivative of order $0<\alpha<1$ \cite{KBOldham1974,
TAMLanglands2005,CLi2012} with error estimate $\mathcal{O}(\Delta x^{2-\alpha})$.
For the case $1<\alpha<2$ the suitable method to discretise the Caputo derivative
is the L2 scheme \cite{KBOldham1974,CLi2012,VELynch2003}, namely,
\begin{equation}
\label{eq:discleft2}
\fl\int\limits_{-L}^{x_i}\frac{f^{(2)}(\zeta,t_j)}{(x_i-\zeta)^{\alpha-1}}\mathrm{
d}\zeta=\displaystyle\sum_{k=1}^i\frac{f^j_{k+1}-2f^j_k+f^j_{k-1}}{(\Delta x)^2}
\int\limits_{x_{k-1}}^{x_k}\frac{1}{(x_i-\zeta)^{\alpha-1}}\mathrm{d}\zeta+
\mathcal{O}(\Delta x)
\end{equation}
for the left derivative, and
\begin{equation}
\label{eq:discright2}
\fl\int\limits_{x_i}^L\frac{f^{(2)}(\zeta,t_j)}{(\zeta-x_i)^{\alpha-1}}\mathrm{d}
\zeta=\displaystyle\sum_{k=i}^{I-1}\frac{f^j_{k+1}-2f^j_k+f^j_{k-1}}{(\Delta x)^2}
\int\limits_{x_k}^{x_{k+1}}\frac{1}{(\zeta-x_i)^{\alpha-1}}\mathrm{d}\zeta+
\mathcal{O}(\Delta x)
\end{equation}
for the right derivative. We note that the truncation error of the L2 scheme is
$\mathcal{O}(\Delta x)$ \cite{VELynch2003,ESousa2010}. For the special case
$\alpha=1$ and $\beta=0$ we approximate the derivative in space with the backward
difference scheme
\begin{equation}
\label{eq:discleftalpa1}
\int\limits_{-L}^{x_i}\frac{f^{(1)}(\zeta,t_j)}{x_i-\zeta}\mathrm{d}\zeta=
\displaystyle\sum_{k=1}^i\frac{f^j_k-f^j_{k-1}}{\Delta x}\frac{2}{2(i-k)+1}
+\mathcal{O}(\Delta x^2)
\end{equation}
for the left derivative, and with a forward difference scheme
\begin{equation}
\label{eq:discrightalpa1}
\int\limits_{x_i}^L\frac{f^{(1)}(\zeta,t_j)}{\zeta-x_i}\mathrm{d}\zeta=
\displaystyle\sum_{k=i}^{I-1}\frac{f^j_k-f^j_{k+1}}{\Delta x}\frac{2}{2(k-i)+1}
+\mathcal{O}(\Delta x^2)
\end{equation}
for the right derivative. We note that here the truncation error is the
order $\mathcal{O}(\Delta x^{2})$. By substitution of equations (\ref{eq:disctime})
to (\ref{eq:discrightalpa1}) into (\ref{eq:ffpe}) we obtain
\begin{equation}
\label{eq:matrixab}
\mathbf{A}f^{j+1}=\mathbf{B}f^j,
\end{equation}
where the coefficients $\mathbf{A}$ and $\mathbf{B}$ have matrix form of
dimension $(I+1)\times(I+1)$ and $j=0,1,2,\ldots,J-1$. In the numerical
scheme for the setup used in our numerical simulations (see section
\ref{frac-FPT} and figure \ref{fig:fig1} below) the initial condition
$f(x,0)=\delta(x-x_0)$ at $x_0=L-d$ is approximated as
\begin{equation}
f(x_i,0)=\left\{\begin{array}{ll}(\Delta x)^{-1},&i=(2L-d)/\Delta x\\
0,& \mbox{otherwise}\end{array}\right..
\end{equation}
In the next step, the time evolution of the PDF is obtained by applying the
absorbing boundary conditions $f_0^j=f_I^j=0$ for all $j$.

\subsection{Langevin dynamics}
\label{Lange}

The fractional diffusion equation (\ref{eq:ffpe}) can be related to the LF
Langevin equation \cite{Dybiec2016,HCFogedby1994,SJespersen1999}
\begin{equation}
\label{eq:langevin}
\frac{\mathrm{d}}{\mathrm{d}t}x(t)=K_\alpha^{1/\alpha}\zeta(t),
\end{equation}
where $\zeta(t)$ is L\'evy noise characterised by the same $\alpha$ and
$\beta$ parameters as the space-fractional operator (\ref{eq:ffpe}) and with
unit scale parameter. The Langevin equation (\ref{eq:langevin}) provides
a microscopic (trajectory-wise) representation of the space-fractional
diffusion equation~(\ref{eq:ffpe}). Therefore, from an ensemble of
trajectories generated from equation (\ref{eq:langevin}), it is possible to
estimate the time-dependent PDF whose evolution is described by equation
(\ref{eq:ffpe}). In numerical simulations, LFs can be described by the
discretised form of Langevin equation
\begin{equation}
\label{eq:discrete}
x(t+\Delta t)=x(t)+K_\alpha^{1/\alpha}(\Delta t)^{1/\alpha}\zeta_t,
\end{equation}
where $\zeta_t$ stands for the sequence of i.i.d. $\alpha$-stable random
variables with unit scale parameter \cite{Samorodnitsky-Taqqu,janicki1994}
and identical index of stability $\alpha$ and skewness $\beta$ as in equation
(\ref{eq:langevin}). Relation (\ref{eq:discrete}) is exactly the Euler-Maruyama
approximation \cite{janicki1996,kloeden2011numerical,maruyama1955continuous} to
a general $\alpha$-stable L\'evy process.

From the trajectories $x(t)$, see equations (\ref{eq:langevin}) and
(\ref{eq:discrete}), it is also possible to estimate the first-passage time
$\tau$ as
\begin{equation}
\tau=\min\{t:|x(t)|\geqslant L\}.
\end{equation}
From the ensemble of first-passage times, it is then possible to obtain the
survival probability $S(t)$, which is the complementary cumulative density of
first-passage times. More precisely, the initial condition is $S(0)=1$, and
at every recorded first-passage event at time $\tau_i$, $S(t)$ is decreased
by the amount $1/N$ where $N$ is the overall number of recorded first-passage
events.

\section{First passage time properties of $\alpha$-stable processes}
\label{frac-FPT}

\begin{figure}
\centering
\includegraphics[width=0.80\textwidth]{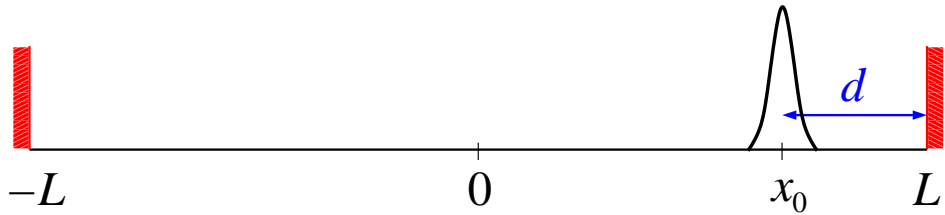}
\caption{Schematic of our setup. In the interval of length $2L$ the initial
condition is given by a $\delta$-distribution located at $x_0$, which is
chosen the distance $d$ away from the right boundary. At both interval
boundaries we implement absorbing boundary conditions, that is, when the
particle hits the boundaries or attempts to move beyond them, it is absorbed.}
\label{fig:fig1}
\end{figure}

For an $\alpha$-stable random process, the survival probability and the
first-passage time are observable statistical quantities characterising the
stochastic motion in bounded domains with absorbing boundary conditions. In
the following, we investigate the properties of the first-passage time
moments in a semi-infinite and finite interval for symmetric and asymmetric
$\alpha$-stable laws underlying the space-fractional diffusion equation. In
addition a comparison with the Langevin approach and with analytical expressions
for the MFPT of LFs in a finite interval is presented. To this end, we use the
setup shown in figure \ref{fig:fig1}, in which the absorbing boundaries are
located at $-L$ and $L$, and the centre point of the initial $\delta$-distribution
is located the distance $d$ away from the right boundary.

The survival probability that up until time $t$ a random walker remains "alive"
within the interval $[-L, L]$ is defined as \cite{SRedner2001,SidneyRedner2014}
\begin{equation}
\label{eq:Survival}
S(t|x_0)=\int\limits_{-L}^{L}P_{\alpha,\beta}(x,t|x_0)\mathrm{d}x,
\end{equation}
Recall that $P_{\alpha,\beta}(x,t|x_0)$ is the PDF of an LF confined to the
interval $[-L,L]$ which starts at $x_0$. The associated first-passage time PDF
reads
\begin{equation}
\label{eq:firstpassage}
\wp(t|x_0)=-\frac{\mathrm{d}S(t|x_0)}{\mathrm{d}t} .
\end{equation}
The first-passage time PDF satisfies in particular the normalisation
\begin{equation}
\int\limits_0^{\infty}\wp(t|x_0)\,\mathrm{d}t=1,
\end{equation}
and the positive integer moments of this random variable are defined as
\begin{equation}
\label{eq:FPT-moments}
\langle\tau^m\rangle(x_0)=\int\limits_0^{\infty}t^m\wp(t|x_0)\mathrm{d}t=
\int\limits_{0}^{\infty}m t^{m-1}S(t|x_0)\mathrm{d}t,\,\,\,\, m=1,2,\dots.
\end{equation}
Employing the Laplace transform,
\begin{equation}
f(t)\div\mathscr{L}\{f(t);s\}=\int\limits_0^{\infty}\e^{-st}f(t)\mathrm{d}t,
\end{equation}
we obtain
\begin{equation}
\label{eq:FPT-moments-Lapl}
\langle\tau^m\rangle(x_0)=(-1)^m\frac{\partial^m}{\partial s^m}\Big[\wp(s|x_0)
\Big]\Big|_{s=0}.
\end{equation}
Conversely, following the procedure suggested in \cite{AZoia2007}, by substitution
of equation (\ref{eq:Survival}) into equation (\ref{eq:FPT-moments}) we get
\begin{equation}
\label{eq:intorderFPT}
\langle \tau^m\rangle(x_0)=\int\limits_0^{\infty}mt^{m-1}\int\limits_{-L}^L
P_{\alpha,\beta}(x,t|x_0)\mathrm{d}x\mathrm{d}t.
\end{equation}
Applying the backward space-fractional Kolmogorov operator $D^{\alpha}_{
x_0}$ in a finite domain\footnote{More precisely, $D^{\alpha}_{x_0}$ is the
generator of LFs killed upon leaving the domain.} (see details in
\ref{DX0derivation}),
\begin{equation}
\label{eq:fraccoefx0}
D^{\alpha}_{x_0}f(x_0)=R_{\alpha,\beta}\, _{-L}D_{x_0}^{\alpha}f(x_0)+L_{\alpha,
\beta}\, _{x_0}D_L^{\alpha}f(x_0),
 \end{equation}
to both sides of equation (\ref{eq:intorderFPT}),
\begin{equation}
D^{\alpha}_{x_0}\langle\tau^m\rangle(x_0)=\int\limits_0^{\infty}m t^{m-1}\int
\limits_{-L}^LD^{\alpha}_{x_0}P_{\alpha,\beta}(x,t|x_0)\mathrm{d}x\mathrm{d}t,
\end{equation}
and using the corresponding backward Kolmogorov equation
\begin{equation}
\frac{\partial P_{\alpha,\beta}(x,t|x_0)}{\partial t}=K_{\alpha}\,D^{\alpha}_{
x_0}P_{\alpha,\beta}(x,t|x_0),
\end{equation}
we get
\begin{equation}
D^{\alpha}_{x_0}\langle\tau^m\rangle(x_0)=\frac{m}{K_\alpha}\int\limits_0^{
\infty}t^{m-1}\frac{\partial}{\partial t}\int\limits_{-L}^LP_{\alpha,\beta}
(x,t|x_0)\mathrm{d}x\mathrm{d}t.
\end{equation}
In the limit $m=1$,
\begin{equation}
D^{\alpha}_{x_0}\langle\tau\rangle(x_0)=\frac{1}{K_\alpha}\left(\int\limits_{-L}^L
P_{\alpha,\beta}(x,\infty|x_0)\mathrm{d}x-\int\limits_{-L}^LP_{\alpha,\beta}(x,0|
x_0)\mathrm{d}x\right).
\end{equation}
Then, by including the initial condition of the density function $P_{\alpha,\beta}
(x,0|x_0)=\delta(x-x_0)$, where $x_0\in[-L, L]$, we get the functional relation
\begin{equation}
\label{eq:MFPToperat}
D^{\alpha}_{x_0}\langle\tau\rangle(x_0)=-\frac{1}{K_\alpha}
\end{equation}
for the MFPT. This result is similar to equation (41) in \cite{AZoia2007}, except
that instead of a symmetric Riesz-Feller operator we here employ a more general
form of the fractional derivative operator $D_{x_0}^{\alpha}$ which is called
backward space-fractional Kolmogorov operator in a finite domain. We note that
in comparison with the forward space-fractional derivative defined by equation
(\ref{eq:fraccoef}) in equation (\ref{eq:fraccoefx0}) the left and right
weight coefficients are exchanged.

For the case $m=2$, we have
\begin{equation}
D^{\alpha}_{x_0}\langle\tau^2\rangle(x_0)=\frac{2}{K_\alpha}\int\limits_0^{
\infty}t\frac{\partial}{\partial t}\int\limits_{-L}^LP_{\alpha,\beta}(x,t|x_0)
\mathrm{d}x\mathrm{d}t.
\end{equation}
Changing the order of integration,
\begin{equation}
D^{\alpha}_{x_0}\langle\tau^2\rangle(x_0)=\frac{2}{K_\alpha}\int\limits_{-L}^L
\int\limits_0^{\infty}t\frac{\partial}{\partial t}P_{\alpha,\beta}(x,t|x_0)
\mathrm{d}t\mathrm{d}x,
\end{equation}
integrating by parts in the inner integral,
\begin{equation}
D^{\alpha}_{x_0}\langle\tau^2\rangle(x_0)=-\frac{2}{K_\alpha}\int\limits_{-L}^L
\int\limits_0^{\infty}P_{\alpha,\beta}(x,t|x_0)\mathrm{d}t\mathrm{d}x,
\end{equation}
and, once again, changing the order of integration, we find
\begin{equation}
D^{\alpha}_{x_0}\langle\tau^2\rangle(x_0)=-\frac{2}{K_\alpha}\int\limits_0^{
\infty}\int\limits_{-L}^LP_{\alpha,\beta}(x,t|x_0)\mathrm{d}x\mathrm{d}t.
\end{equation}
Calling on equation (\ref{eq:intorderFPT}) with $m=1$, we obtain the functional
relation
\begin{equation}
\label{eq:socondmfptoperat}
D^{\alpha}_{x_0}\langle\tau^2\rangle(x_0)=-\frac{2}{K_\alpha}\langle\tau\rangle(x_0)
\end{equation}
for the second moment of the first-passage time PDF.

More generally, by using this recursion relation one can write
\begin{equation}
\label{eq:frac-intorderFPT}
D^{\alpha}_{x_0}\langle\tau^m\rangle(x_0)=-\frac{m}{K_\alpha}\langle\tau^{m-1}
\rangle(x_0),\,\,\, m=1,2,\ldots.
\end{equation}
By applying $D^{\alpha}_{x_0}$ on both sides,
\begin{equation}
(D^{\alpha}_{x_0})^{2}\langle\tau^m\rangle(x_0)=-\frac{m}{K_\alpha}D^{\alpha}_{
x_0}\langle\tau^{m-1}\rangle(x_0),
\end{equation}
and with equation (\ref{eq:frac-intorderFPT}) we have
\begin{equation}
(D^{\alpha}_{x_0})^2\langle\tau^m\rangle(x_0)=\frac{m(m-1)}{{K_\alpha}^2}\langle
\tau^{m-2}\rangle(x_0).
\end{equation}
By repeating this procedure, we derive
\begin{equation}
\label{eq:Zoiamethodmoments}
(D^{\alpha}_{x_0})^m\langle\tau^m\rangle(x_0)=\frac{(-1)^m\Gamma(1+m)}{{K_\alpha}^m}.
\end{equation}
This equation is the generalisation of the result obtained in \cite{AZoia2007}
for symmetric LFs (see equation (44) there).

\section{First passage time properties of LFs in a semi-infinite domain}
\label{mean semi-infi}

In this section, we investigate the first-passage time properties of LFs
in a semi-infinite domain. The motion starts at $x_0$, and the boundary
is located at $x=L$, in such a way that in our setup $L-x_0=d$. In order to
reproduce numerically the results for semi-infinite domain with the scenario
shown in figure \ref{fig:fig1}, we employ $L$ as well as $x_0$, as large as
possible in order to allow a constant $d$ ($L=10^{12}$ in our simulations).

\subsection{Symmetric LFs in a semi-infinite domain}
\label{mean sym semi-infi}

For a semi-infinite domain with an absorbing boundary condition, as said
above it is well known that the first-passage time density for any symmetric
jump length distribution in a Markovian setting has the universal Sparre
Andersen scaling $\wp(t) \simeq t^{-3/2}$ \cite{SRedner2001,ESparre1953,
ESparre1954}. In the theory of a general class of L\'evy processes, that is,
homogeneous random processes with independent increments, there exists a
theorem, that provides an analytical expression for the PDF of first-passage
times in a semi-infinite interval, often referred to as the Skorokhod theorem
\cite{Gikhman-Skorokhod1975,Skorokhod1964}. Based on this theorem the asymptotic
expression for the first-passage time PDF of symmetric $\alpha$-stable laws is
\cite{TKoren2007}
\begin{equation}
\label{eq:symFPT-PDF}
\wp(t)\sim\frac{d^{\alpha/2}}{\alpha\sqrt{\pi K_{\alpha}}\Gamma{(\alpha/2)}}t^{
-3/2},
\end{equation}
which specifies an exact expression for the prefactor in the Sparre Andersen
scaling law. The existence of this long-time tail leads to the divergence of
the MFPT ($\langle\tau\rangle$ in equation (\ref{eq:FPT-moments})). This
means that the LF will eventually cross the boundary $d$ with unit probability,
but the expected time that this takes is infinite. For Brownian motion ($\alpha
=2$), the PDF for the first-passage time has the well known L\'evy-Smirnov form
\cite{Feller1971}
\begin{equation}
\label{eq:LevySmirnovFPT}
\wp(t)=\frac{d}{\sqrt{4\pi K_2t^3}}\exp{\left(-\frac{d^{2}}{4 K_{2}t}\right)},
\end{equation}
which is exact for all times \cite{Feller1971,SRedner2001} and whose asymptote
coincides with result (\ref{eq:symFPT-PDF}) for the appropriate limit $\alpha=2$.

For the moments of Brownian motion ($\alpha=2$) we have
\begin{equation}
\langle\tau^q\rangle=\int\limits_0^{\infty}\frac{t^qd}{\sqrt{4\pi K_2t^3}}\exp{
\left(-\frac{d^2}{4K_2t}\right)}\mathrm{d}t,
\end{equation}
where by change of variables $u=d^2/4K_2t$ and using the integral form of the
Gamma function,
\begin{equation}
\label{eq:GammaintP}
\Gamma(z)=\int\limits_0^{\infty}\zeta^{z-1}\e^{-\zeta}\mathrm{d}\zeta,\,\,\,
\mathrm{Re}(z)>0,
\end{equation}
we get (see page 84 in \cite{SRedner2001})
\begin{equation}
\langle\tau^q\rangle=\frac{\Gamma(\frac{1}{2}-q)}{2^{2q}\sqrt{\pi}}\frac{d^{2q}}{
K_2^q}=\frac{\Gamma(1-2q)}{\Gamma(1-q)}\frac{d^{2q}}{K_2^q},\,\,\,-\infty<q<1/2.
\label{eq:Brown-Semi-Moment}
\end{equation}
In the last step we used the duplication rule $2^{2z}\Gamma(z)\Gamma(z+1/2)=2
\sqrt{\pi}\Gamma(2z)$.

\begin{figure}
\centering
\includegraphics[width=0.49\textwidth]{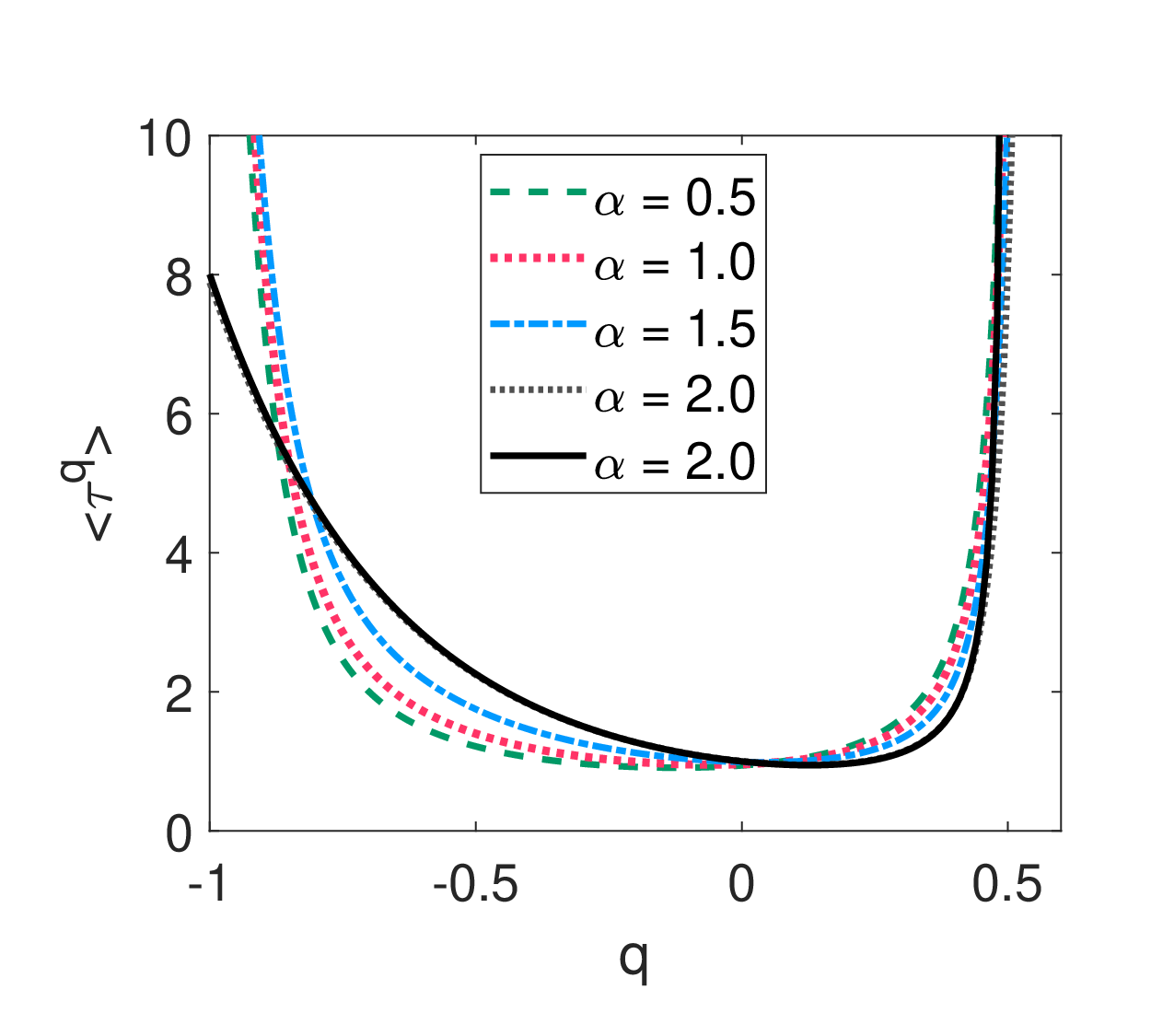}
\includegraphics[width=0.49\textwidth]{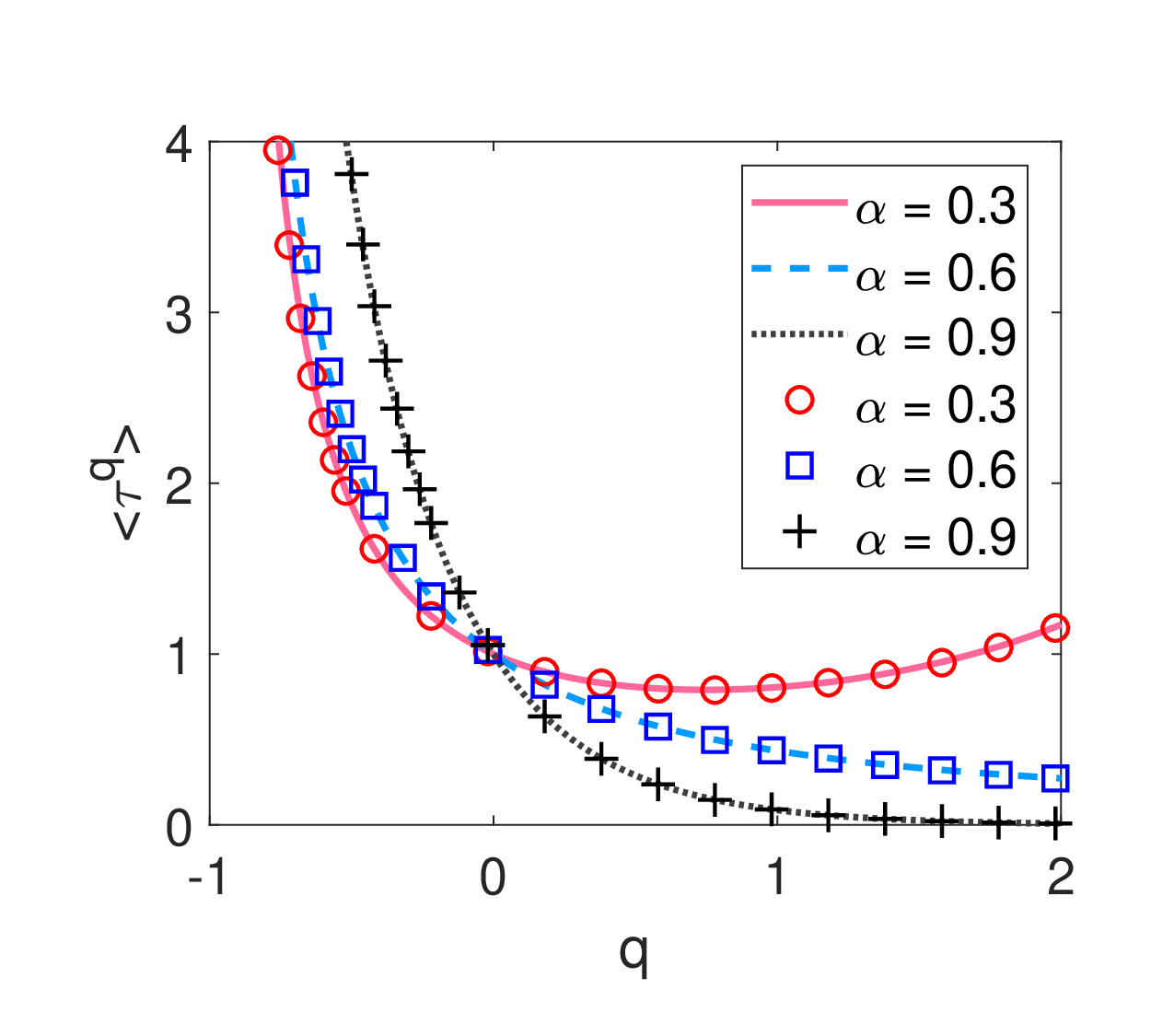}\\
\includegraphics[width=0.49\textwidth]{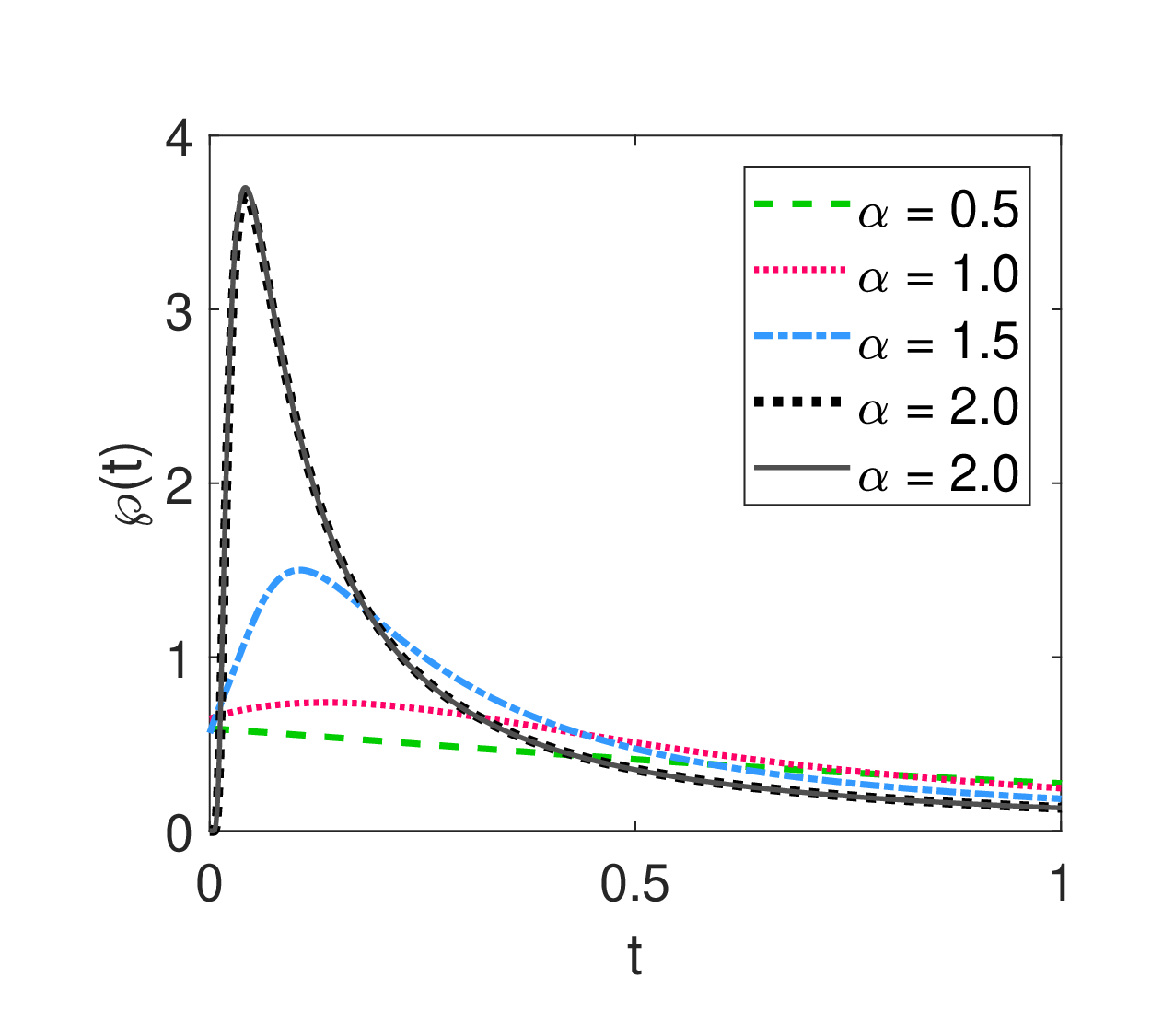}
\includegraphics[width=0.49\textwidth]{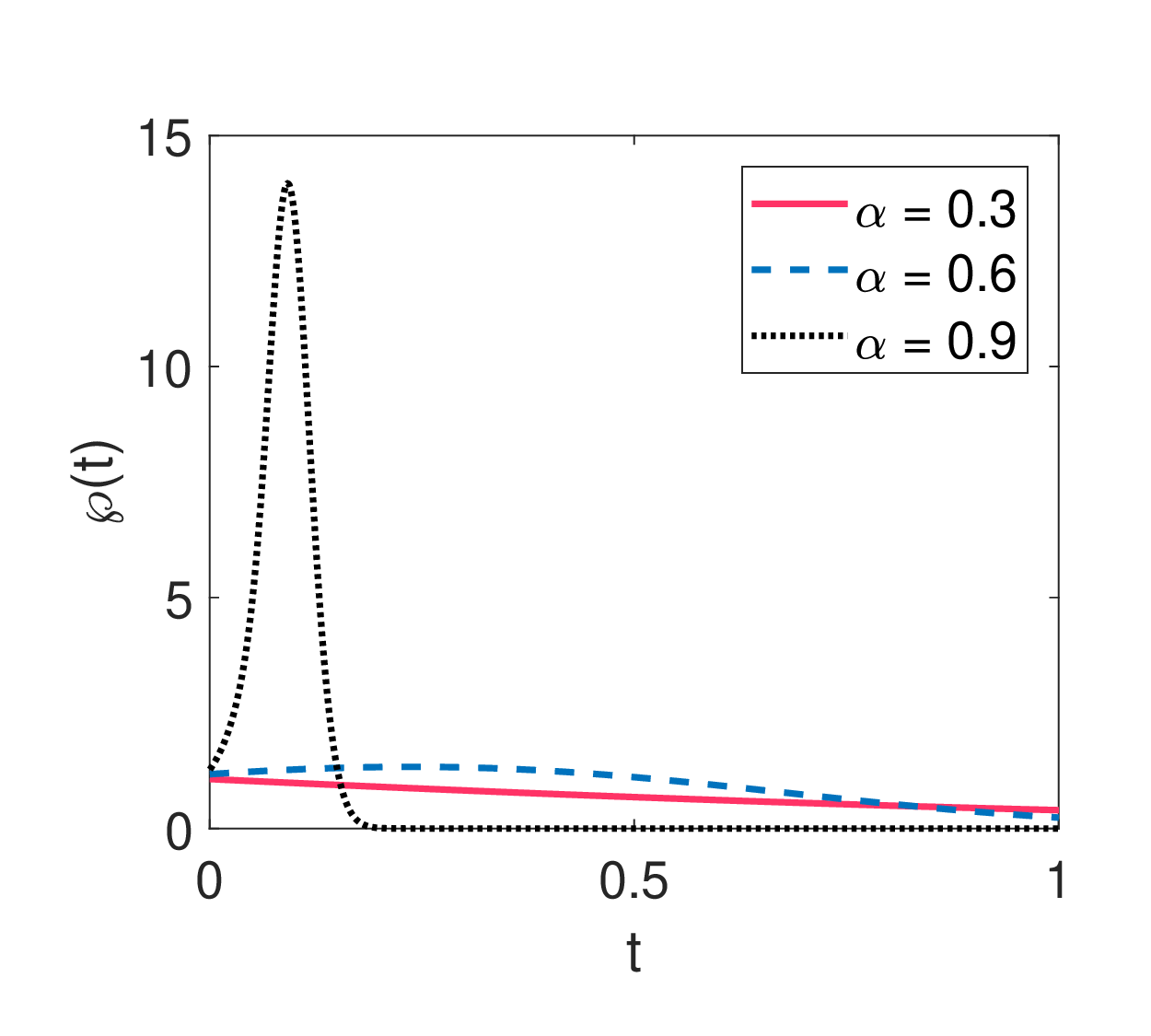}
\caption{Top left: fractional order moments of the first-passage time PDF for
symmetric ($0<\alpha\leq2$, $\beta=0$) $\alpha$-stable laws in a semi-infinite
domain with $d=0.5$. Here and in the following we set $K_{\alpha}=1$.
Results are shown for the case when $L$ is sufficiently
large (here we used $L=10^{12}$). Dashed lines represent the numerical solution
of the space-fractional diffusion equation and the solid line shows the analytical
result (\ref{eq:Brown-Semi-Moment}) for Brownian motion. Top right: fractional
order moments of the first-passage time PDF for one-sided ($0<\alpha<1$, $\beta
=1$) $\alpha$-stable laws in a semi-infinite domain. Symbols show the numerical
solution of the space-fractional diffusion equation with $d=0.5$, $\Delta
x=0.01$, and $\Delta t=0.001$, and lines represent the analytic result
(\ref{eq:mfptoneside}). Bottom left: first-passage time PDF of symmetric
$\alpha$-stable laws with $0<\alpha\leq2$ and $\beta=0$. Lines correspond to
the numerical solution of the space-fractional diffusion equation and the solid
line shows result (\ref{eq:LevySmirnovFPT}). Bottom right: first-passage time
PDF of one-sided ($0<\alpha<1$ with $\beta=1$) $\alpha$-stable laws obtained by
numerical solution of the space-fractional diffusion equation.}
\label{fig:fig2}
\end{figure}

To find a closed form of the first-passage time PDF of LFs based on general
symmetric $\alpha$-stable probability laws ($0<\alpha<2$) remains an unsolved
problem. We show the short time behaviour for symmetric LFs in figure
\ref{fig:fig2}, bottom left panel. As can be seen, only for the case of
Brownian motion ($\alpha=2$) the PDF has value zero at $t=0$, while for LFs
with $\alpha<2$ the first-passage time PDF exhibits a non-zero value at $t=0$,
thus demonstrating that LFs can instantly cross the boundary with their first
jump away from their initial position $x_0$. The magnitude of $\wp(t\to0)$ can
be estimated from the survival probability, as shown by equations (3) and (A.5)
in \cite{VVPalyulin2019} for symmetric LFs and here by equation (\ref{limit-pdf})
in section \ref{semi-inf-Gen} below for asymmetric LFs with
$\alpha\in(0,2]$ and $\beta\in(-1,1]$ (excluding $\alpha=1$ with $\beta\neq0$).
Of course, in the case of symmetric LFs ($\beta=0$) equation (\ref{limit-pdf})
coincides with equation (3) in \cite{VVPalyulin2019}. The values of the first-passage
time PDF at $t=0$ obtained by numerical solution of the space-fractional diffusion
equation are in perfect agreement with those obtained from equation (\ref{limit-pdf}). 
Fractional moments of the first-passage time PDF for symmetric $\alpha$-stable laws
in a semi-infinite domain for different ranges of the stability index $\alpha$ are
shown in the top left panel of figure \ref{fig:fig2}. As can be seen the
fractional moments are finite only for $-1<q<1/2$, as expected from the Sparre
Andersen universal scaling with exponent $3/2$.

\subsection{Asymmetric LFs in a semi-infinite domain}
\label{mean Asym semi-infi}

\subsubsection{One-sided $\alpha$-stable processes with $0<\alpha<1$
and $\beta=1$.}
\label{moment semi-infi one-sided}

By applying the Skorokhod theorem, it can be shown that the first-passage time
PDF of one-sided $\alpha$-stable laws has the exact form \cite{TKoren2007}
\begin{equation}
\label{eq:fptoneside}
\wp(t)=\frac{\xi}{d^\alpha}M_{\alpha}\left(\frac{\xi t}{d^\alpha}\right)
\end{equation}
with
\begin{equation}
\label{eq:xiparameter}
\xi=\frac{K_{\alpha}}{\cos{(\alpha\pi(\rho-1/2))}}, \,\,\,
\rho=\frac{1}{2}+\frac{1}{\alpha\pi}\arctan{(\beta\tan{(\alpha\pi/2)})},
\end{equation}
which connects to our case here via
\begin{equation}
\xi=\frac{K_{\alpha}}{\cos{(\alpha\pi/2)}}, \,\,\,
\rho=1.
\end{equation}
Here $M_{\alpha}(z)$ is the Wright $M$-function \cite{Podlubny1999,FMainardi2010}
(also sometimes called Mainardi function) with the integral representation
\cite{FMainardi2010} (page 241)
\begin{equation}
\label{eq:intMfun}
M_\alpha(z)=\frac{1}{2\pi i}\int\limits_{Ha}e^{\sigma-z\sigma^{\alpha}}\frac{
\mathrm{d}\sigma}{\sigma^{1-\alpha}},\,\,\,z\in\mathbb{C}, \,\,\,0<\alpha<1,
\end{equation}
where the contour of integration Ha (the Hankel path) is the loop starting and
ending at $-\infty$ and encircling the disk $|\sigma|\leq|z|^{1/\alpha}$
counterclockwise, i.e., $|\arg(\sigma)|\leq\pi$ on Ha. Here and below for the
asymptotic behavior of the first-passage time PDFs we refer the reader to our
recent paper \cite{AminP2019}. The asymptotics of the $M$-function at short and
long times is presented in appendix E of \cite{AminP2019}. The long-time asymptotics
of the PDF (\ref{eq:fptoneside}) is given by equations (31) and (32) of
\cite{AminP2019}, while the short-time asymptotics of (\ref{eq:fptoneside}) is
given by equation (33) of \cite{AminP2019} (or equivalently, equation
(\ref{limit-pdf}) below with $\rho=1$). By definition
(\ref{eq:FPT-moments}) of the moments of the first-passage time PDF and the
first-passage time PDF (\ref{eq:fptoneside}) of one-sided stable laws, we find
\begin{eqnarray}
\nonumber
\langle\tau^q\rangle&=&\int\limits_0^{\infty}t^q\frac{\xi}{d^\alpha}M_{\alpha}
\left(\frac{\xi t}{d^\alpha}\right)\mathrm{d}t\\
\nonumber
&=&\frac{\xi}{d^\alpha}\int\limits_0^{\infty}t^q\frac{1}{2\pi i}\int\limits_{Ha}
\sigma^{\alpha-1}\e^{\sigma-\xi t( \sigma/d)^\alpha}\,\mathrm{d}\sigma\mathrm{d}t\\
&=&\frac{\xi}{d^\alpha}\frac{1}{2\pi i}\int\limits_{Ha}\sigma^{\alpha-1}\e^{\sigma}
\int\limits_0^{\infty}t^q\e^{-\xi t(\sigma/d)^\alpha}\mathrm{d}t\mathrm{d}\sigma.
\end{eqnarray}
By change of variables $u=\xi t(\sigma/d)^\alpha $ in the inner integral and with
the help of equation (\ref{eq:GammaintP}) we get
\begin{equation}
\langle \tau^{q} \rangle=\frac{d^{q\alpha}\Gamma(1+q)}{\xi^{q}}\frac{1}{2\pi i}
\int\limits_{Ha}\sigma^{-q\alpha-1}\e^{\sigma}\,\mathrm{d}\sigma.
\end{equation}
Using Hankel's contour integral
\begin{equation}
\label{eq:GammaintN}
\frac{1}{\Gamma(z)}=\frac{1}{2\pi i}\int\limits_{Ha}\zeta^{-z}\e^{\zeta}\mathrm{d}
\zeta,\,\,\, z\in\mathbb{C},
\end{equation}
we then obtain the fractional order moments of the first-passage time PDF for
one-sided $\alpha$-stable laws with $0<\alpha<1$ and $\beta=1$,
\begin{equation}
\label{eq:mfptoneside}
\langle\tau^q\rangle=\frac{\Gamma(1+q)}{\Gamma(1+q\alpha)}\frac{d^{q\alpha}}{\xi^q},
\,\,\, q>-1.
\end{equation}
The MFPT ($q=1$) for one-sided $\alpha$-stable process was derived in
\cite{TKoren2007,SCPort1970}. Also, from equation (\ref{eq:FPT-moments-Lapl})
and the Laplace transform of the first-passage time PDF, which has the form
of the Mittag-Leffler function \cite{TKoren2007}, it is possible to find all
moments explicitly. In the right panel of figure \ref{fig:fig2} we show the
results for the fractional order moments of one-sided $\alpha$-stable laws
obtained by numerically solving the space-fractional diffusion equation, along
with the analytical results of equation (\ref{eq:mfptoneside}).

\subsubsection{One-sided $\alpha$-stable processes with $0<\alpha<1$ and
$\beta=-1$.}

One-sided $\alpha$-stable laws with the stability index $0<\alpha<1$ and
skewness parameter $\beta=-1$ satisfy the non-positivity of their increments.
Therefore, the random walker never crosses the right boundary $d$. In the
semi-infinite domain therefore the survival probability remains unity ($S(t)
=1$) and the first-passage time PDF $\wp(t)=0$. Therefore, the fractional moments
read
\begin{equation}
\langle \tau^q\rangle=\left\{\begin{array}{ll}
0,&q< 0\\1,&q=0\\\infty,&q>0\end{array}\right..
\end{equation}
Due to normalisation of the first-passage time PDF, $\langle\tau^q\rangle=1$
when $q=0$.

\subsubsection{Extremal two-sided $\alpha$-stable processes with $1<\alpha<2$
and $\beta=-1$.}
\label{moment semi-infi two-sided}

Stable laws with stability index $1<\alpha<2$ and skewness $\beta=1$ or $\beta=-1$
are called extremal two-sided skewed $\alpha$-stable laws \cite{AVSkorokhod1954}.
Let us first consider the case $1<\alpha<2$, $\beta=-1$. By applying the Skorokhod
theorem it can be shown that the first-passage time PDF of extremal two-sided
$\alpha$-stable laws with $1<\alpha<2$ and $\beta=-1$ has the following exact
form \cite{AminP2019}
\begin{equation}
\label{eq:fptPDFtwobn}
\wp(t)=\frac{t^{-1-1/\alpha}d}{\alpha\xi^{1/\alpha}}M_{1/\alpha}\left(\frac{d}{
(\xi t)^{1/\alpha}}\right),
\end{equation}
in terms of the Wright $M$-function $M_{1/\alpha}$. The long-time asymptotic
of the PDF (\ref{eq:fptPDFtwobn}) is given by equation (41) of \cite{AminP2019}
or, equivalently, equation (\ref{eq:generalasym}) below with $\rho=1/\alpha$.
Respectively, the short-time asymptotic of equation (\ref{eq:fptPDFtwobn}) is
given by equation (39) of \cite{AminP2019}, or by equation (\ref{limit-pdf})
below with $\rho=1/\alpha$.

\begin{figure}
\centering
\includegraphics[width=0.49\textwidth]{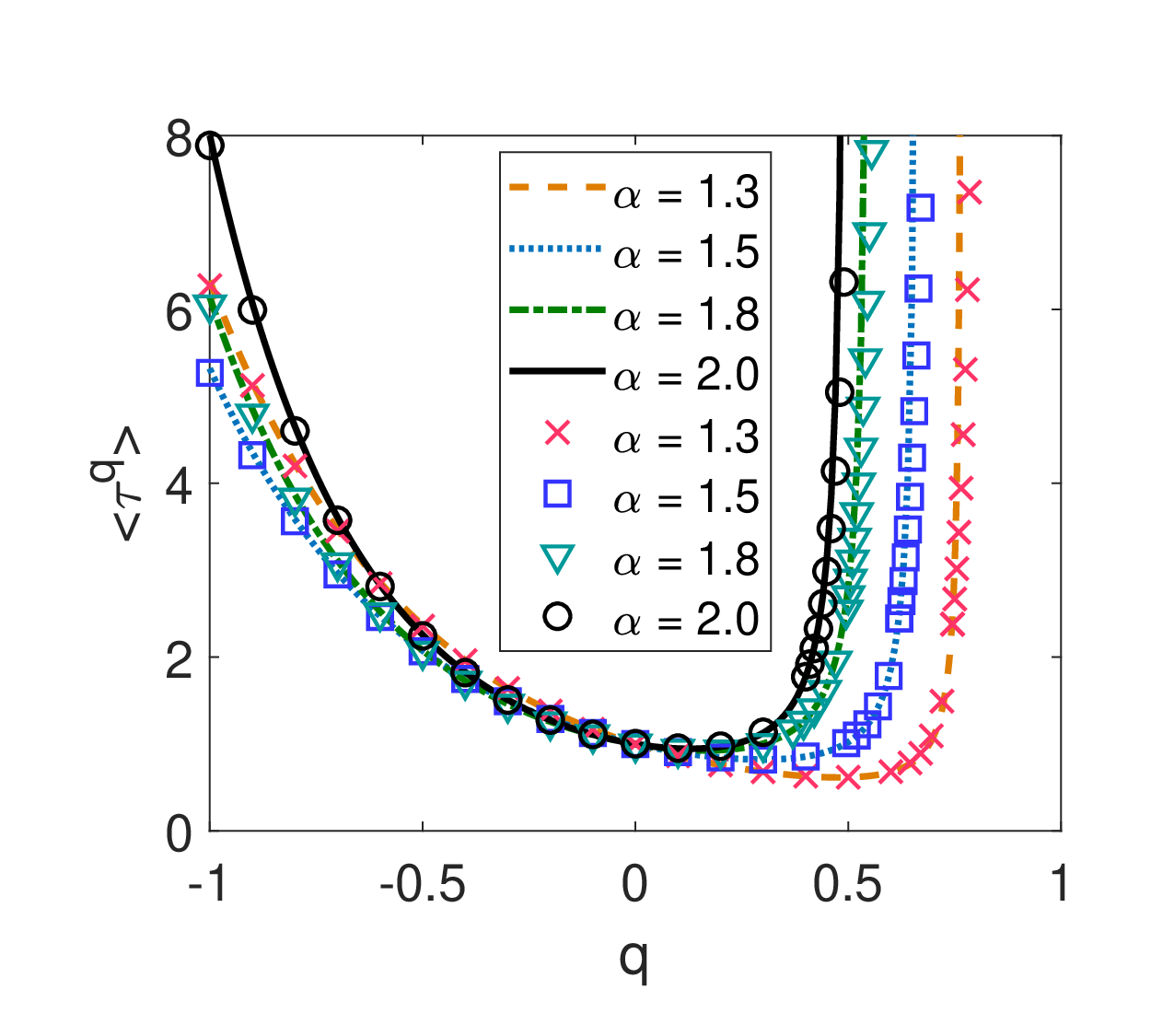}
\includegraphics[width=0.49\textwidth]{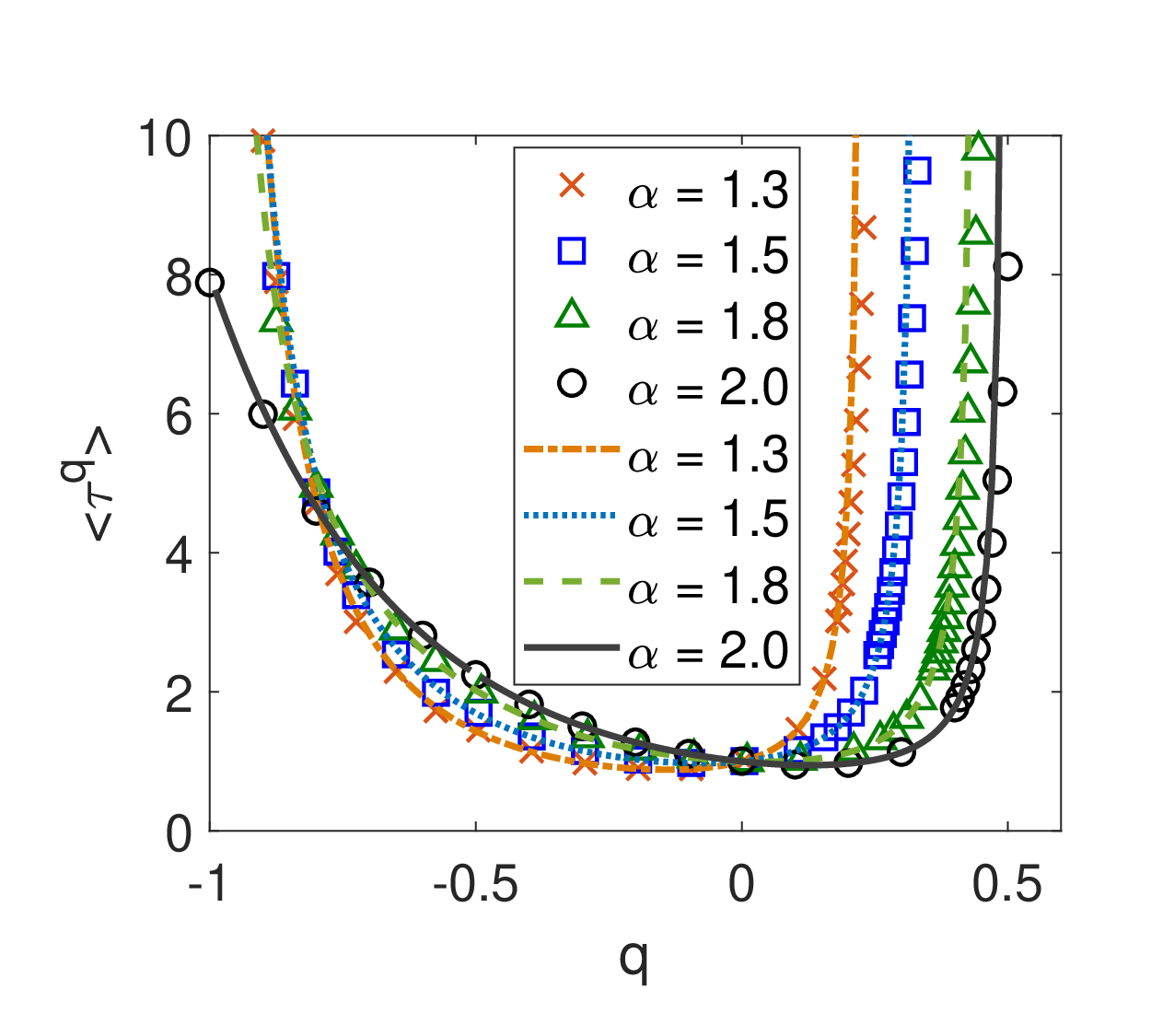}
\caption{Left: Fractional order moments of the first-passage time PDF for
extremal two-sided $\alpha$-stable laws in a semi-infinite domain with
stability index $1<\alpha\leq2$ and skewness $\beta=-1$. Symbols represent
the numerical solution of the space-fractional diffusion equation and lines
correspond to the analytic result (\ref{eq:Extr-Two--1-Semi-moment}). Right:
same as in the left panel but with skewness $\beta=1$. Lines show results of
equation (\ref{eq:Extr-Two-1-Semi-moment}). In both panels, we used $d=0.5$
and $L=10^{12}$.}
\label{fig:fig3}
\end{figure}

For the considered case of extremal two-sided $\alpha$-stable laws with $1<\alpha
<2$ and $\beta=-1$ by recalling the integral representation (\ref{eq:intMfun}) of
the $M$-function, the first-passage time PDF moments become
\begin{eqnarray}
\nonumber
\langle\tau^q\rangle&=&\frac{d}{\alpha\xi^{1/\alpha}}\int\limits_0^{\infty}t^{q-1
-1/\alpha}M_{1/\alpha}\left(\frac{d}{(\xi t)^{1/\alpha}}\right)\mathrm{d}t\\
\nonumber
&=&\frac{d}{\alpha\xi^{1/\alpha}}\int\limits_0^{\infty}t^{q-1-1/\alpha}\frac{1}{2\pi
i}\int\limits_{Ha}\e^{\sigma-d(\sigma/\xi t)^{1/\alpha}}\frac{\mathrm{d}\sigma}{
\sigma^{1-1/\alpha}}\mathrm{d}t\\
&=&\frac{d}{\alpha\xi^{1/\alpha}}\frac{1}{2\pi i}\int\limits_{Ha}\sigma^{1/\alpha-1}
\e^{\sigma}\int\limits_0^{\infty}t^{q-1-1/\alpha}\e^{-d(\sigma/\xi t)^{1/\alpha}}
\mathrm{d}t\mathrm{d}\sigma.
\end{eqnarray}
Changing variables, $u=d (\sigma/\xi t)^{1/\alpha}$ in the inner integral and
with the help of equation (\ref{eq:GammaintP}), we find
\begin{equation}
\label{eq:Extr-Two--1-Semi-moment}
\fl\langle\tau^q\rangle=\frac{d^{q\alpha}\Gamma(1-q\alpha)}{\xi^q}\frac{1}{2\pi
i}\int\limits_{Ha}^{}\sigma^{q-1}\e^{\sigma}\mathrm{d}\sigma=\frac{\Gamma(1-q
\alpha)}{\Gamma(1-q)}\frac{d^{q\alpha}}{\xi^q},\,\,\,-\infty<q<1/\alpha,
\end{equation}
where in the last equality we used equation (\ref{eq:GammaintN}) to get the
desired result. In the limit $\alpha=2$ we recover the fractional moments of
the first-passage time PDF (\ref{eq:Brown-Semi-Moment}) for a Gaussian process.
The left panel of figure \ref{fig:fig3} shows the results of equation
(\ref{eq:Extr-Two--1-Semi-moment}) along with numerical solutions of the
space-fractional diffusion equation. As can be seen the fractional order
moments $-\infty<q<1/\alpha$ are finite, as they should.

\subsubsection{Extremal two-sided $\alpha$-stable processes with $1<\alpha<2$
and $\beta=1$.}

Applying the Skorokhod theorem it can be shown that the first-passage time PDF
of the extremal two-sided $\alpha$-stable law with stability index $1<\alpha<2$
and skewness $\beta=1$ has the following series representation \cite{AminP2019}
(see equation (D.73))
\begin{equation}
\label{other}
\wp(t)=\frac{t^{-2+1/\alpha}d^{\alpha-1}}{\alpha \xi^{1-1/\alpha}}
\sum_{n=0}^{\infty}\frac{(d^\alpha/{\xi t})^n}{\Gamma{(\alpha n+
\alpha-1)}\Gamma{(-n+1/\alpha)}}.
\end{equation}
Now, with the help of Euler's reflection formula $\Gamma(1-z)\Gamma(z)
\sin{(\pi z)}=\pi$ and the relation $\sin{\pi(z-n)}=(-1)^{n}\sin{(\pi z)}$
we rewrite this expression in the form
\begin{equation}
\label{rewrite}
\wp(t)=\frac{\sin{(\pi/\alpha)}t^{-2+1/\alpha}d^{\alpha-1}}{\pi\alpha\xi^{1
-1/\alpha}}\sum_{n=0}^{\infty}\frac{\Gamma{(n+1-1/\alpha)}(-d^\alpha/{\xi t})^
n}{\Gamma{(\alpha n+\alpha-1)}}.
\end{equation}
To obtain the long-time asymptotics of the PDF we take $n=0$ in equation
(\ref{rewrite}) and arrive at the power-law decay given by equation (43) of
\cite{AminP2019} or, equivalently, equation (\ref{eq:generalasym}) below with
$\rho=1-1/\alpha$.

To calculate the moments of the first-passage time we use the relation between
the Wright generalised hypergeometric function and the $H$-function \cite{Mathai}
(see equations (1.123) and (1.140)). We arrive at 
\begin{equation}
\wp(t)=\frac{\sin{(\pi/\alpha)}t^{-2+1/\alpha}d^{\alpha-1}}{\pi\alpha\xi^{1-1/
\alpha}}H_{2,2}^{1,2}\left[\frac{d^\alpha}{\xi t}\left|\begin{array}{l}(0,1),
(1/\alpha, 1)\\(0,1),(2-\alpha,\alpha)\end{array}\right.\right].
\end{equation}
Further, with the help of the inversion property of the $H$-function
\cite{Mathai} (Property 1.3, equation (1.58)), we have
\begin{equation}
\label{eq:FPT-twoside-Hfun}
\wp(t)=\frac{\sin{(\pi/\alpha)}t^{-2+1/\alpha}d^{\alpha-1}}{\pi \alpha\xi^
{1-1/\alpha}}H_{2,2}^{2,1}\left[\frac{\xi t}{d^\alpha}\left|\begin{array}{l}
(1,1),(\alpha-1,\alpha)\\ (1,1),(1-1/\alpha,1)\end{array}\right.\right].
\end{equation}
At short times the $H$-function representation of the first-passage PDF leads to
equation (44) of \cite{AminP2019} or, equivalently, equation (\ref{limit-pdf})
below with $\rho=1-1/\alpha$. Substitution of equation (\ref{eq:FPT-twoside-Hfun})
into (\ref{eq:FPT-moments}) yields
\begin{equation}
\langle\tau^q\rangle=\int\limits_0^{\infty}\frac{\sin{(\pi/\alpha)}t^{q-2+1/
\alpha}d^{\alpha-1}}{\pi\alpha \xi^{1-1/\alpha}}H_{2,2}^{2,1}\left[\frac{\xi
t}{d^\alpha}\left|\begin{array}{l}(1,1),(\alpha-1,\alpha)\\(1,1),(1-1/\alpha,
1)\end{array}\right.\right]\mathrm{d}t.
\end{equation}
Recalling the Mellin transform of the $H$-function \cite{Mathai} (page 47,
equation (2.8)), we find
\begin{equation}
\langle\tau^q\rangle=\frac{\sin{(\pi/\alpha)}\Gamma(1-1/\alpha-q)\Gamma(q+1/
\alpha)\Gamma{(q)}}{\pi\alpha\Gamma(q\alpha)}\frac{d^{q\alpha}}{\xi^q}.
\end{equation}
Using Euler's reflection formula $\Gamma(1-z)\Gamma(z)\sin{(\pi z)}=\pi$, we
finally get
\begin{equation}
\label{eq:Extr-Two-1-Semi-moment}  
\langle\tau^q\rangle=\frac{\sin{(\pi/\alpha)}}{\sin{(\pi(q+1/\alpha))}}\frac{
\Gamma(1+q)}{\Gamma(1+q\alpha)}\frac{d^{q\alpha}}{\xi^q},\,\,\,-1<q<1-1/\alpha,
\end{equation}
where $\xi$ is given by equation (\ref{eq:xiparameter}).
The same result with a different method was given in dimensionless form in
\cite{TSimon2011} (see proposition 4). For $\alpha=2$, we again consistently
recover result (\ref{eq:Brown-Semi-Moment}). In the right panel of figure
\ref{fig:fig3} we plot the numerical result for the space-fractional
diffusion equation and the analytic result corresponding to equation
(\ref{eq:Extr-Two-1-Semi-moment}). As expected, moments of order $-1<q<1-1/
\alpha$ are finite.

\begin{figure}
\includegraphics[width=0.49\textwidth]{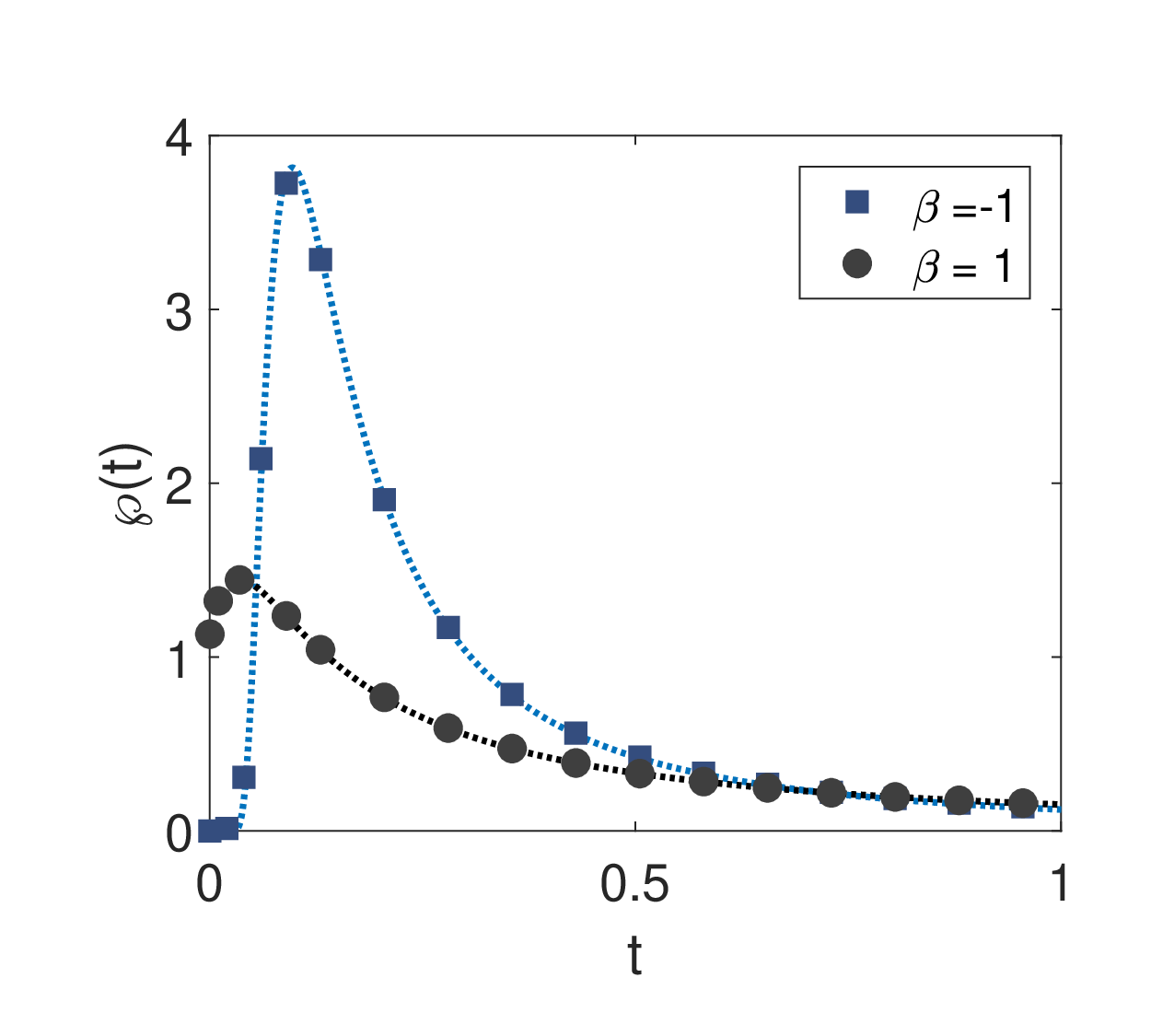}
\includegraphics[width=0.49\textwidth]{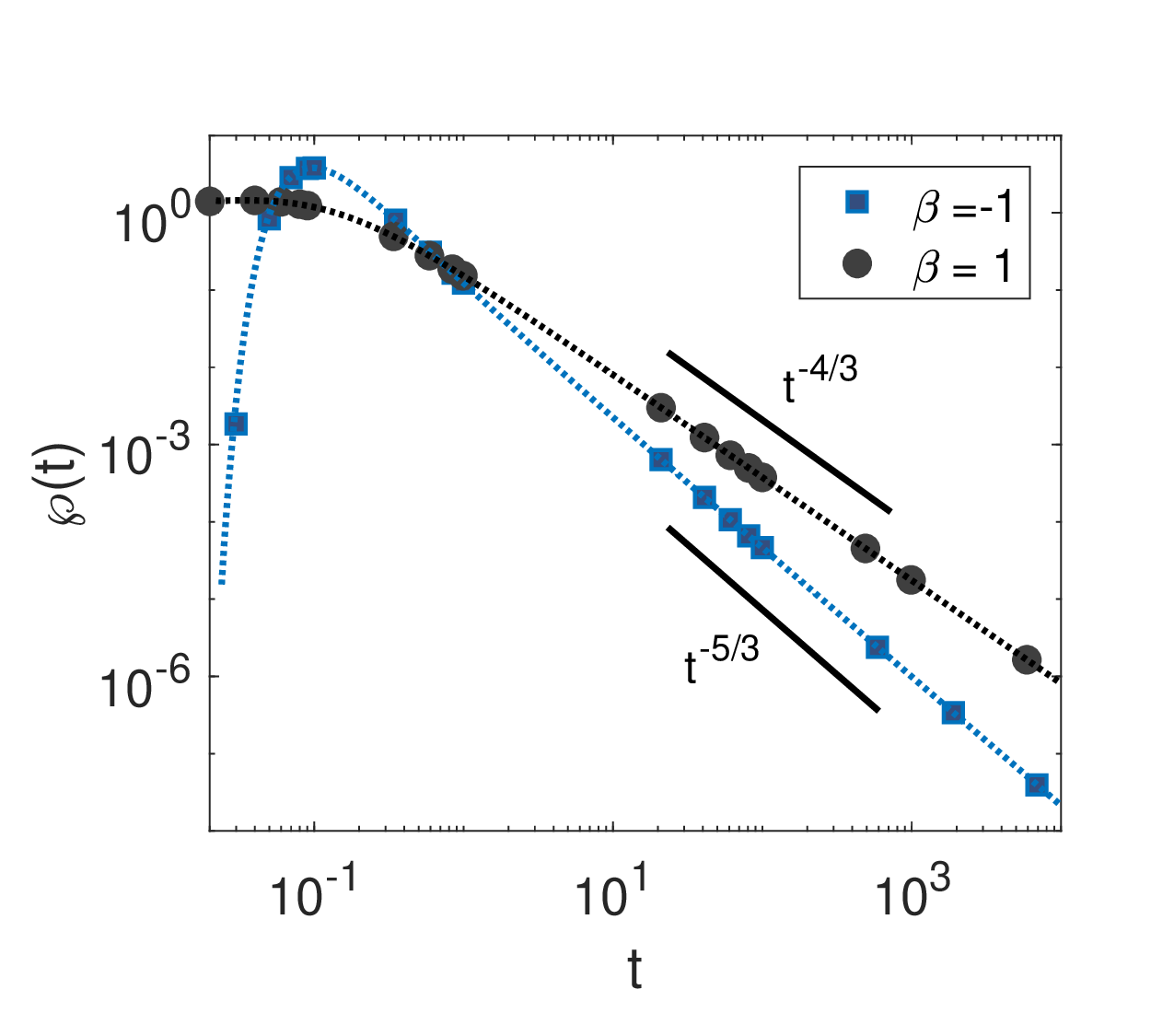}
\centering
\caption{Left: first-passage time PDF at short times for the extremal
$\alpha$-stable processes in the semi-infinite domain with stability index
$\alpha=1.5$. Right: long-time behaviour of the same PDF on log-log. The
black lines show the asymptotic behaviour of the PDFs. In both panels $d=0.5$,
symbols represent the numerical solution of the space-fractional diffusion
equation, and the dotted lines show the analytical results namely, equation
(\ref{eq:fptPDFtwobn}) for $\beta=-1$ and equation (\ref{other}) for $\beta=1$.}
\label{fig3a}
\end{figure}

For completeness in figure \ref{fig3a} we also provide a comparison of the
first-passage time PDFs for the extremal two-sided $\alpha$-stable processes in
the semi-infinite domain with $\beta=-1$ and $\beta=1$. One can see (left panel)
that in the limit $t\to0$ the first-passage time PDF tends to zero for $\beta=-1$
and attains a finite value for $\beta=1$. Respectively, in the long-time limit
(right panel) the PDFs decay differently, faster for $\beta=-1$ (like $\simeq t^{
-1-1/\alpha}$) and slower for $\beta=1$ (like $\simeq t^{-2+1/\alpha}$). 

\subsubsection{General asymmetric form of $\alpha$-stable processes}
\label{semi-inf-Gen}

\begin{figure}
\centering
\includegraphics[width=0.49\textwidth]{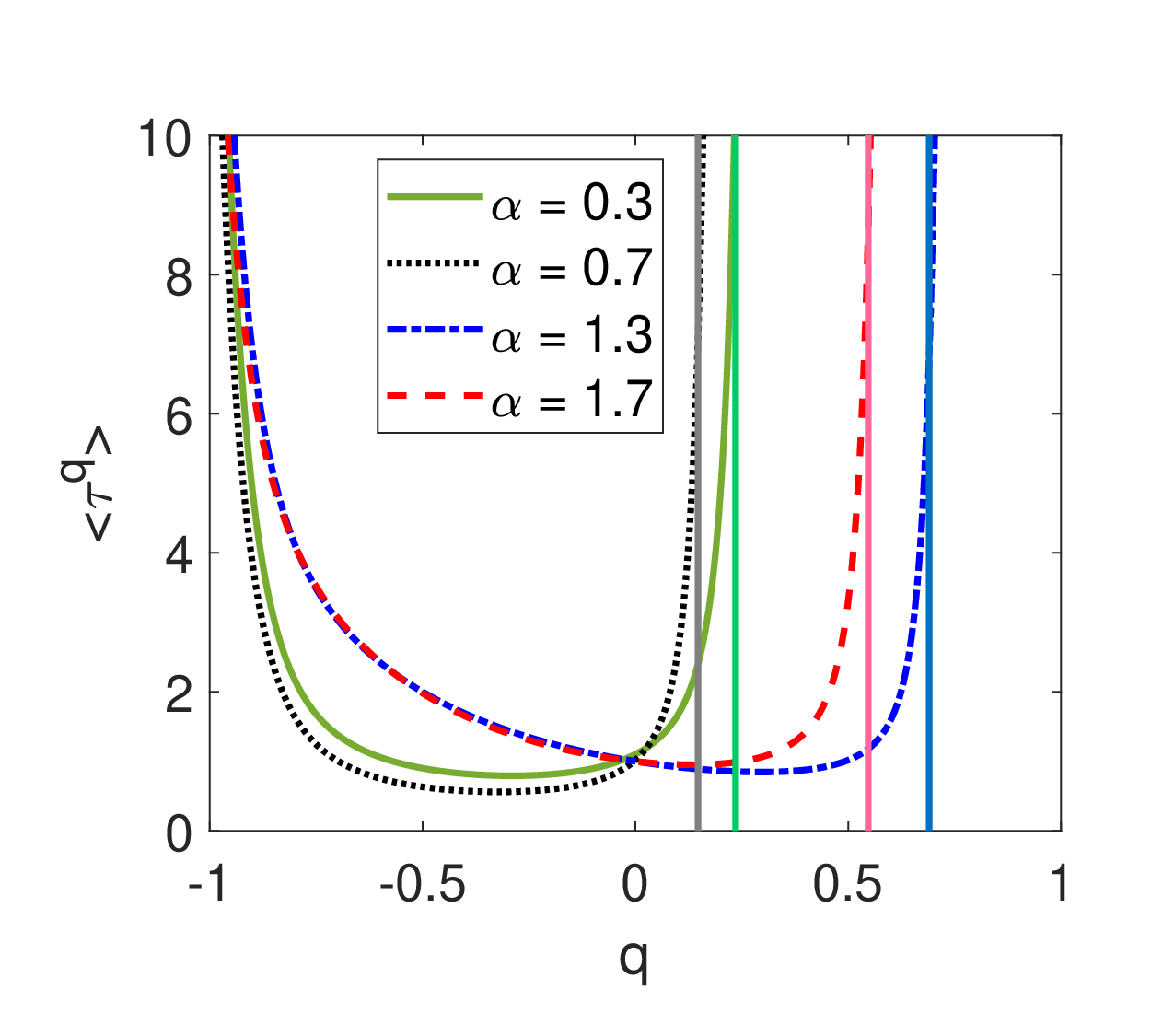}
\includegraphics[width=0.49\textwidth]{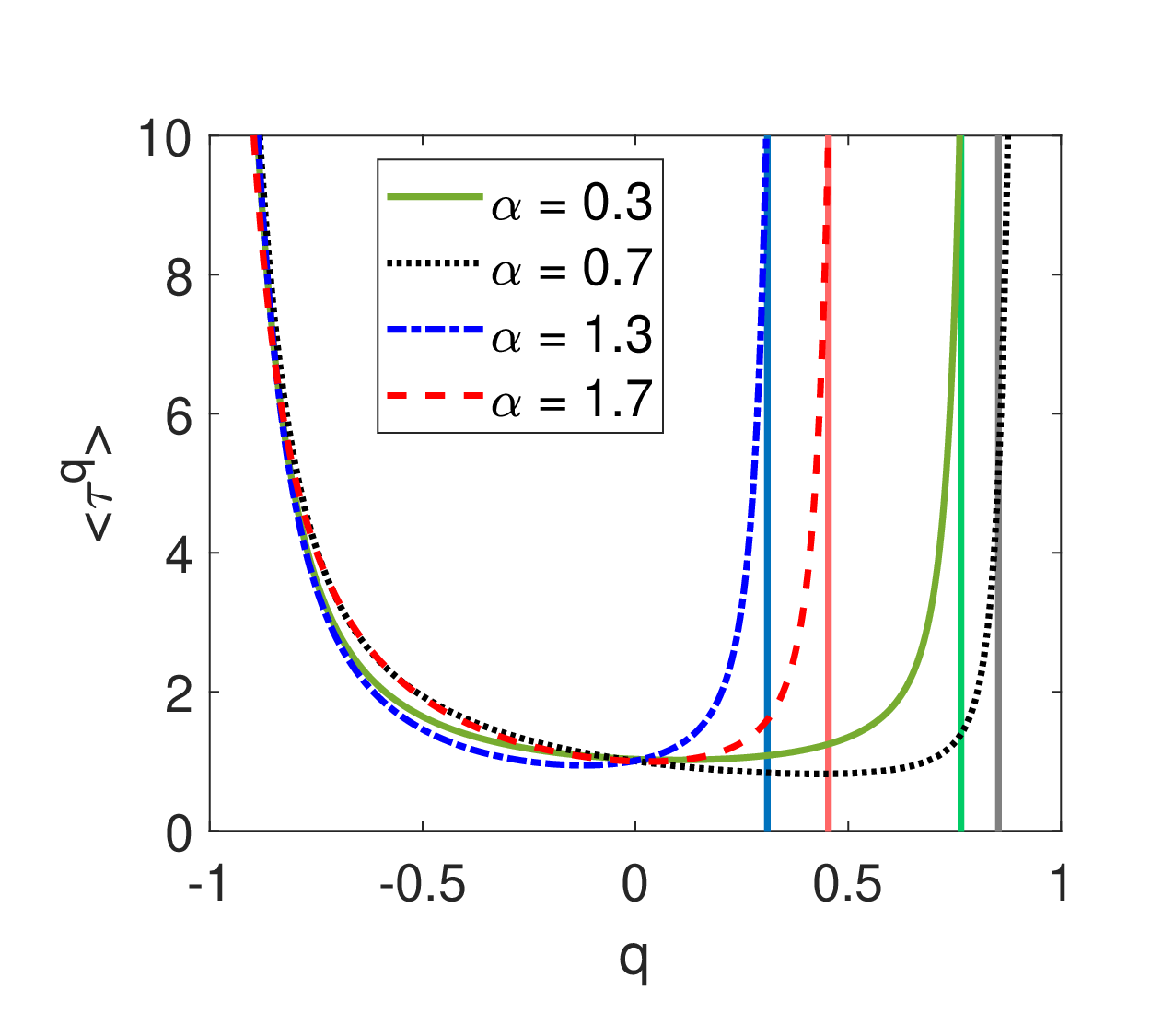}\\
\includegraphics[width=0.49\textwidth]{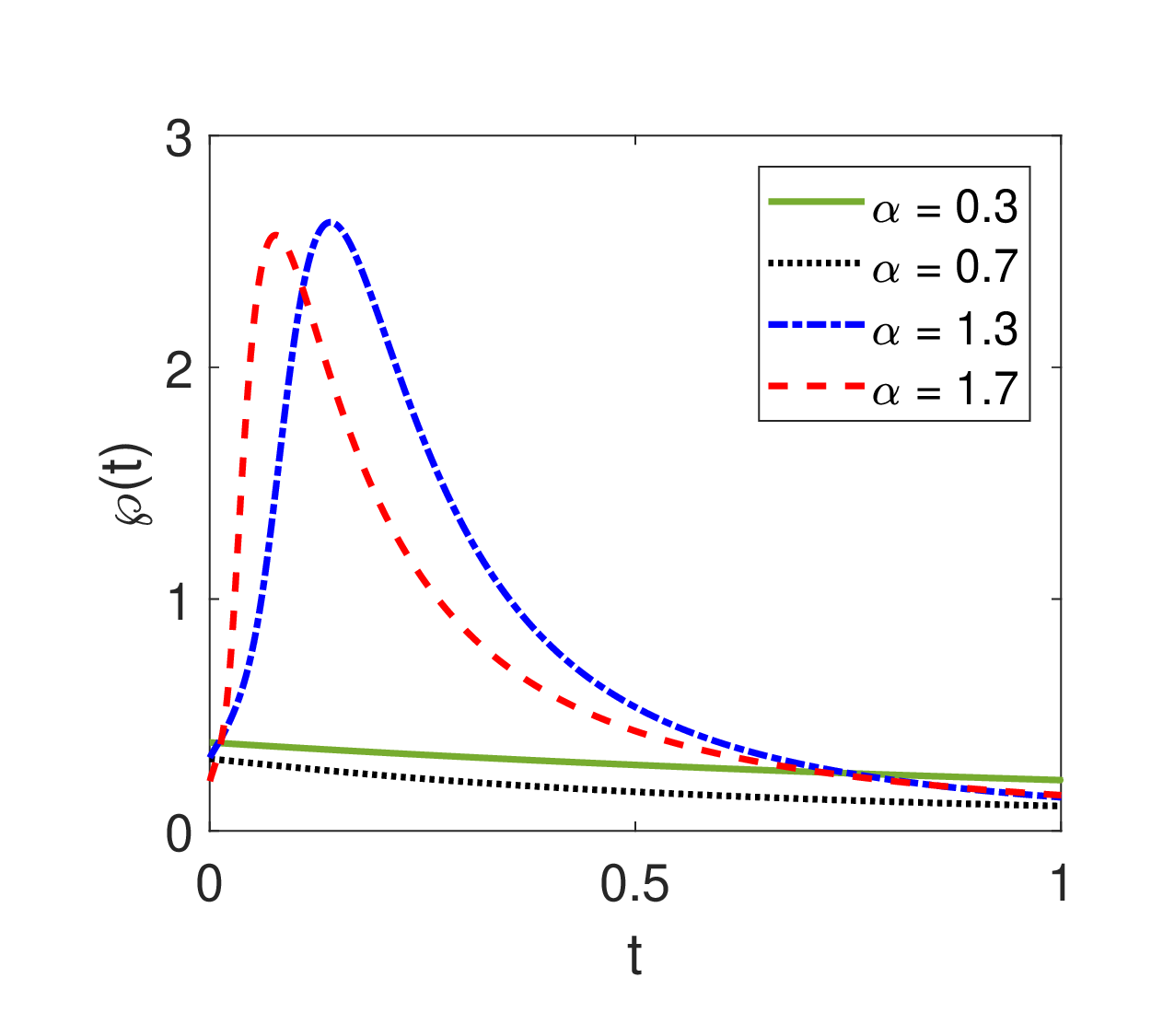}
\includegraphics[width=0.49\textwidth]{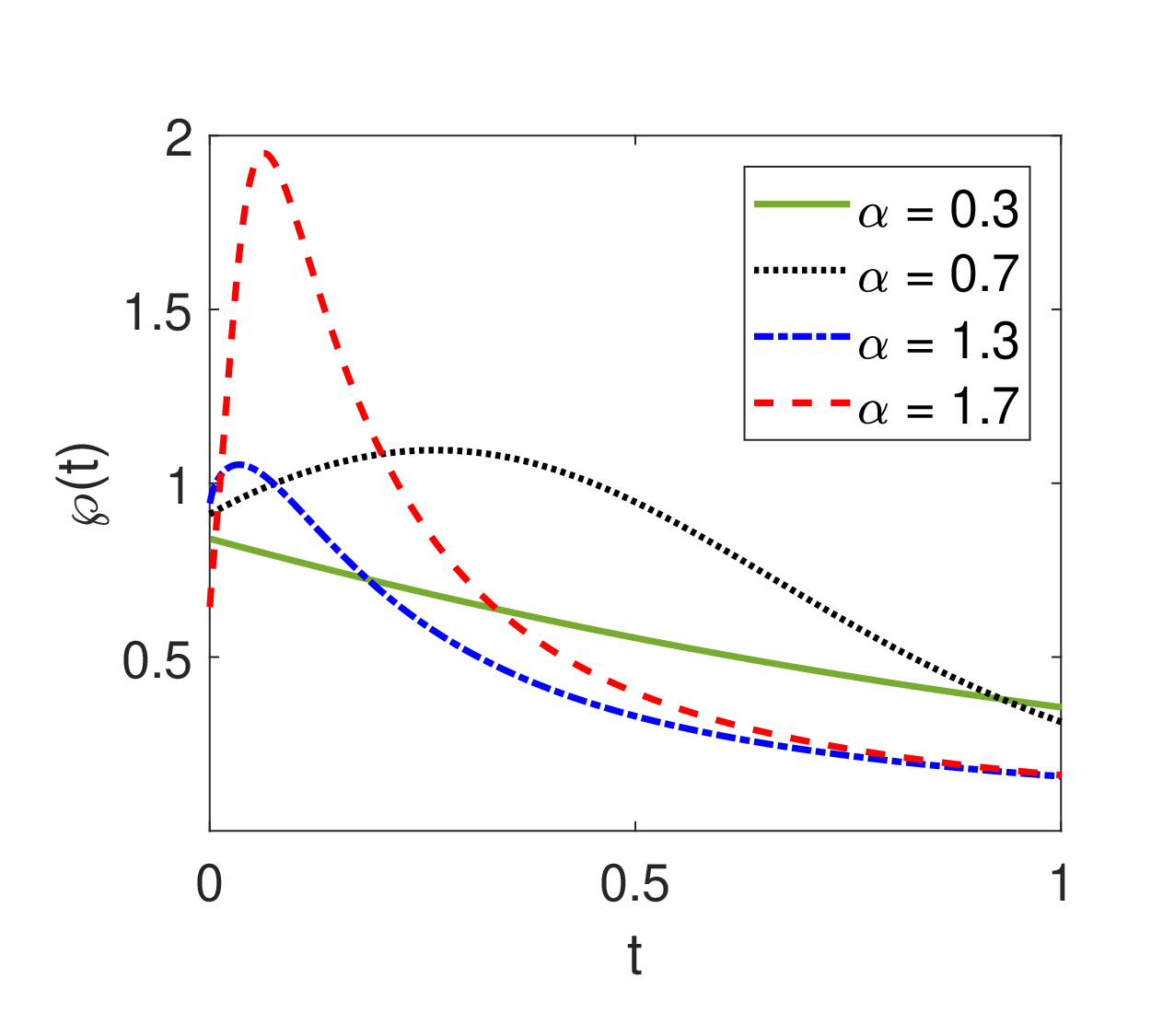}
\caption{Left: fractional order moments of the first-passage time PDF (top)
and the first-passage time PDF (bottom) of $\alpha$-stable laws in the
semi-infinite domain with skewness $\beta=-0.5$. Right: the same but for
skewness $\beta=0.5$. For all panels we used $d=0.5$ and $L=10^{12}$. The
lines represent numerical solutions of the space-fractional diffusion equation
and vertical lines in the top panels represent the limit $q=\rho$.}
\label{fig:fig4}
\end{figure}

In this section we present the first-passage properties of $\alpha$-stable
processes in general form. By applying the Skorokhod theorem for $\alpha\in(
0,1)$ with $\beta\in(-1,1)$, for $\alpha=1$ with $\beta=0$, as well as for
$\alpha\in(1,2]$ with $\beta\in[-1,1]$, it was shown that the first-passage
time PDF has the following power-law decay \cite{AminP2019}
\begin{equation}
\label{eq:generalasym}
\fl\wp(t)\sim \frac{\rho(K_\alpha(1+\beta^2\tan^2{(\alpha\pi/2)})^{1/2})^{-\rho}
d^{\alpha\rho}}{\Gamma(1-\rho)\Gamma(1+\alpha\rho)}t^{-\rho-1}=\frac{1}{\alpha
\Gamma(1-\rho)\Gamma(\alpha\rho)}\frac{d^{\alpha\rho}}{\xi^{\rho}}t^{-\rho-1},
\end{equation}
where $\xi$ and $\rho$ are defined in equation (\ref{eq:xiparameter}).
It is obvious that the corresponding integral (\ref{eq:FPT-moments}) is finite for
moments $q<\rho$, otherwise the integral diverges. To estimate the behaviour of
the first-passage time PDF at short times, we employ the asymptotic expression of
LFs for large $x$. For the purpose of this derivation, we follow the method
introduced in \cite{VVPalyulin2019} and assume that the starting position is at
$x_0=0$ while the boundary is located at $x=d$, which is identical to our setting
in a semi-infinite domain. Therefore, the survival probability at short times reads
\begin{equation}
S(t|0)=\int\limits_{-\infty}^dP_{\alpha,\beta}(x,t|0)\mathrm{d}x=1-\int\limits_d^{
\infty}P_{\alpha,\beta}(x, t|0)\mathrm{d}x,
\label{eq:shorttimesurvai}
\end{equation}
where the $\alpha$-stable law with the stability index $\alpha\in(0,2]$ ($\alpha
\neq1$) and skewness $\beta\in(-1,1]$ in the limit $x\to\infty$ is given by
\cite{AVSkorokhod1954}
\begin{eqnarray}
\nonumber
P_{\alpha,\beta}(x, t|0)&\sim&\pi^{-1}(1+\beta^2\tan^2{(\alpha\pi/2)})^{1/2}\sin{
(\alpha\pi\rho)}\Gamma(1+\alpha)\frac{K_{\alpha}t}{x^{1+\alpha}}\\
&=&\pi^{-1}\sin(\alpha\pi\rho)\Gamma(1+\alpha)\frac{\xi t}{x^{1+\alpha}}.
\end{eqnarray}
By substitution into equation (\ref{eq:shorttimesurvai}) and recalling equation
(\ref{eq:firstpassage}) we arrive at
\begin{equation}
\label{limit-pdf}
\wp(t\to0)=\pi^{-1}\sin(\alpha\pi\rho)\Gamma(\alpha)\frac{\xi}{d^{\alpha}}.
\end{equation}
It is easy to check with the use of equation (\ref{eq:xiparameter})
that the first-passage time PDF is only zero for Brownian
motion ($\alpha=2$ and $\rho=1/2$) at short times. Otherwise, the boundary is
crossed immediately with a finite probability on the first jump. To support our
conclusion regarding the existence of fractional order moments of the
first-passage time PDF for general asymmetric form of the $\alpha$-stable law,
we plot the fractional order moments and the first-passage time PDF for two
sets of the skewness, $\beta=0.5$ and $-0.5$, and different values of the
stability index $\alpha$. The results are shown in figure \ref{fig:fig4}, and
it can be seen moments with $-1<q<\rho$ are finite. The lower bound ($-1<q$),
arising due to the finite jump in the first-passage time PDF at $t\to0$,
can be seen in the bottom panels of figure \ref{fig:fig4}.
Similar to the symmetric case with $\beta=0$ shown in figure \ref{fig:fig2} the
values of the first-passage time PDF at $t\to0$ obtained by numerical solution of
the space-fractional diffusion equation are in perfect agreement with the behaviour
provided by equation (\ref{limit-pdf}). We also note that in \cite{RADoney2004}
(theorem 2) presented a sufficient condition for the finiteness of the moments of
the first-passage time of the general L\'evy process which is in agreement with
our results for LFs in general asymmetric form.

\section{First passage time properties of LFs in a bounded domain}
\label{mean bound}

In this section we consider an LF in the interval $[-L, L]$ with initial
point $x_0$ and absorbing boundary conditions at both interval borders
(figure \ref{fig:fig1}). Eventually, the LF is absorbed, and our basic
goal is to characterise the time dependence of this trapping phenomenon.
From equation (\ref{eq:MFPToperat}) and with the space-fractional operators
(\ref{eq:lcapu}) and (\ref{eq:rcapu}) we find
\begin{equation}
\fl K_{\alpha}\left(\frac{R_{\alpha,\beta}}{\Gamma(n-\alpha)}\int\limits_{-L}^{
x_0}\frac{\langle\tau\rangle^{(n)}(\zeta)}{(x_0-\zeta)^{
\alpha-n+1}}\,\mathrm{d}\zeta+\frac{L_{\alpha,\beta}(-1)^n}{\Gamma(n-\alpha)}
\int\limits_{x_0}^L\frac{\langle\tau\rangle^{(n)}(\zeta)}{(\zeta-
x_0)^{\alpha-n+1}}\,\mathrm{d}\zeta\right)=-1.
\label{eq:boundmfpt}
\end{equation}
Applying the boundary condition $\langle\tau\rangle(\pm L)=0$ and the fact that
$_0D_{L\pm x_0}^{\alpha}(L\pm x_0)^{\alpha-n+1}=const$ \cite{SGSamko1993} (page
626, theorem 30.7) leads us to a solution of equation (\ref{eq:boundmfpt}) in
the following form
\begin{equation}
\label{eq:mfptboundgensolution}
\langle\tau\rangle(x_0)=C_{\alpha,\beta}(L-x_0)^{\mu}(L+x_0)^{\nu},
\end{equation}
where $C_{\alpha,\beta}$ is a normalisation factor. First we consider the case
$0<\alpha<1$ ($n=1$). After substitution of equation (\ref{eq:mfptboundgensolution})
into (\ref{eq:boundmfpt}) and some calculations (see details in \ref{appendixMFPT})
we obtain
\begin{equation}
\mu=\alpha\rho,\,\,\,\nu=\alpha-\alpha\rho
\label{eq:muandnu}
\end{equation}
and
\begin{equation}
\label{eq:Cnormalfactor}
C_{\alpha,\beta}=\frac{\cos{(\alpha\pi(\rho-1/2))}}{\Gamma(1+\alpha)K_\alpha}
=\frac{1}{\Gamma(1+\alpha)\xi}.
\end{equation}
For the case $1<\alpha\leq2$ ($n=2$) a similar procedure leads to the same result,
and formulas (\ref{eq:muandnu}) and (\ref{eq:Cnormalfactor}) are valid for all
$\alpha\in(0,2]$ with $\beta\in[-1,1]$ (excluding the case $\alpha=1$, $\beta\neq
0$). Finally, the MFPT for LFs in a bounded domain $[-L, L]$ reads
\begin{equation}
\label{eq:MFPTBoundedGenral}
\langle\tau\rangle=\frac{(L-x_0)^{\alpha\rho}(L+x_0)^{\alpha-\alpha\rho}}{
\Gamma(1+\alpha)\xi},
\end{equation}
where $\rho$ and $\xi$ are given in expression (\ref{eq:xiparameter}).
We note that in \cite{CProfeta2016} from the Green's function of a L\'evy stable
process \cite{AEKyprianou2014} the MFPT of LFs in the interval (-1, 1) in the
dimensionless Z-form of the characteristic function ($K_{\alpha}^{Z}=1$) is
given (see Remark 5 in \cite{CProfeta2016}). To see the equivalence between
equation (\ref{eq:MFPTBoundedGenral}) and the result in \cite{CProfeta2016}
we note that the following relation between the parameters in the A- and Z-forms
is established and reads (see equation (A.11) in \cite{AminP2019})
\begin{equation}
\rho=\frac{1}{2}+\frac{1}{\alpha\pi}\arctan\left(\beta_{A}\tan\left(\frac{\alpha
\pi}{2}\right)\right),\quad K_{\alpha}^{Z}=\frac{K_{\alpha}^{A}}{\cos{(\alpha\pi
(\rho-1/2))}}.
\end{equation}
Here, we use the standard A-from parameterisation for the characteristic function.

\subsection{Symmetric $\alpha$-stable processes}
\label{mean sym bound}

For symmetric $\alpha$-stable processes in a bounded domain, the MFPT for
stability index $0<\alpha\leq2$ and $|x_0|<L$ in $N$-dimension is given by
\cite{RKGetoor1961}
\begin{equation}
\langle\tau\rangle=K(\alpha,N)(L^2-x_0^2)^{\alpha/2},
\end{equation}
where
\begin{equation}
K(\alpha,N)=\Gamma\left(\frac{N}{2}\right)\left[2^{\alpha}\Gamma\left(1+\frac{
\alpha}{2}\right)\Gamma\left(\frac{N+\alpha}{2}\right)\right]^{-1}.
\end{equation}
In one dimension by using the duplication rule $2^{2z}\Gamma(z)\Gamma(z+1/2)=2
\sqrt{\pi}\Gamma(2z)$ this equation reads \cite{SVBuldyrev2001-2,AZoia2007}
\begin{equation}
\label{eq:MFPTsym}
\langle \tau \rangle=\frac{(L^2-x_0^2)^{\alpha/2}}{\Gamma(1+\alpha)}.
\end{equation}
For the setup in figure \ref{fig:fig1}, $x_0=L-d$ and by defining $l=d/L$, in
dimensional variables the MFPT yields in the form
\begin{equation}
\label{eq:mfptsym}
\langle\tau\rangle=\frac{(d(2L-d))^{\alpha/2}}{\Gamma(1+\alpha)K_{\alpha}}=
\frac{L^{\alpha}(l(2-l))^{\alpha/2}}{\Gamma(1+\alpha)K_{\alpha}}.
\end{equation}
This result is consistently recovered from the general formula
(\ref{eq:MFPTBoundedGenral}) by setting $\rho=1/2$ (or, equivalently,
$\beta=0$).

The second moment of the first-passage time PDF for symmetric $\alpha$-stable
process with stability index $0<\alpha\leq2$ and $|x_0|<L$ in $N$ dimensions
was derived in \cite{RKGetoor1961},
\begin{equation}
\langle\tau^2\rangle=\alpha L^{\alpha}K(\alpha,N)^{2}\int\limits_{x_0^2}^{L^2}
\left(s-x_0^2\right)^{\alpha/2-1}F\left(-\frac{\alpha}{2};\frac{N}{2};\frac{N+
\alpha}{2};sL^{-2}\right)\mathrm{d}s,
\label{eq:second-moment-sym}
\end{equation}
where $F$ is the Gauss hypergeometric function defined in equation
(\ref{eq:hypergeometricfunc}). Analogous to the MFPT we set $N=1$, $x_0=L-d$,
and in order to make time dimensional, equation (\ref{eq:second-moment-sym})
has to be divided by $K_{\alpha}^2$. Equation (\ref{eq:second-moment-sym}) is
reduced to a simple form for Brownian motion only \cite{SRedner2001},
\begin{equation}
\langle\tau^2\rangle=\frac{L^4}{12K_{\alpha}^2}(l^2-2l)(l^2-2l-4),
\label{eq:Brown-Bound-secmomen}
\end{equation}
where $l=d/L$. The behaviour of arbitrary-order moments is similar and reads
$\langle\tau^m\rangle\propto L^{m\alpha}/K_{\alpha}^m$ (see figure \ref{fig:fig8}),
for the case when we start the process at the centre of the interval $[-L,L]$.

\begin{figure}
\centering
\includegraphics[width=0.49\textwidth]{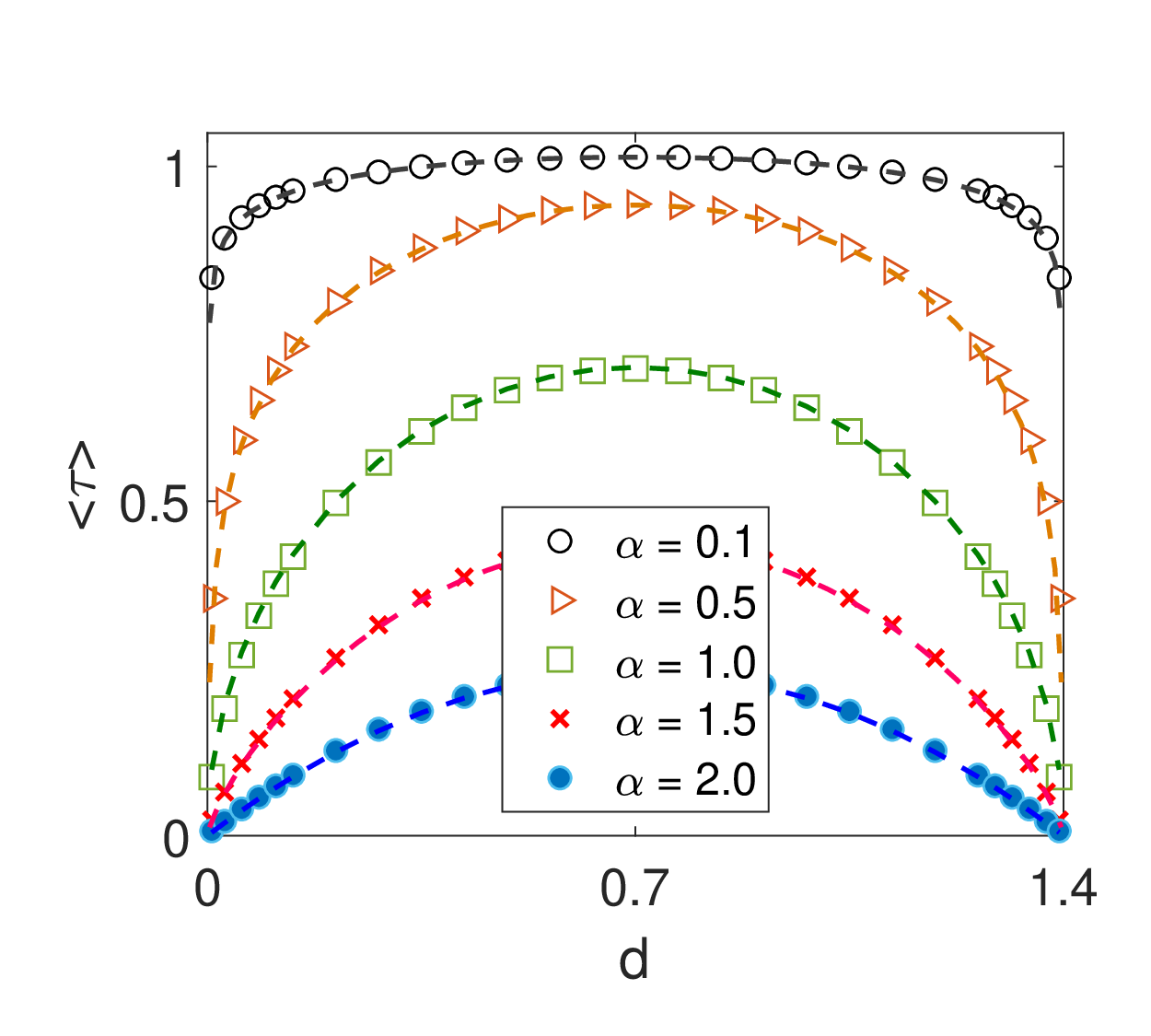}
\includegraphics[width=0.49\textwidth]{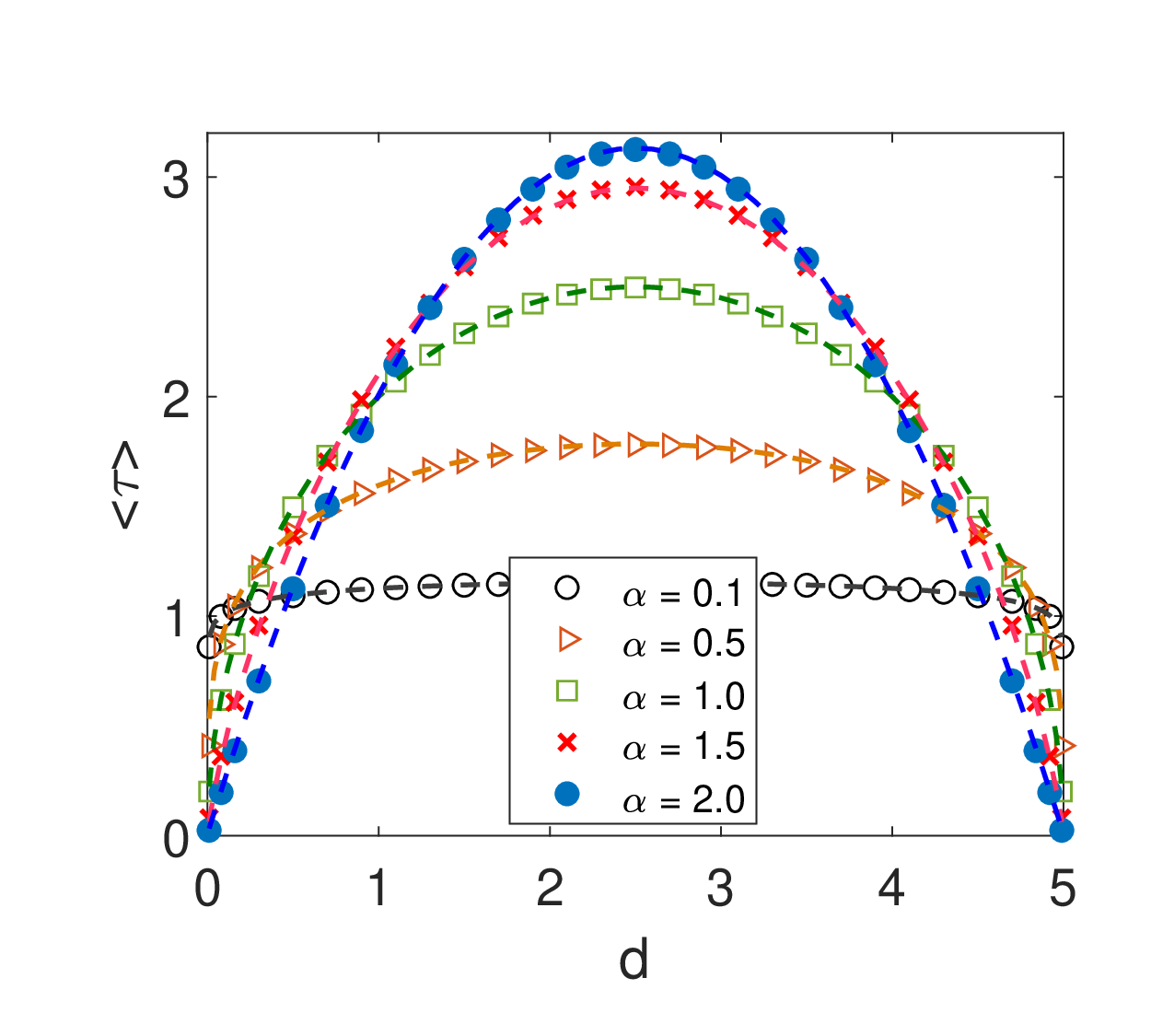}
\caption{MFPT versus distance $d$ of the initial point of the random
process from the right side boundary for symmetric $\alpha$-stable
processes ($\beta=0$) and different sets of the stability index
$\alpha$. Left: $L=0.7$. Right: $L=2.5$. Dashed lines show the analytic
solution (\ref{eq:mfptsym}) and symbols represent numerical solutions
of the space-fractional diffusion equation.}
\label{fig:fig5}
\end{figure}

In figure \ref{fig:fig5} we study the MFPT for symmetric $\alpha$-stable
processes with varying initial position. We employ two different interval
lengths and plot the MFPT versus $d$ for different sets of the stability
index $\alpha$. As can be seen, for interval length of $L=0.7$,
regardless of the starting point of the random walker the MFPT is always
longer for smaller $\alpha$. In contrast, for interval length $L=2.5$,
when the starting point of the random walker is close to the centre of the
interval, for larger $\alpha$ the MFPT is longer. When the starting point
gets closer to the boundaries, the behaviour is opposite. These observations
are in line with the fact that LFs have a propensity for long but rarer jumps,
a phenomenon becoming increasingly pronounced when the value of $\alpha$
decreases. Conversely, LFs have short relocation events with a higher
frequency for values $\alpha$ close to 2. Therefore, for small intervals
(left panel of figure \ref{fig:fig5}) it is easier to cross the boundaries
when short relocation events happen with a high frequency, corresponding to
L\'evy motion with $\alpha$ closer to 2. In the opposite case, LFs with
low-frequency large jumps ($\alpha\to 0$) can escape more efficiently from
large intervals (right panel of figure \ref{fig:fig5}), except for initial
positions close to the boundaries. We also note that in both panels of figure
\ref{fig:fig5}, when the stability index $\alpha$ gets closer to $0$, the
MFPT becomes flatter away from the boundaries. This result implies that with
different starting points the random walker crosses the interval by a single
jump---concurrently, the MFPT has a small variation.

\subsection{Asymmetric $\alpha$-stable processes}
\label{mean Asym bound}

\subsubsection{One-sided $\alpha$-stable processes with $0<\alpha<1$ and
$\beta=1$.}
\label{moment bound one-sided}

This type of jump length distribution is defined on the
positive axis. Therefore the situation for this process in semi-infinite and
bounded domains is similar and moments for the first-passage time PDF turn out
to be exactly the same as in equation (\ref{eq:mfptoneside}) obtained above.
Another method to find the moments of the first-passage time PDF is to employ
relation (\ref{eq:Zoiamethodmoments}), addressed originally in \cite{AZoia2007}
for symmetric $\alpha$-stable laws. The space-fractional operator for one-sided
$\alpha$-stable laws ($0<\alpha<1$ and $\beta=1$) reads
\begin{equation}
D_{x_0}^{\alpha}=-\frac{1}{\cos{(\alpha\pi/2)}}\,_{x_0}D_L^{\alpha}.
\end{equation}
We apply the space-fractional integration operator $D_{x_0}^{-m\alpha}$ on both
sides of equation (\ref{eq:Zoiamethodmoments}) and get (see \ref{appc} for
details)
\begin{equation}
\label{eq:Zoiamethodmoment-oneside}
\langle\tau^m\rangle(x_0)=\, _{x_0}D_L^{-m\alpha}{\frac{\cos^{m}{(\alpha\pi/2)}
\Gamma(1+m)}{{K_\alpha}^m}},
\end{equation}
where the sequential rule was used, namely, $(D^{\alpha}_{x_0})^{m}=D^{m\alpha}_{
x_0}$ \cite{Podlubny1999} (page 86, equation (2.169)). The space-fractional
integration operator $_{x_0}D_L^{-m\alpha}$ used here is defined as
\cite{Podlubny1999} (page 51, equation (2.40))
\begin{equation}
\label{eq:frac-integral}
_{x_0}D_L^{-m\alpha}{f(x)}=\frac{1}{\Gamma(m\alpha)}\int\limits_{x_0}^{L}\frac{
f(\zeta)}{(\zeta-x_0)^{1-m\alpha}}\,\mathrm{d}\zeta.
\end{equation}
By substitution of equation (\ref{eq:frac-integral}) with $f(\zeta)={\bf 1}$ into
equation (\ref{eq:Zoiamethodmoment-oneside}) we arrive at
\begin{equation}
\label{eq:moments-oneside-bound}
\fl\langle\tau^m\rangle(x_0)=\frac{\cos^{m}{(\alpha \pi/2)}\Gamma(1+m)}{
\Gamma(m\alpha){K_\alpha}^m}\int\limits_{x_0}^{L}(\zeta-x_0)^{m\alpha-1}
\mathrm{d}\zeta=\frac{ \Gamma(1+m)}{\Gamma(1+m\alpha)}\frac{d^{m\alpha}}{\xi^m}.
\end{equation}
This result is the same as equation (\ref{eq:mfptoneside}) with parameter
$\xi$ defined in (\ref{eq:xiparameter}) and $d=L-x_0$. The same result for
$m=1$ is also shown in \cite{TKoren2007,SCPort1970}. Moreover from equation
(\ref{eq:MFPTBoundedGenral}) by setting $\rho=1$ or $\beta=1$ ($0<\alpha<1$)
we arrive at above expression with $m=1$. The left panels of figure
\ref{fig:fig6} show the MFPT of one-sided LFs ($0<\alpha<1$ and $\beta=1$)
for different values of the stability index $\alpha$ for two interval lengths
(top: $L=0.7$, bottom: $L=2.5$). For interval length $L=0.7$, smaller $\alpha$
values lead to longer MFPTs for different initial positions, except for the
situations when the LF starts really close to the left boundary. This
observation is due to the lower frequency of long-range jumps compared to
high-frequency shorter-range jumps for larger $\alpha$ values, similar to
the above. For interval length $L=2.5$, when the initial position of the
random walker is located a distance $d<2$ away from the right boundary,
for smaller $\alpha$ it takes longer to cross the right boundary. For
larger $d$ values the smaller $\alpha$ values overtake the LFs with the
intermediate stable index $\alpha=0.5$. Note, however, that the MFPT for
$\alpha=0.9$ remains shorter than for LFs with the smaller stable index. For
increasing interval length low-frequency long jumps will eventually win out
unless the particle is released close to an absorbing boundary, compare also the
discussion in \cite{vlad1,vlad2}. Thus, the crossing of curves with different
$\alpha$ values in the left panel of figure \ref{fig:fig6} has a simple physical
meaning: it reflects the growing role of long jumps with smaller $\alpha$ when
the distance $d$ to the right boundary (respectively, the interval length $L$)
increases.

\subsubsection{One-sided $\alpha$-stable processes, $0<\alpha<1$, $\beta=-1$.}

\begin{figure}
\centering
\includegraphics[width=0.49\textwidth]{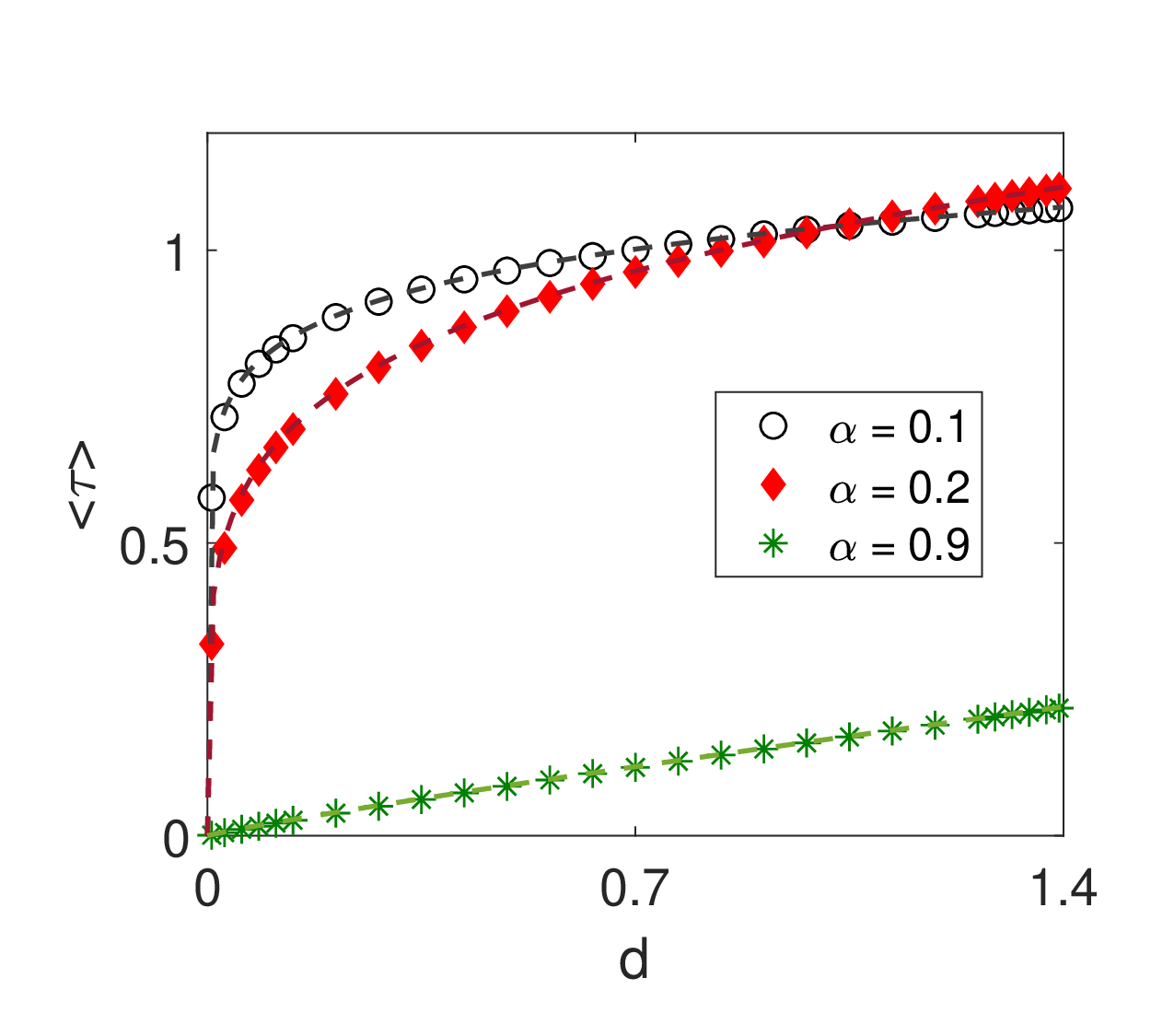}
\includegraphics[width=0.49\textwidth]{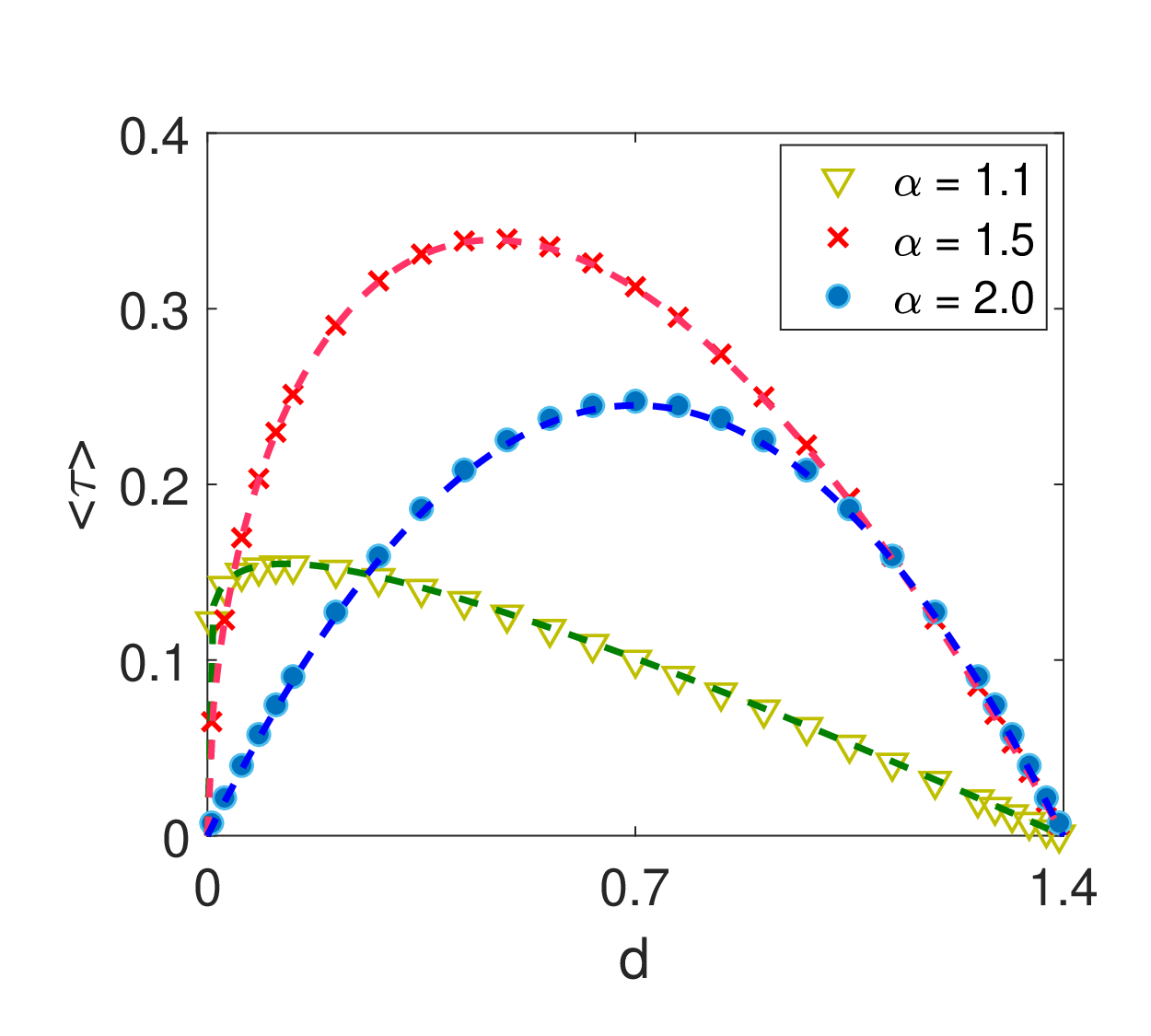}\\
\includegraphics[width=0.49\textwidth]{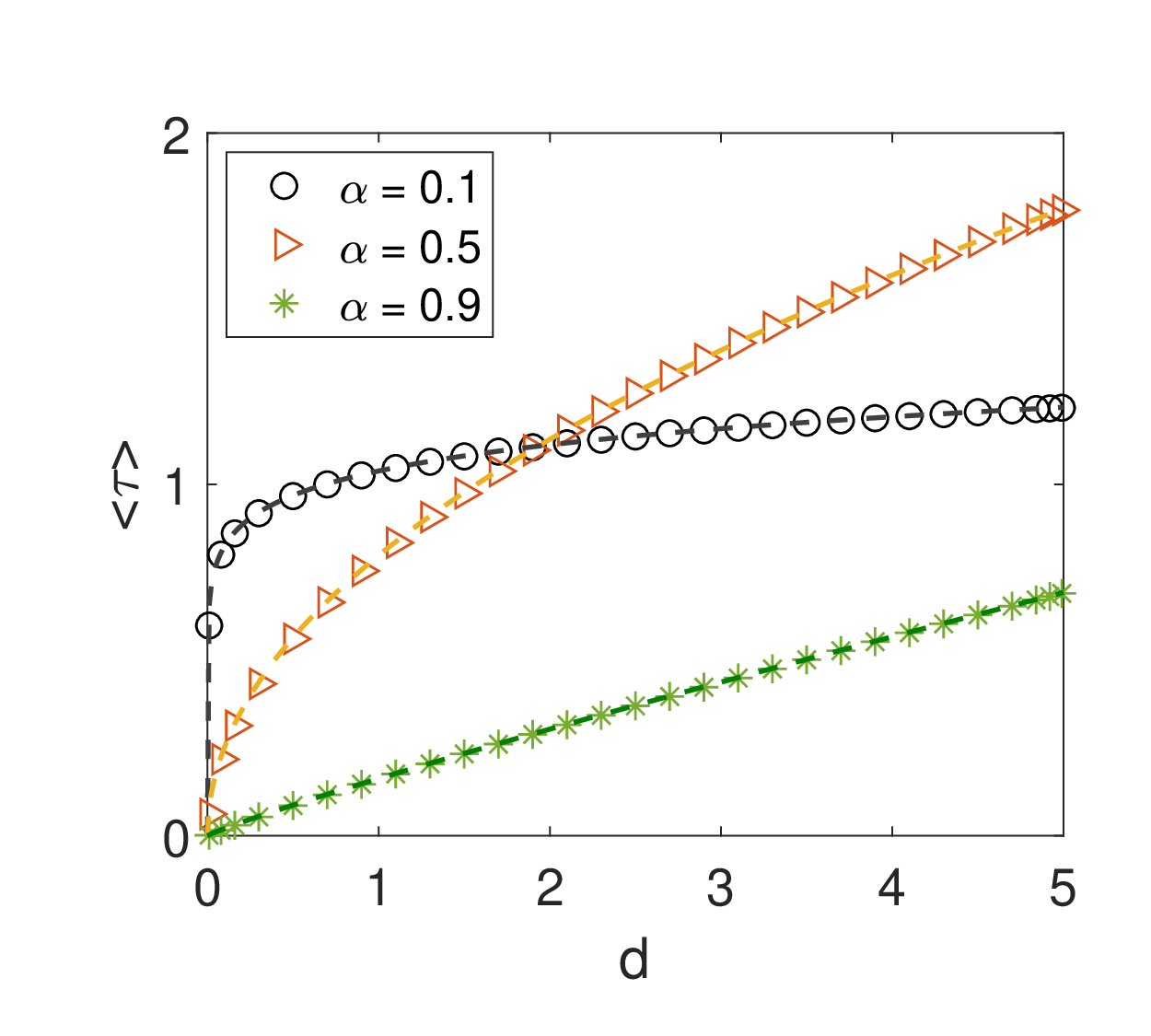}
\includegraphics[width=0.49\textwidth]{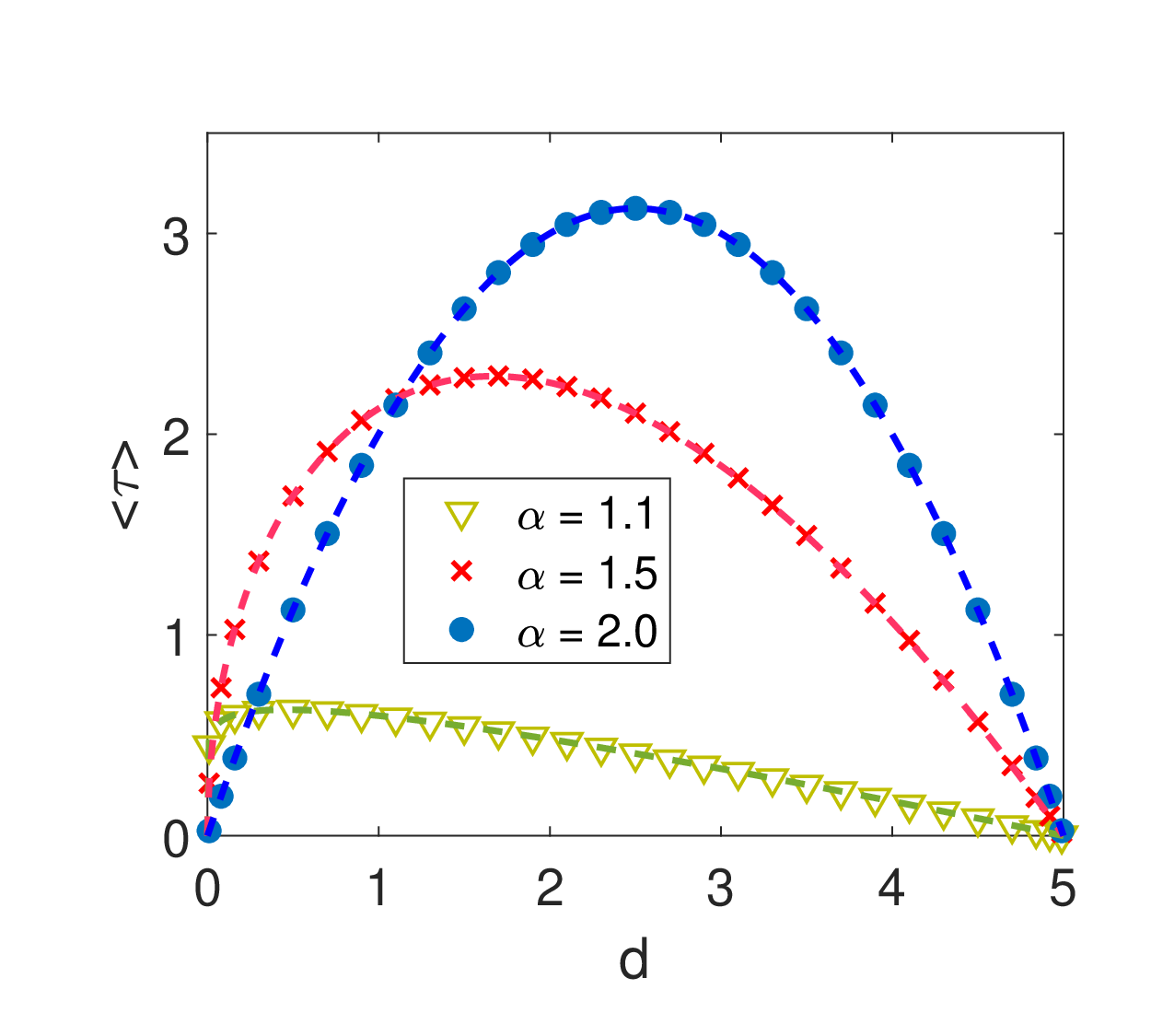}
\caption{MFPT versus distance $d$ of the initial position of the LF from the
right boundary. Top panels: interval length $L=0.7$. Bottom panels: interval
length $L=2.5$. Left panels: extremal one-sided $\alpha$-stable processes
with $\beta=1$ and different sets of the stability index $\alpha$. Dashed
lines represent the analytic result (\ref{eq:moments-oneside-bound}) while
symbols represent the numerical solution of the space-fractional diffusion
equation. Right panels: MFPT for extremal two-sided $\alpha$-stable processes with
skewness $\beta=1$. Dashed lines show the analytic result (\ref{eq:mfpttwoside1})
and symbols represent the numerical solution of the space-fractional diffusion
equation.}
\label{fig:fig6}
\end{figure}

For one-sided $\alpha$-stable processes with $0<\alpha<1$ and $\beta=-1$ the
space-fractional operator reads
\begin{equation}
D_{x_0}^{\alpha}=-\frac{1}{\cos{(\alpha\pi/2)}}\, _{-L}D_{x_0}^{
\alpha},
\end{equation}
and following a similar procedure as for the case $0<\alpha<1$ with $\beta=1$,
we obtain
\begin{equation}
\fl\langle\tau^m\rangle(x_0)=\frac{\cos^m{(\alpha \pi/2)}\Gamma(1+m)}{\Gamma(m
\alpha){K_\alpha}^m}\int\limits_{-L}^{x_0}(\zeta-x_0)^{m\alpha-1}\mathrm{d}\zeta
=\frac{(2L-d)^{m\alpha}}{\xi^m}\frac{ \Gamma(1+m)}{\Gamma(1+m\alpha)}.
\end{equation}
By setting $\rho=0$ or $\beta=-1$ ($0<\alpha<1$) for $m=1$ we recover the same
result as in equation (\ref{eq:MFPTBoundedGenral}). The behaviour of the MFPT
for this section is similar to the left panels of figure \ref{fig:fig6}, apart
from substituting $d$ for $2L-d$.

\subsubsection{Extremal two-sided $\alpha$-stable processes with $1<\alpha<2$
and $\beta=-1,1$.}
\label{moment bound two-sided}

For extremal two-sided $\alpha$-stable processes with stability index $1<\alpha<2$,
when the initial position is the distance $d$ away from the right boundary
and for skewness $\beta=-1$ (or $\rho=1/\alpha$) in (\ref{eq:MFPTBoundedGenral})
we obtain the MFPT
\begin{equation}
\label{eq:mfpttwoside-1}
\langle\tau\rangle=\frac{d(2L-d)^{\alpha-1}}{\Gamma(1+\alpha)\xi},
\end{equation}
where $\xi$ defied in equation (\ref{eq:xiparameter}). For the case $\beta=1$,
by setting $\rho=1-1/\alpha$ in equation (\ref{eq:MFPTBoundedGenral}) the
following result yields,
\begin{equation}
\label{eq:mfpttwoside1}
\langle\tau\rangle=\frac{d^{\alpha-1}(2L-d)}{\Gamma(1+\alpha)\xi}.
\end{equation}
In contrast to the completely one-sided cases above, in results
(\ref{eq:mfpttwoside-1}) and (\ref{eq:mfpttwoside1}) two factors appear
that include the distances $d$ and $2L-d$. As a direct consequence, we
recognise the completely different functional behaviour in the right
panels of figure \ref{fig:fig6}. Namely, the MFPT decays to zero at both
interval boundaries.

For completely asymmetric LFs the first-passage of the two-sided exit problem
was addressed in \cite{SCPort1970,LTakacs1966,NHBingham1975,JBertoin1996,
ALambert2000,FAvram2004}. A different expression (instead of $d^{\alpha-1}$
in equation (\ref{eq:mfpttwoside1}) it is $d^\alpha$) for the MFPT of completely
asymmetric LFs with $1<\alpha<2$ and $\beta=1$ in dimensionless form was derived
with the help of the Green's function method in \cite{SCPort1970} (see equation
(1.8)). In \cite{JBertoin1996} the distribution of the first-exit time from a
finite interval for extremal two-sided $\alpha$-stable probability laws with
$1<\alpha<2$ and $\beta=-1$ was reported in the Laplace domain.

In the right panels of figure \ref{fig:fig6} we show the MFPT for extremal
$\alpha$-stable processes with skewness $\beta=1$ for two different interval
lengths as function of the initial distance $d$ from the right boundary. To
compare the MFPT of extremal two-sided LFs with arbitrary $\alpha\in(1,2)$
and $\beta=1$ with that of Brownian motion, we employ equation
(\ref{eq:mfpttwoside1}) and obtain
\begin{equation}
\langle\tau\rangle|_{\alpha=2}-\langle\tau\rangle|_{\alpha}=0,
\end{equation}
with $\xi$ defined in equation (\ref{eq:xiparameter}). By solving for $d$,
we find
\begin{equation}
\label{eq:paradtwoside}
d=\left(\frac{2\cos{(\alpha\pi(1/2-1/\alpha))}}{\Gamma(1+\alpha)}\right)^{1/
(2-\alpha)}.
\end{equation}
For $\alpha=1.1$ and $\alpha=1.5$ the MFPT is equal with the Brownian case for
$d=0.261$ and $d=1.132$, respectively. The right side panels of figure \ref{fig:fig6}
indeed demonstrate that as long as the distance $d$ of the initial position of the
LF is within the range $0<d<0.261$ from the right boundary for $\alpha=1.1$ and in
the range $0<d<1.132$ for $\alpha=1.5$, Brownian motion has a shorter MFPT,
otherwise the LF is faster. In general, if $d$ is less than the term on the right
hand side of equation (\ref{eq:paradtwoside}) for arbitrary $\alpha\in(1, 2)$,
Brownian motion is faster on average. In the opposite case, long-range
relocation events and left direction effective drift of LFs with positive
skewness parameter lead to shorter MFPTs.

\subsubsection{General asymmetric $\alpha$-stable processes}
\label{mean gen-asym bound}

\begin{figure}
\centering
\includegraphics[width=0.49\textwidth]{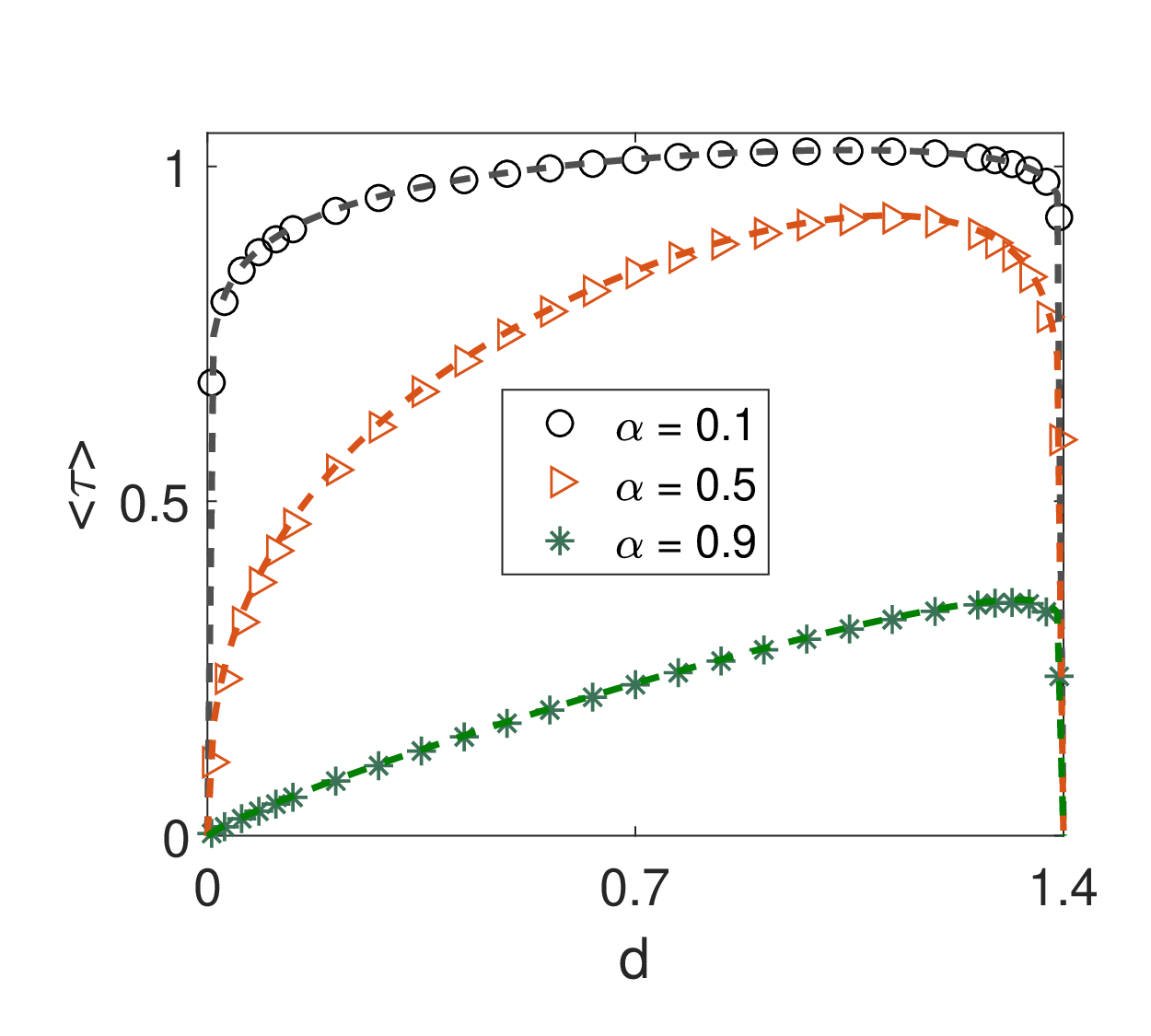}
\includegraphics[width=0.49\textwidth]{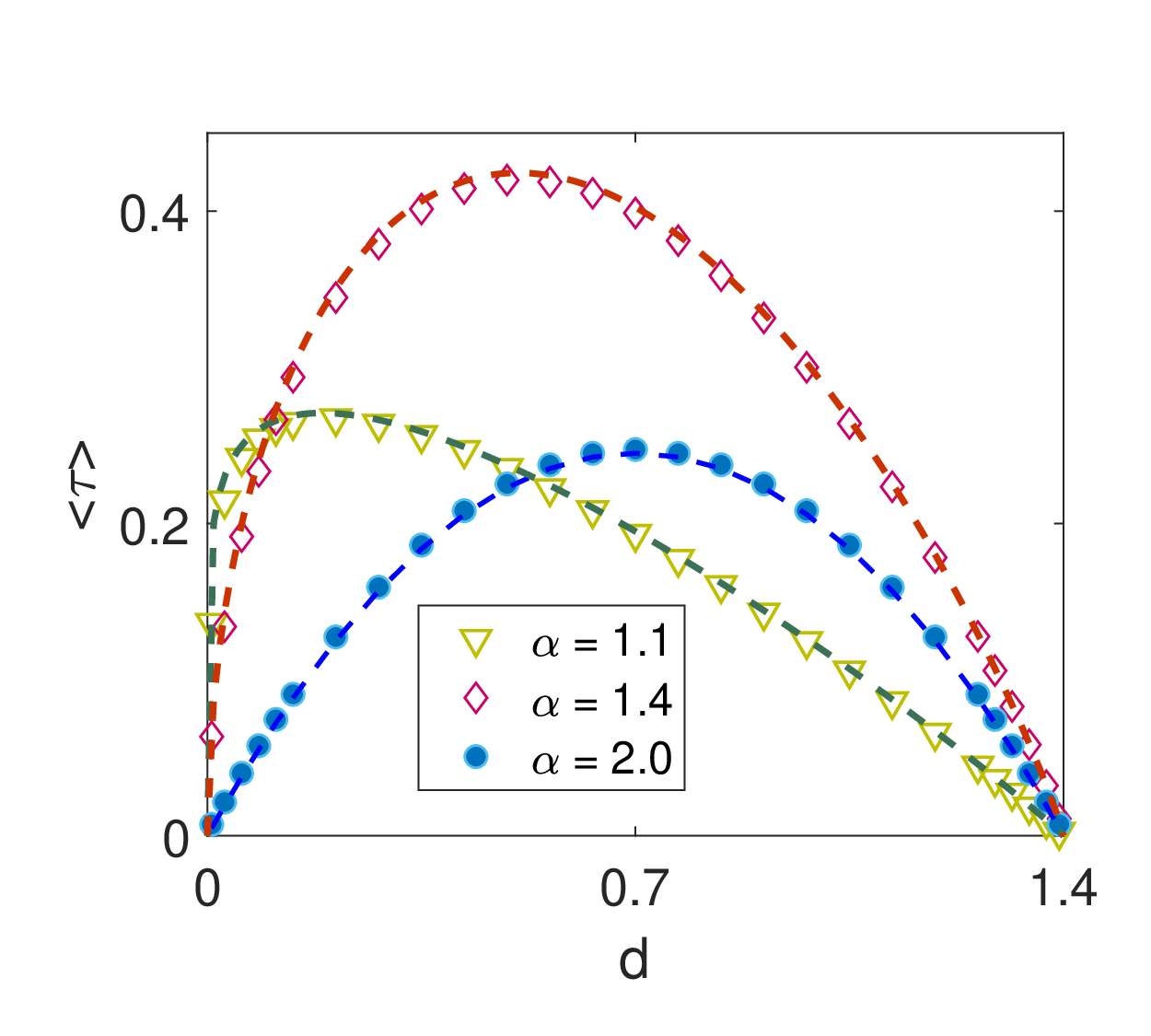}\\
\includegraphics[width=0.49\textwidth]{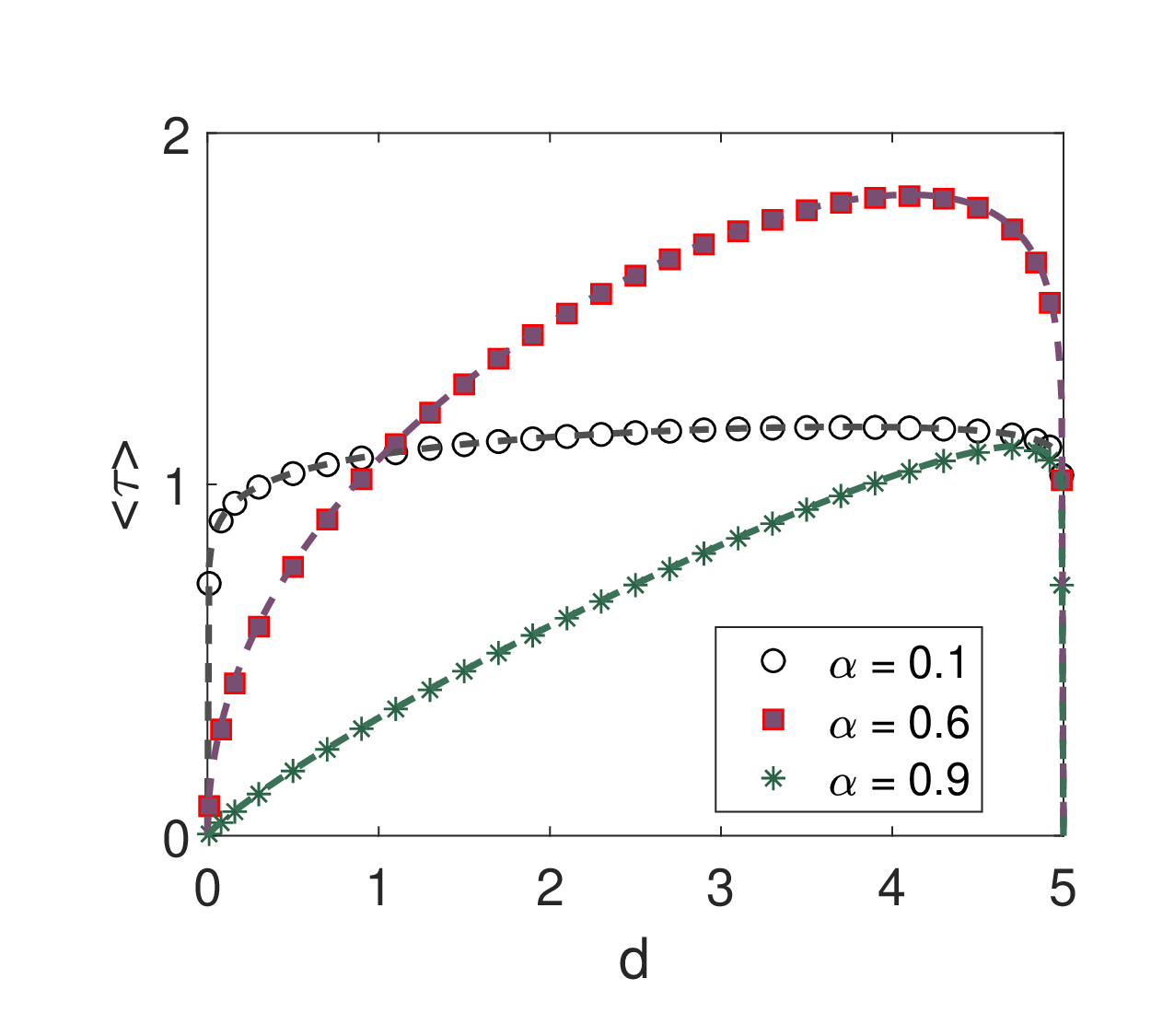}
\includegraphics[width=0.49\textwidth]{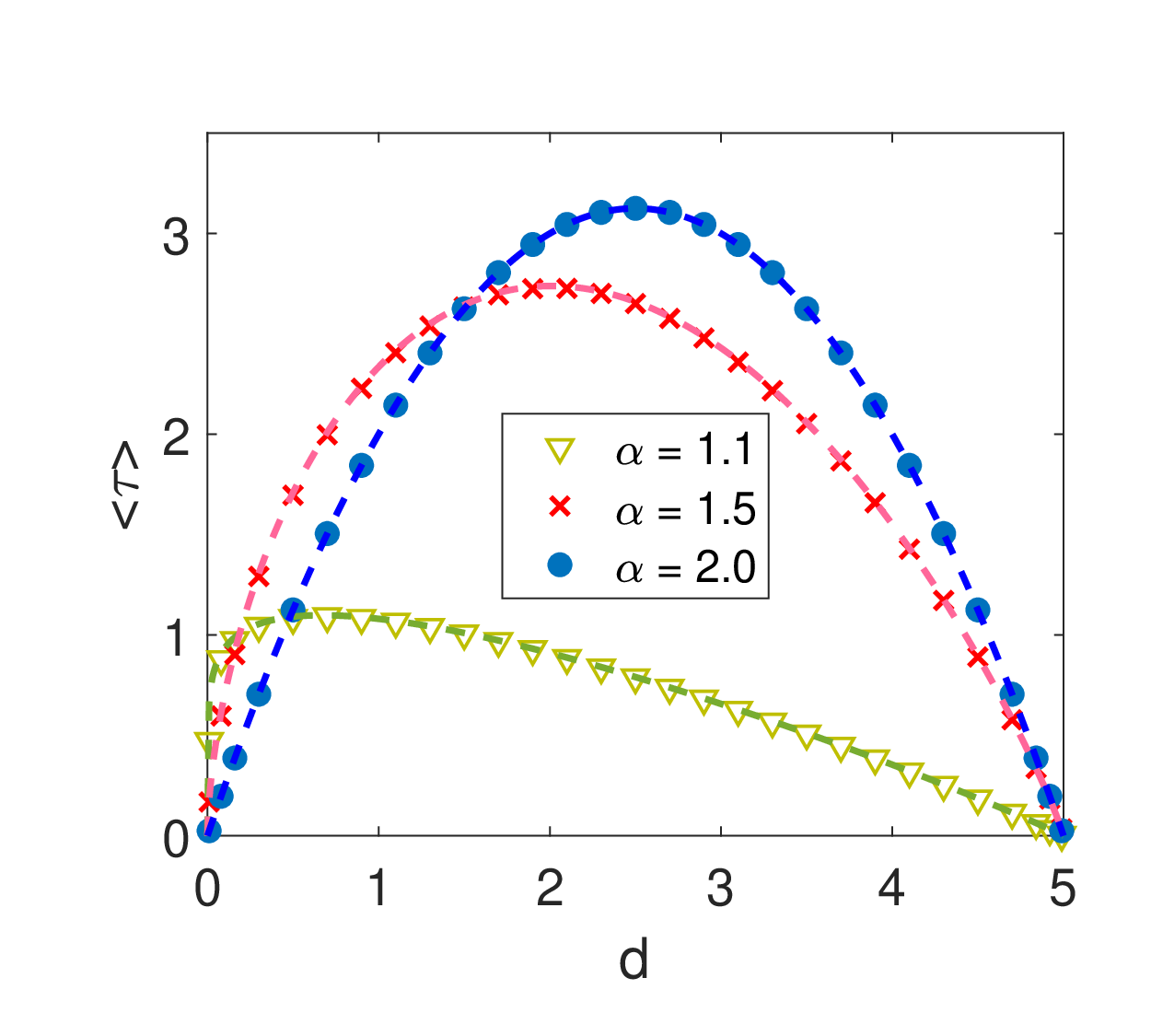}
\caption{Left: MFPT of a general asymmetric $\alpha$-stable process versus
initial distance $d$ from the right boundary for skewness $\beta=0.5$
and $\alpha\in(0,1)$. Right: the same for $\alpha\in(1,2)$. Top: interval length
$L=0.7$. Bottom: interval length $L=2.5$. Symbols are numerical solutions of
the space-fractional diffusion equation, the dashed lines represent equation
(\ref{eq:mfptgeneralasy}).}
\label{fig:fig7}
\end{figure}

We finally show the result for the first-passage time moments of asymmetric
$\alpha$-stable processes with arbitrary skewness $\beta$. The corresponding
result for the MFPT with $\alpha\in(0,2]$ and $\beta\in[-1, 1]$ (excluding
the case $\alpha=1$ and $\beta\neq0$) has the following expression
\begin{equation}
\langle\tau\rangle=\frac{(L-x_0)^{\alpha\rho}(L+x_0)^{\alpha-\alpha\rho}}{
\Gamma(1+\alpha)\xi}.
\end{equation}
Setting $d=L-x_0$ and $2L-d=L+x_0$ we find
\begin{equation}
\label{eq:mfptgeneralasy}
\langle\tau\rangle=\frac{d^{\alpha\rho}(2L-d)^{\alpha-\alpha\rho}}{\Gamma(1
+\alpha)\xi}
\end{equation}
with $\rho$ and $\xi$ defined in equation (\ref{eq:xiparameter}). In figure
\ref{fig:fig7}, analogous to figure \ref{fig:fig6}, we show the MFPT
for $\alpha$-stable processes with skewness $\beta=0.5$ and two different
interval lengths ($L=0.7$ and $L=2.5$). The left panels of figure
\ref{fig:fig7} show the MFPT versus the distance $d$ from the right
boundary for $\alpha$-stable processes with $0<\alpha<1$ and skewness
$\beta=0.5$, for the two different lengths. As can be seen for the smaller
interval, increasing $\alpha$ from $0.1$ to $0.9$, regardless of the initial
position the MFPT decreases. This result can be explained as follows. An
$\alpha$-stable process with stability index $0<\alpha<1$ and skewness
$\beta=0.5$, has a longer tail on the positive axis and a shorter tail on
the negative axis. Moreover, with increasing $\alpha$ from $0.1$ to $0.9$,
the process experiences a larger effective drift to the \emph{right boundary}.
Concurrently, when $\alpha$ decreases (increases), larger (shorter) jumps
are possible with lower (higher) frequency. Therefore, with increasing
$\alpha$ the possibility of shorter jumps with higher frequency and a
larger effective drift toward the right side absorbing boundary arises and
leads to shorter MFPTs. The decay of the MFPT around $d=1.4$, when the initial
position is close to the left boundary, shows us the effect of small jumps of
the negative short tail of the underlying $\alpha$-stable law. The behaviour
of the MFPT in the larger interval is more complicated. For initial positions
with distance $d<1$ from the right boundary increasing $\alpha$ leads to
decreasing MFPTs. This is due to the dominance of an effective drift to the
right and a higher frequency of long jumps when $\alpha$ changes from $0.1$
to $0.9$. Conversely, when $d>1$ we observe two scenarios. First, for $0.1<
\alpha<0.6$, with increasing $\alpha$ MFPT increases. We can explain this
result as follows. By increasing $\alpha$ in the range $(0.1,0.6)$ the long
relocation events dominate the effective drift and higher frequency events
with shorter jump length. Second, for $0.6<\alpha<0.9$, with increasing
$\alpha$ the MFPT decreases. This is now due to the dominance of the effective
drift and higher frequency of shorter jump events against long-range jumps
in the range $0.6<\alpha<0.9$.

$\alpha$-stable processes with $1<\alpha<2$ and $\beta=0.5$, have a heavier
tail on the positive axis and a resulting effective drift to the left. Based
on the above properties, the behaviour of MFPT is quite rich, as can be seen
in the right panels of figure \ref{fig:fig7}. For instance, for interval length
$L=0.7$, when $\alpha\in(1.4, 2)$, regardless of the initial position,
Brownian motion always has a shorter MFPT, whereas for $\alpha\in(1,1.4)$ it
does depends on the initial position. For the interval length $L=2.5$, when the
initial position is located in $2.5<d<5$, smaller $\alpha$ always has a shorter
MFPT. Otherwise, the superiority of LFs over the Brownian particle depends on
its initial position.

\begin{figure}
\centering
\includegraphics[width=0.49\textwidth]{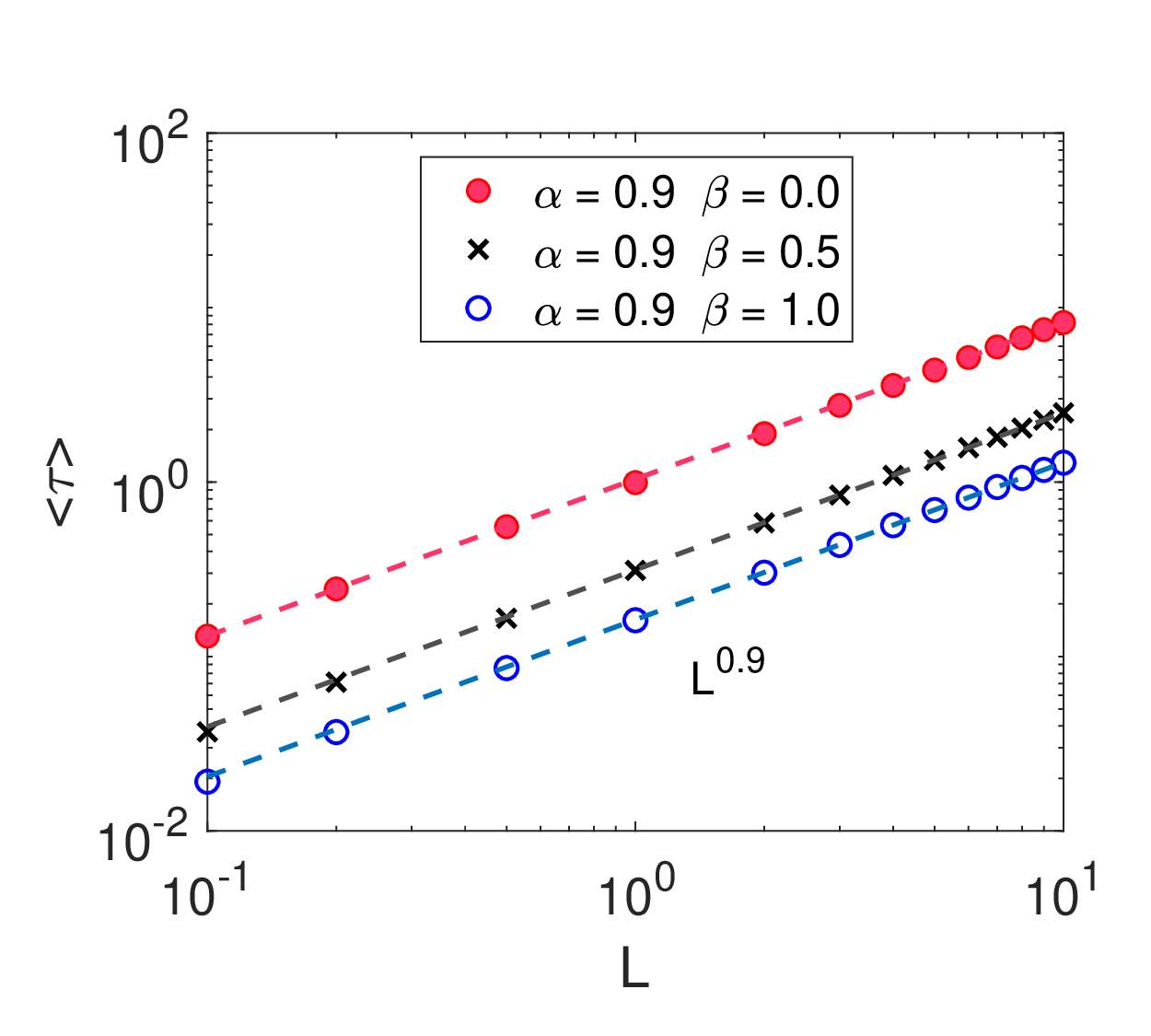}
\includegraphics[width=0.49\textwidth]{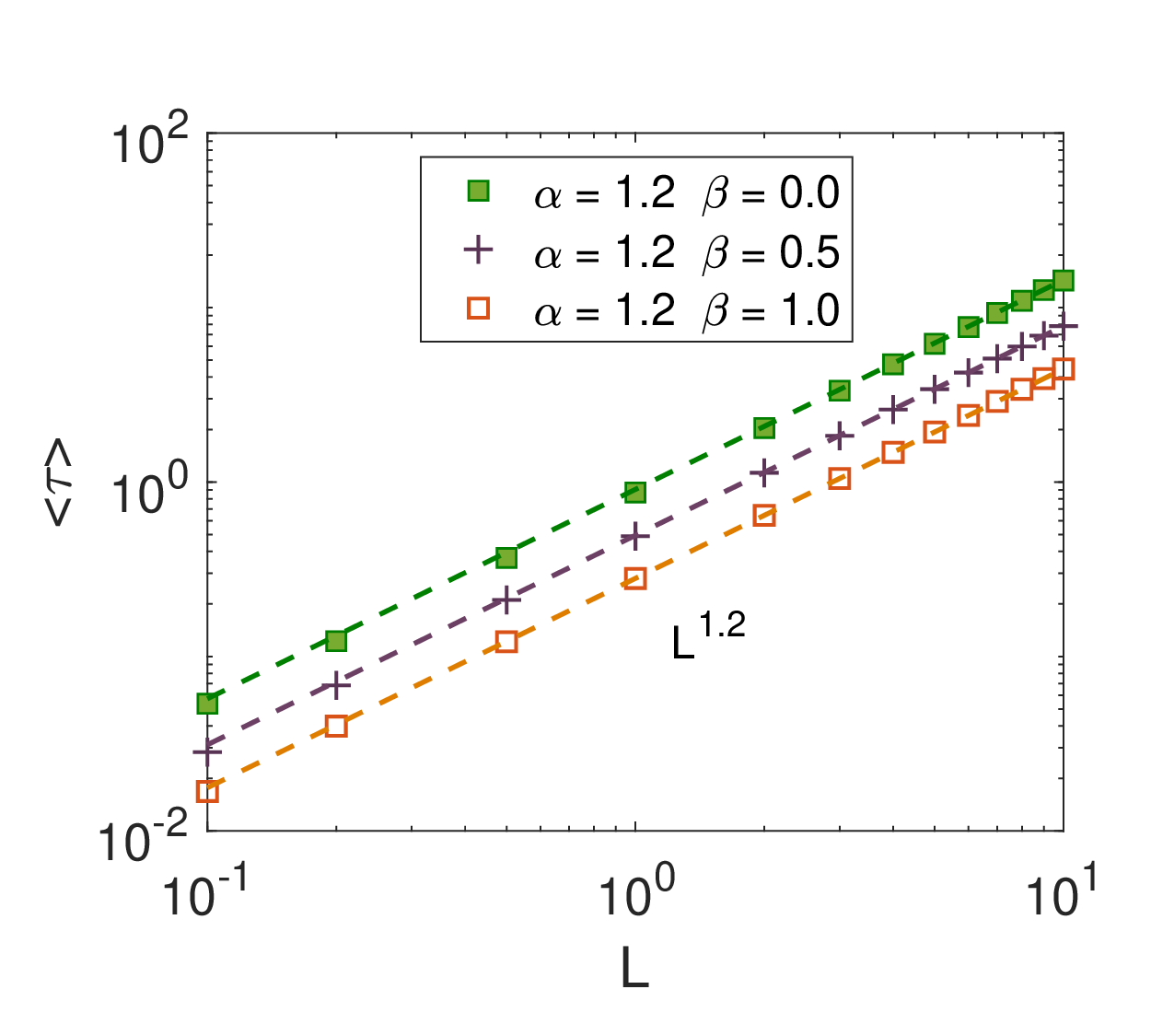}\\
\includegraphics[width=0.49\textwidth]{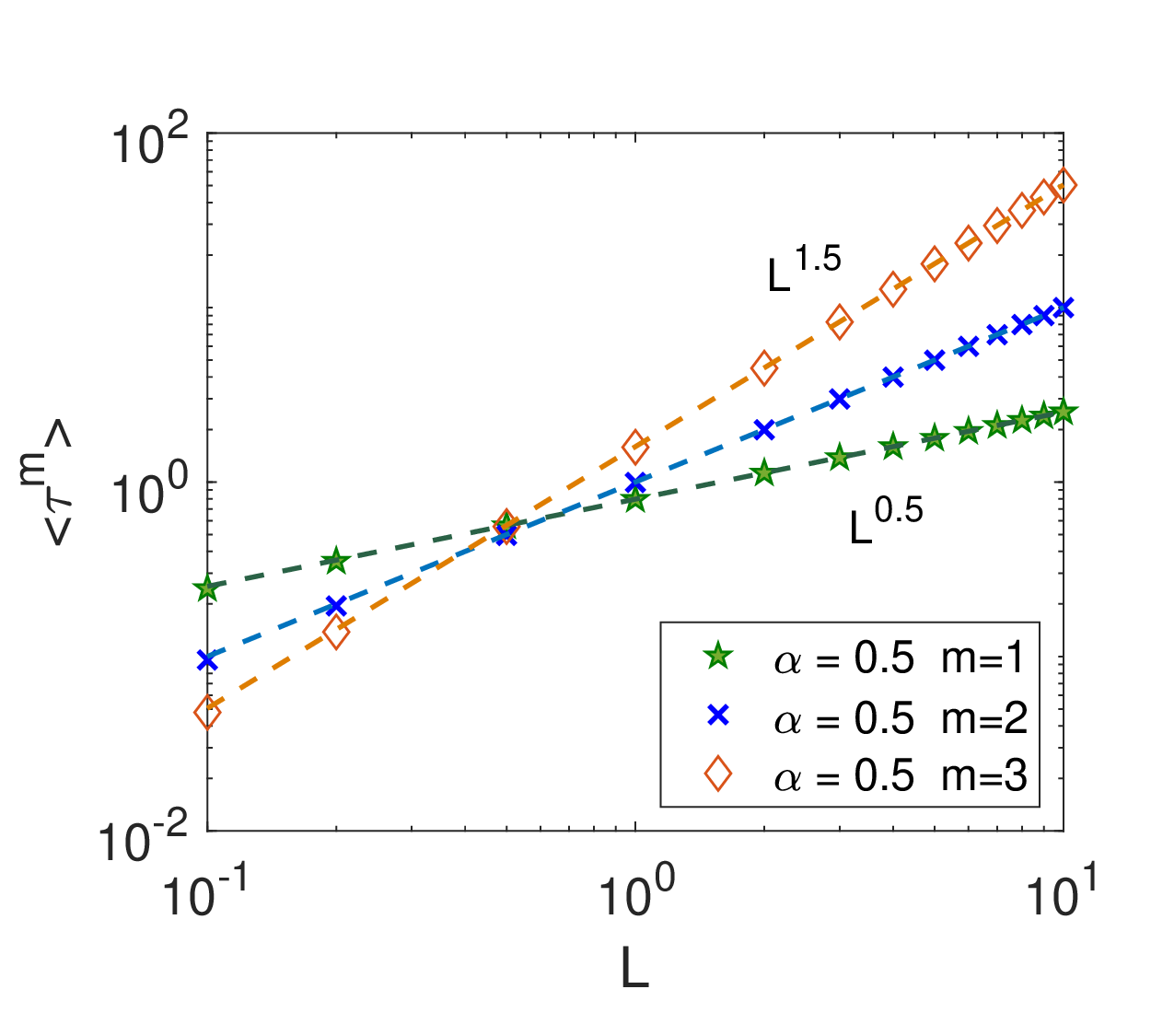}
\includegraphics[width=0.49\textwidth]{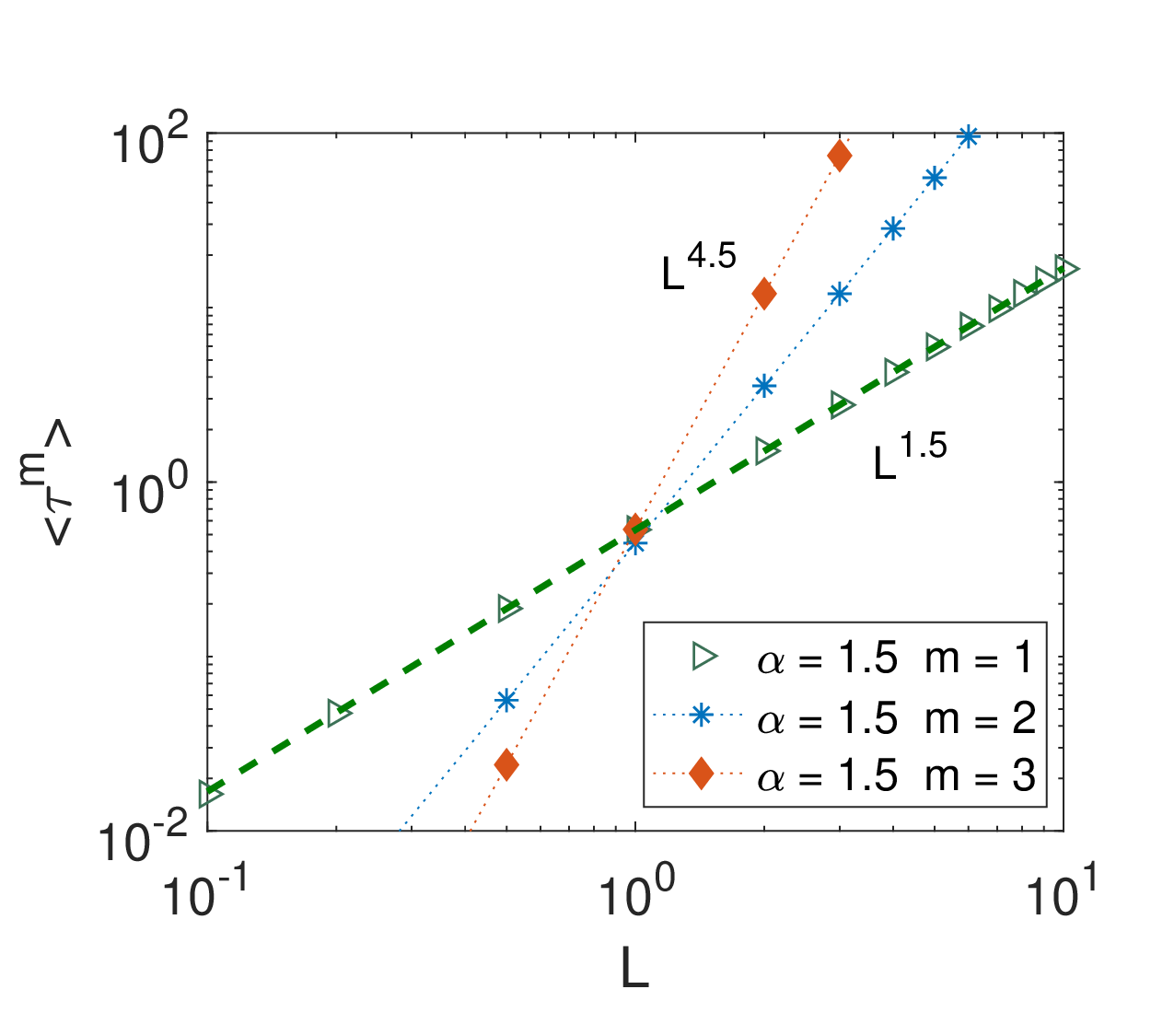}
\caption{Top: MFPT versus interval length $L$ when the initial point is
in the centre of the interval ($d=L$) for different values of the skewness
$\beta$ in log-log scale. Symbols show numerical solutions of the
space-fractional diffusion equation and dashed lines represent
equation (\ref{eq:mfptgeneralasy}). Bottom: higher order moments of the
first-passage time PDF versus interval length $L$ for $\beta=1$ and
two values of the stability index $\alpha$ in log-log scale. Symbols show
the numerical solutions of the space-fractional diffusion equation and
dashed lines are equations (\ref{eq:mfptoneside}) and (\ref{eq:mfpttwoside1}).}
\label{fig:fig8}
\end{figure}

When the initial point of the random process is kept at the centre of the
interval ($x_0=0$), we show results for the MFPT and higher moments of the
first-passage time PDF for different stability $\alpha$ as function of the
interval length for symmetric and asymmetric $\alpha$-stable processes in
figure \ref{fig:fig8}. As can be seen, the moments of the first-passage time
PDF scale like $\langle\tau^m\rangle\sim L^{m\alpha}/K_{\alpha}^m$
independent of the skewness $\beta$.

\subsection{Further properties of the MFPT}

\begin{figure}
\centering
\includegraphics[width=0.48\textwidth]{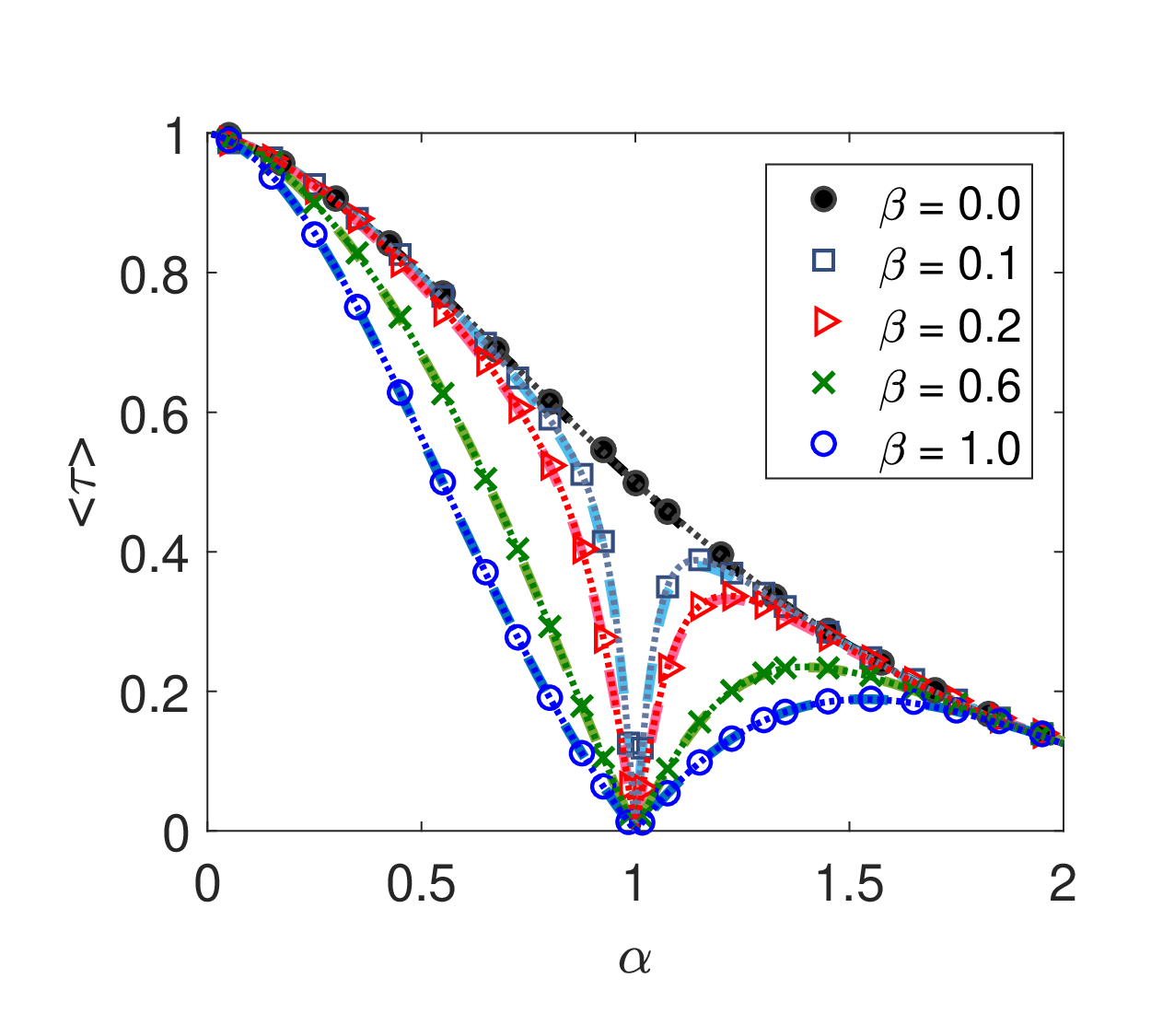}
\includegraphics[width=0.48\textwidth]{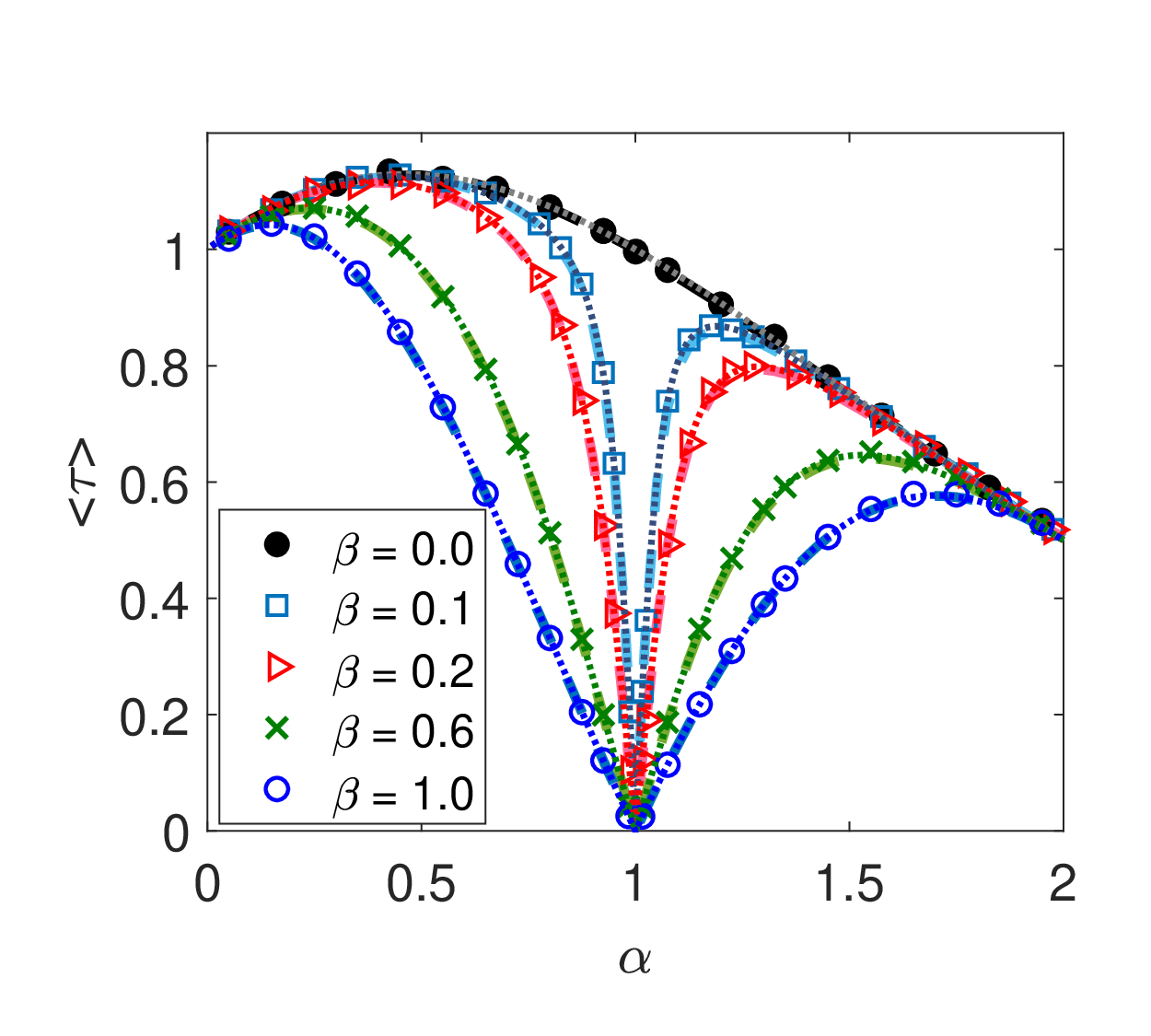}\\
\includegraphics[width=0.48\textwidth]{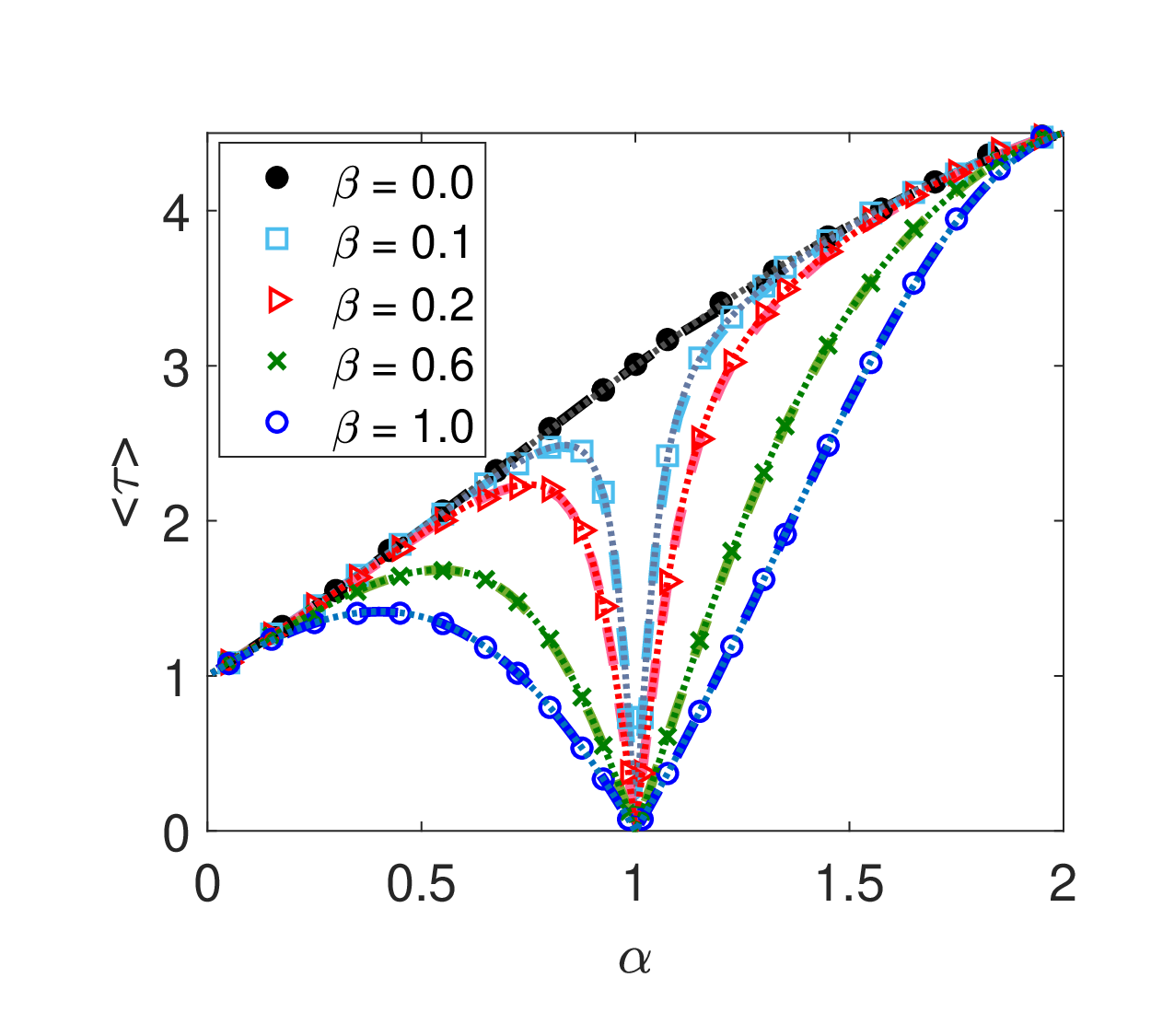}
\caption{MFPT of an asymmetric $\alpha$-stable process for $L=0.5$ (top left),
$L=1.0$ (top right), and $L=3$ (bottom) versus $\alpha$ when the initial
position is in the centre of the interval ($d=L$ in figure \ref{fig:fig1}).
Symbols show results of Langevin equation simulations, dashed lines are based
on the numerical solution of the space-fractional diffusion equation, and the
dotted lines show the analytic solution (\ref{eq:mfptgeneralasy}).}
\label{fig:fig9}
\end{figure}

In this section, we study the MFPT versus the index of stability $\alpha$. In
figure \ref{fig:fig9} we fix the initial position of the random process to
the centre of the interval ($d=L$ in figure \ref{fig:fig1}) and plot the MFPT
versus the stability index $\alpha$ for different skewness $\beta$, for three
different interval lengths $L$. As can be seen, there is a perfect agreement
between the results based on the space-fractional diffusion equation and the
Langevin dynamic approach with the analytic result (\ref{eq:mfptgeneralasy}).
To elucidate the behaviour of the MFPT in figure \ref{fig:fig9} we remind the
reader of some properties of $\alpha$-stable laws. First, $\alpha$-stable
laws with smaller $\alpha$ have a heavier tail and the associated frequency
of long-range relocation events is smaller compared to laws with larger
$\alpha$, for which short jumps with higher frequency are dominant. Second,
symmetric $\alpha$-stable probability laws have the same tail on both sides.
Third, $\alpha$-stable laws with $0<\alpha<1$ and skewness $\beta>0$ have an
effective drift to the right and a longer tail on the positive axis. Moreover,
when $\alpha\to1_-$ with $\beta>0$, the effective drift to the right
direction increases. Conversely, $\alpha$-stable laws with $1<\alpha<2$ and
skewness $\beta>0$ have an effective drift to the left and a longer tail on
the positive axis (see the bottom panel of figure 3 in \cite{AminP2019}).
When $\alpha\to1_+$ with $\beta>0$, the effective drift to the left increases.

For a small interval length ($L=0.5$, top left panel of figure \ref{fig:fig9}),
short relocation events with higher frequency (larger $\alpha$) of symmetric LFs
cross the boundaries quite quickly (full black circles), whereas in large
intervals ($L=3$, bottom panel of figure \ref{fig:fig9}), long-range relocation
events of symmetric LFs lead to shorter MFPTs (full black circles). For
intermediate interval length ($L=1$, top right panel in figure \ref{fig:fig9}),
by increasing $\alpha$ from $0$ to $\approx0.46$ the MFPT increases, but for
$\alpha\in(0.46,2]$ this behaviour reverts. This observation is due to the
tipping balance between long jumps with low frequency and short jumps with
high frequency for $\alpha$ less and larger than $0.46$, respectively.

Conversely, as can be seen from all panels in figure \ref{fig:fig9}, on converging
to the limit $\alpha\to1$ from both sides with skewness $\beta\neq0$, the MFPT tends
to zero, which is in agreement with the analytical result (\ref{eq:mfptgeneralasy}).
To explain this phenomenon we follow \cite{Samorodnitsky-Taqqu,Zolotarev1986} and
first rewrite the characteristic function (\ref{eq:charecA}) and (\ref{eq:charecA1})
of the LFs as
\begin{equation}
\ell_{\alpha,\beta}(k,t)=\exp\left(K_{\alpha}t\left[-|k|^{\alpha}+ik\omega(k,\alpha,
\beta)\right]+i\mu kt\right),
\end{equation}
where
\begin{equation}
\omega(k,\alpha,\beta)=\left\{\begin{array}{ll}|k|^{\alpha-1}\beta\tan(\pi
\alpha/2),&\alpha\neq1\\-(2/\pi)\beta\ln|k|,&\alpha=1\end{array}\right..
\end{equation}
The function $\omega(k,\alpha,\beta)$ is not continuous at $\alpha=1$ and $\beta
\neq0$. However, setting
\begin{equation}
\mu_1=\left\{\begin{array}{ll}\mu+\beta K_{\alpha}\tan(\pi\alpha/2),&\alpha\neq1\\
\mu,&\alpha=1\end{array}\right.
\end{equation}
yields the expression
\begin{equation}
\ell_{\alpha,\beta}=\exp\left(K_{\alpha}t\left[-|k|^{\alpha}+ik\omega_1(k,\alpha,
\beta)\right]+i\mu_1kt\right),
\end{equation}
where
\begin{equation}
\omega_1(k,\alpha,\beta)=\left\{\begin{array}{ll}\beta\left(|k|^{\alpha-1}-1\right)
\tan(\pi\alpha/2),&\alpha\neq1\\-(2/\pi)\beta\ln|k|,&\alpha=1\end{array}\right.
\end{equation}
is a function that is continuous in $\alpha$. Thus for $\beta\neq0$, as the L{\'e}vy
index $\alpha$ approaches unity, the absolute value of the effective drift
$\beta K_{\alpha}\tan(\pi\alpha/2)$ tends to infinity. For $\beta>0$, as seen in
figure \ref{fig:fig9}, the effective drift is directed to the right as $\alpha$
approaches unity from below, $\alpha\to1^-$, and, respectively, to the left as
$\alpha\to1^+$.

\begin{figure}
\centering
\includegraphics[width=0.49\textwidth]{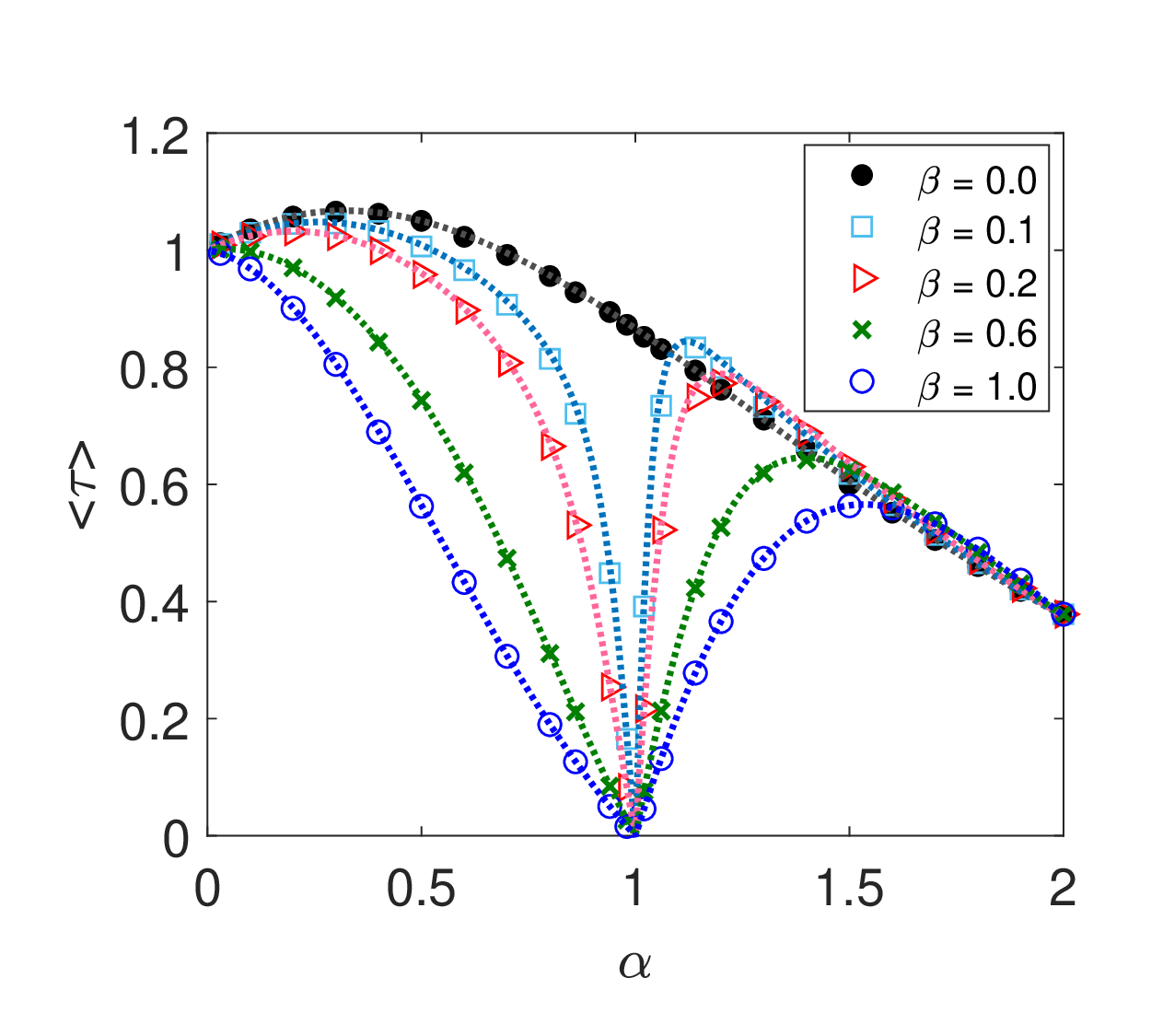}
\includegraphics[width=0.49\textwidth]{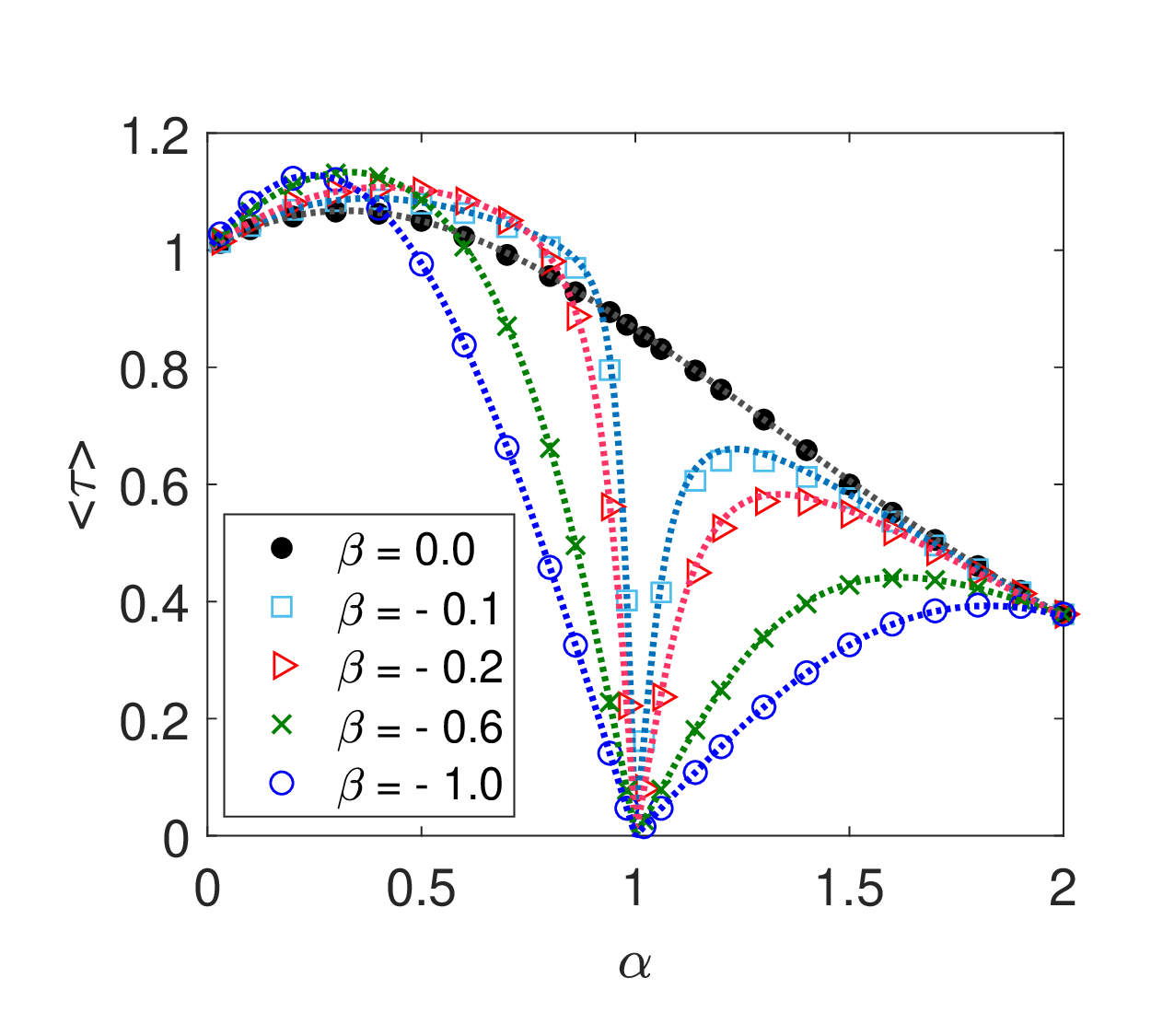}\\
\includegraphics[width=0.49\textwidth]{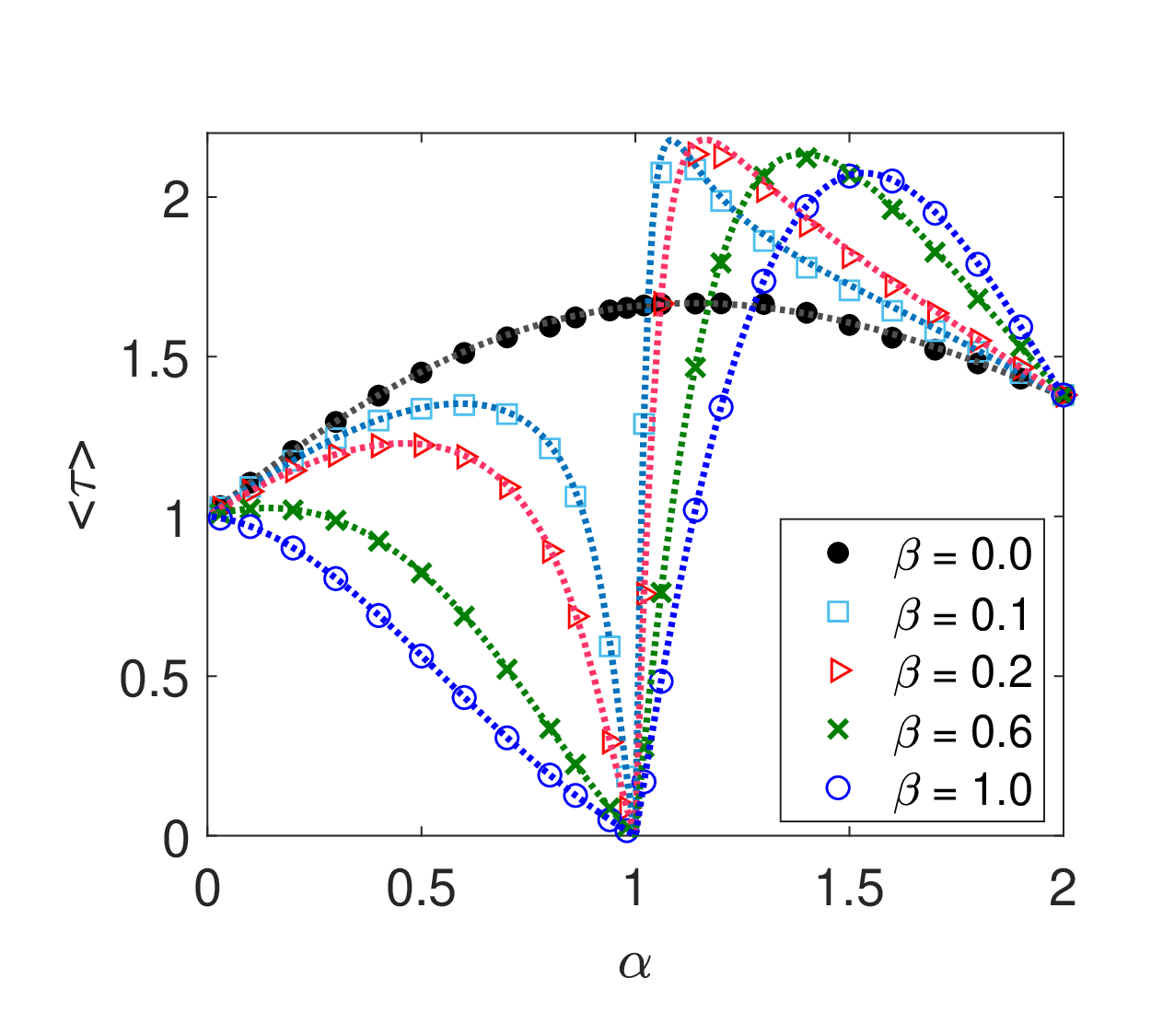}
\includegraphics[width=0.49\textwidth]{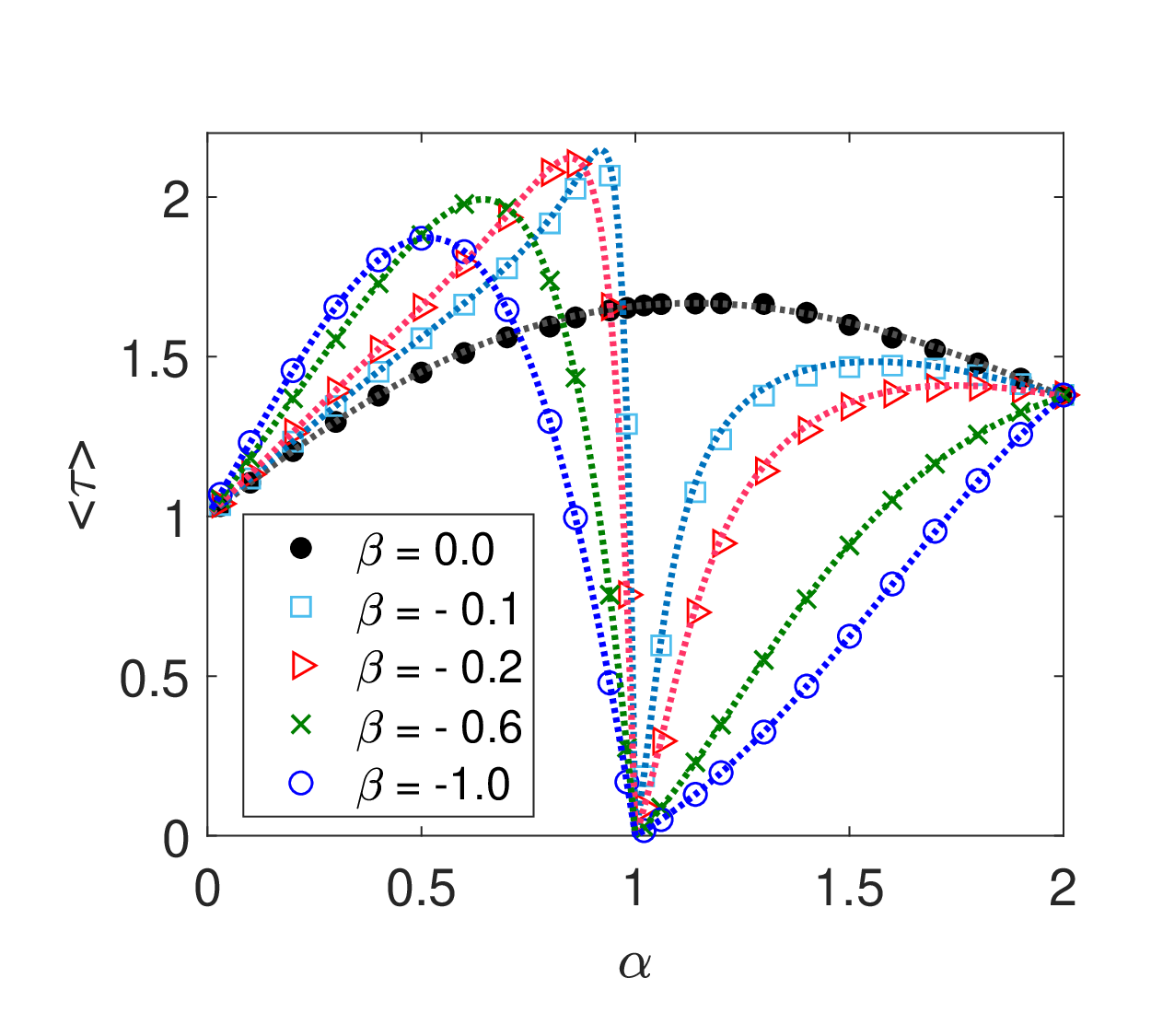}
\caption{MFPT versus $\alpha$ with $d=0.5$ for two interval lengths. Top:
$L=1$. Bottom: $L=3$. Dotted lines show the result (\ref{eq:mfptgeneralasy})
and symbols are the numerical solution of the space-fractional diffusion
equation.}
\label{fig:fig10}
\end{figure}

\begin{figure}
\centering
\includegraphics[width=0.48\textwidth]{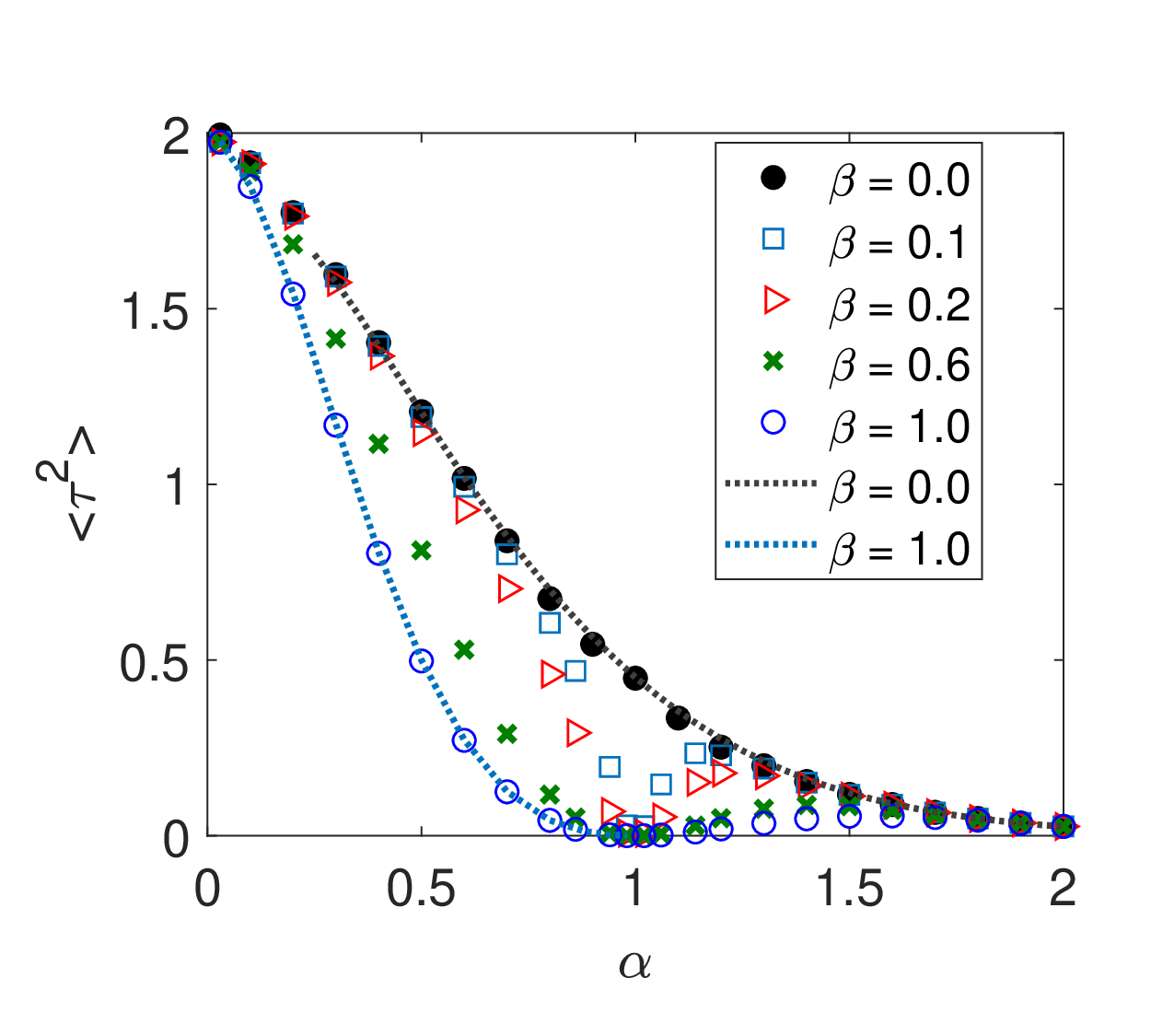}
\includegraphics[width=0.48\textwidth]{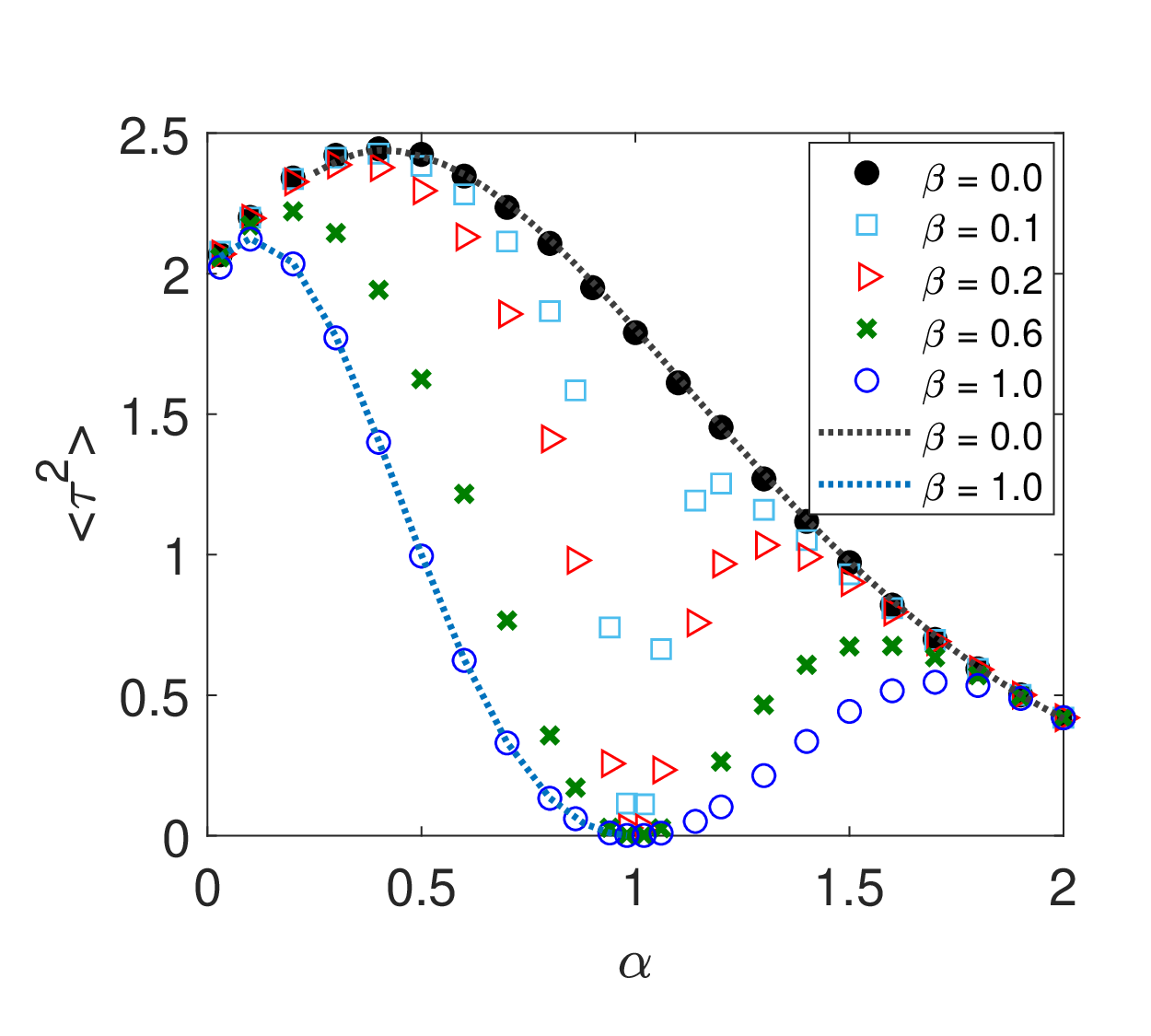}\\
\includegraphics[width=0.48\textwidth]{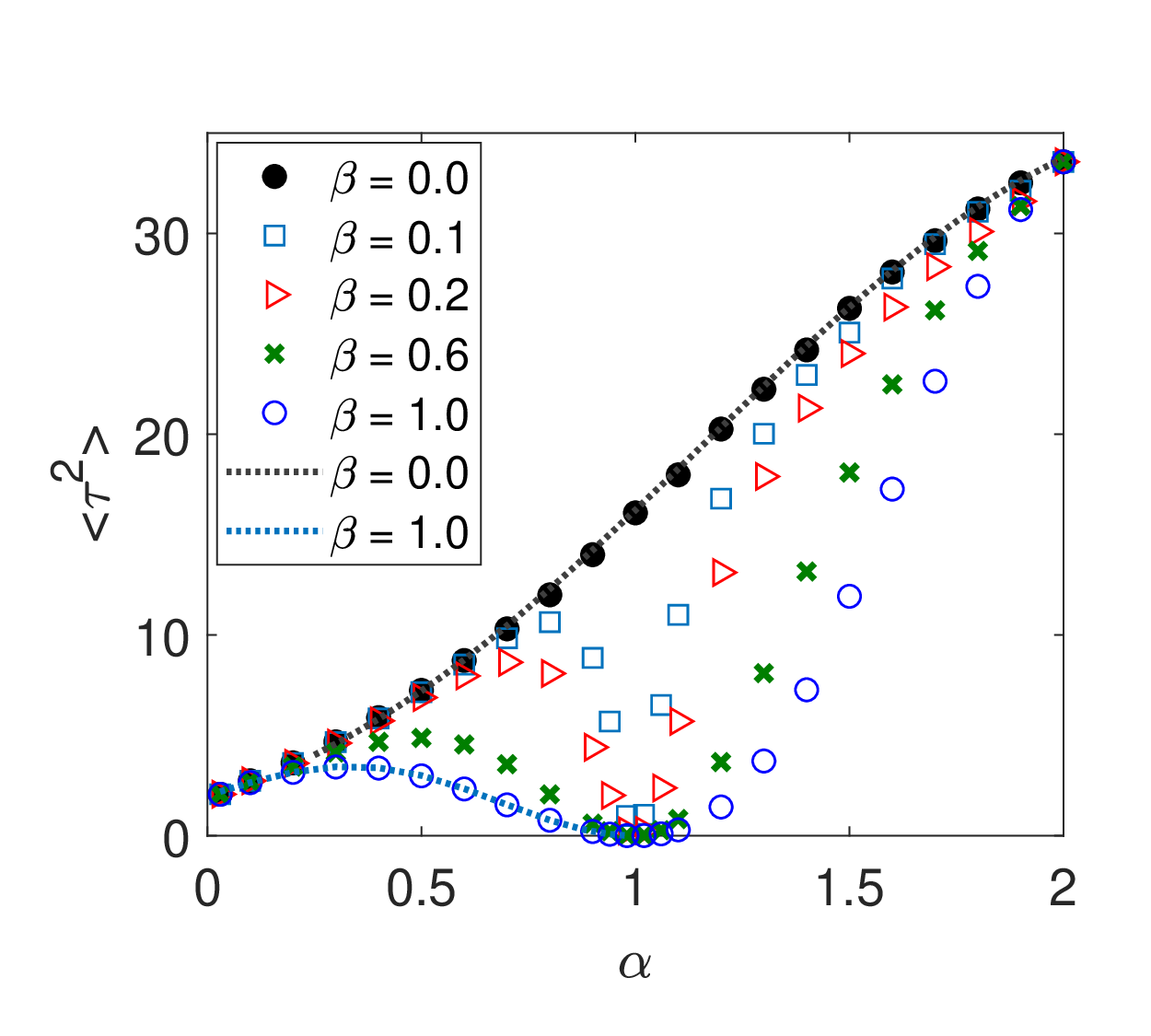}
\caption{Second moment of the first-passage time PDF for interval length $L=0.5$
(top left), $L=1.0$ (top right) and $L=3$ (bottom) versus $\alpha$. The initial
position is in the centre of the interval ($d=L$). Symbols show the numerical
solution of the space-fractional diffusion equation and dotted lines show the
analytic solution (\ref{eq:second-moment-sym}) for the symmetric case ($\beta=0$)
and (\ref{eq:moments-oneside-bound}) with $m=2$ (one-sided $0<\alpha<1,\beta=1$).}
\label{fig:fig11}
\end{figure}

We now change the scenario and set the initial position at a distance $d=0.5$
away from the right boundary. Figure \ref{fig:fig10} analyses the MFPT versus
$\alpha$ and different skewness $\beta$ for two different interval lengths
($L=1$ and $L=3$). As can be seen, there is a perfect agreement between the
results based on the numerical solution of the space-fractional diffusion equation
and the analytic solution (\ref{eq:mfptgeneralasy}). In comparison with the
symmetric initial position of the random process in figure \ref{fig:fig9},
for positive values of the skewness parameter and when $\alpha\in(0,1)$,
since the initial point is closer to the right boundary and the effective
drift is in direction of the positive axis, the MFPT decreases. For $\alpha
\in(1,2)$ and positive skewness, the effective drift is toward the left, and the
MFPT increases rapidly. The opposite behaviour is observed when the skewness
is negative (figure \ref{fig:fig10}, right panels): for $\alpha\in(0,1)$ and
$\alpha\in(1,2)$ with $\beta<0$, the effective drift is to the left and right
directions, respectively.

In figure \ref{fig:fig11}, analogous to figure \ref{fig:fig9}, we show the
results for the second moment of the first-passage time PDF versus the stability
index $\alpha$ for different sets of the skewness parameter $\beta$ when the
initial position is in the centre of the interval ($d=L$).

\begin{figure}
\centering
\includegraphics[width=0.49\textwidth]{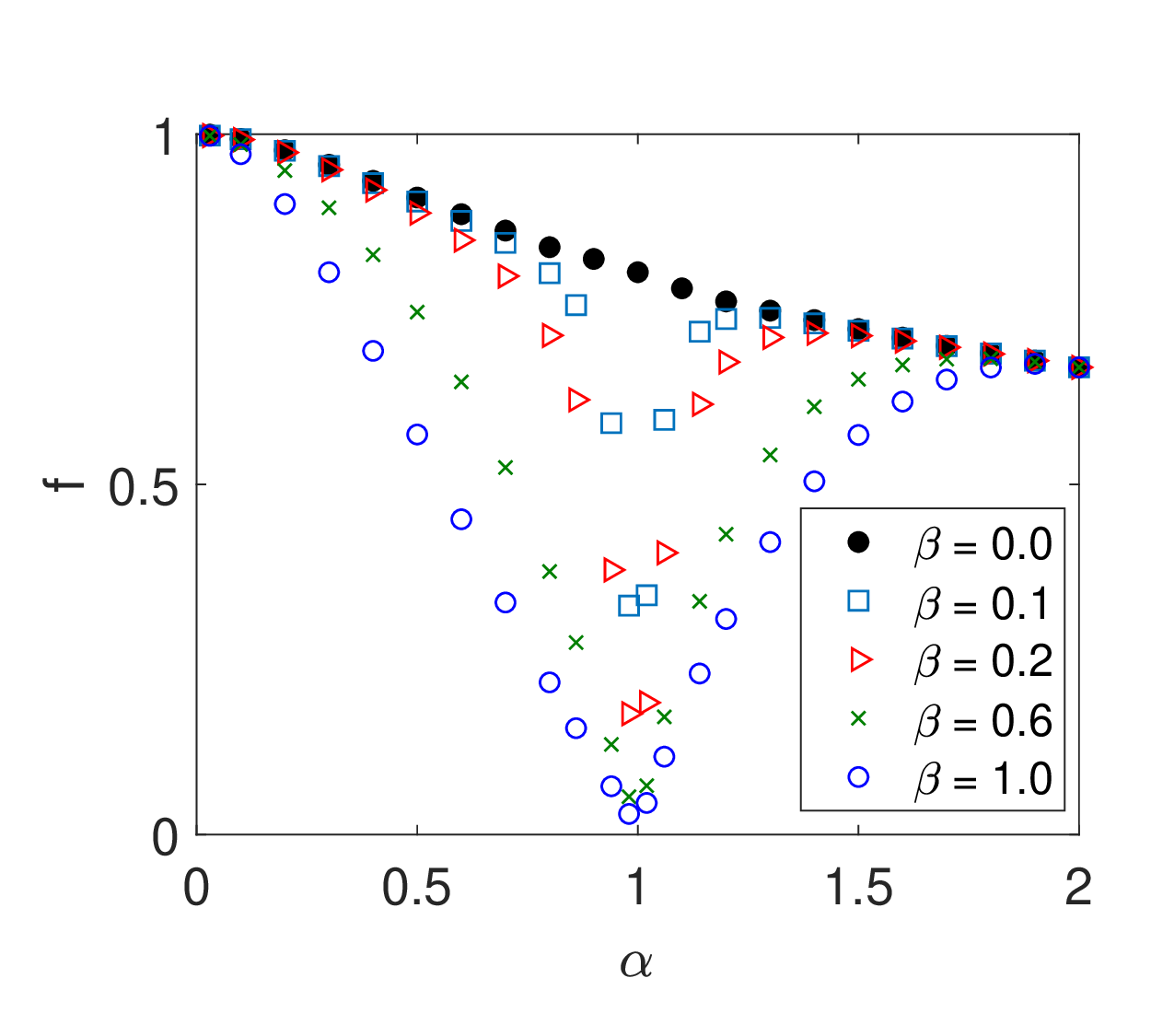}
\caption{Coefficient of variation $f$ versus $\alpha$ for the first-passage
time PDF with initial distance $L=d$.}
\label{fig:fig12}
\end{figure}

Finally, in figure \ref{fig:fig12} we show the coefficient of variation
\begin{equation}
\label{f-parameter}
f=\frac{\langle\tau^2\rangle-{\langle\tau\rangle}^2}{{\langle\tau\rangle}^2}.
\end{equation}
When $f>1$ the underlying distribution
is broad and we need to study higher order moments to get the complete information
of the first-passage time PDF. When $f<1$, the distribution is narrow and
higher order moments are not needed. For the one-sided $\alpha$-stable process
($0<\alpha<1$ and $\beta=1$) recalling equation (\ref{eq:moments-oneside-bound}),
the coefficient of variation reads
\begin{equation}
f=\frac{\frac{\Gamma(1+2)}{\Gamma(1+2\alpha)}\frac{d^{2\alpha}}{\xi^{2}}-\left(
\frac{\Gamma(1+1)}{\Gamma(1+\alpha)}\frac{d^{\alpha}}{\xi}\right)^2}{\left(\frac{
\Gamma(1+1)}{\Gamma(1+\alpha)}\frac{d^{\alpha}}{\xi}\right)^2}=\frac{2\Gamma{(1+
\alpha)}^2}{\Gamma{(1+2\alpha)}}-1,
\end{equation}
which is always less than one, compare also figure \ref{fig:fig12}. Thus, the MFPT
is a fairly good measure for the first-passage process.

\section{Discussion and unsolved problems}
\label{concl}

LFs are relevant proxy processes to study the efficiency and spatial exploration
behaviour of random search processes, from animals ("movement ecology") and humans
to robots and computer algorithms. Apart from the MFPT such processes can be
studied in terms of the mean inverse first-passage time $\langle1/\tau\rangle$ as
well as fractional order moments. Here we quantified the first-passage dynamics of
symmetric and asymmetric LFs in both semi-infinite and finite domains and obtained
the moments of the associated first-passage time PDF. These moments were analyses
as functions of the process parameters, the stable index $\alpha$ and skewness
$\beta$, as well as the system parameters, the initial distance $d$ and the
interval length $L$ (if not infinite). As seen in the results the behaviour for
different parameters can be quite rich and requires careful interpretation. Table
\ref{tab} summarises the main features.

\setlength{\tabcolsep}{14pt}
\begin{table}
\centering
\begin{tabular}{|c|c|c|c|}
\hline\hline
$\alpha$ & $\beta$ & Semi-infinite domain & Bounded domain\\
\hline \hline
\multirow{2}{*}{2} & \multirow{2}{*}{Irrelevant} & \multirow{2}{*}{(\ref{eq:Brown-Semi-Moment}), \,\,\, $-\infty<q<1/2$ \cite{SRedner2001}} & $\langle \tau \rangle \to$ (\ref{eq:mfptsym}) \cite{RKGetoor1961, SVBuldyrev2001-2,SRedner2001}\\
& & &  $\langle \tau^2 \rangle \to$ (\ref{eq:Brown-Bound-secmomen}) \cite{RKGetoor1961,SRedner2001}\\
\hline \hline
\multirow{2}{*}{(0, 2)} & \multirow{2}{*}{0} & \multirow{2}{*}{Unknown, \,\,  $-1<q<1/2$ } & $\langle \tau \rangle \to$ (\ref{eq:mfptsym}) \cite{RKGetoor1961, SVBuldyrev2001-2, AZoia2007}\\
& & &  $\langle \tau^2 \rangle \to$ (\ref{eq:second-moment-sym}) \cite{RKGetoor1961}\\
\hline \hline
(0,1) & 1 & \multicolumn{2}{|c|}{(\ref{eq:mfptoneside}), (\ref{eq:moments-oneside-bound}), \, $-1<q<\infty$, ($q=m=1$ \cite{TKoren2007, SCPort1970})}\\
\hline\hline
\multirow{2}{*}{(1, 2)} & -1& (\ref{eq:Extr-Two--1-Semi-moment}), $-\infty<q<1/\alpha$ & $\langle \tau \rangle \to$ (\ref{eq:mfpttwoside-1})\\
\cline{2-4}& 1 &  (\ref{eq:Extr-Two-1-Semi-moment}), $-1<q<1-1/\alpha$ & $\langle \tau \rangle \to$ (\ref{eq:mfpttwoside1})\\
\hline\hline
(0, 1) & \multirow{2}{*}{(-1, 1)}& \multirow{2}{*}{Unknown, $-1<q<\rho$ \cite{RADoney2004}} & \multirow{2}{*}{$\langle \tau \rangle \to$ (\ref{eq:mfptgeneralasy})}\\
\cline{1-1}(1, 2) & & &\\
\hline\hline
\end{tabular}
\caption{First-passage time PDF moments for different $\alpha$ and skewness $\beta$.}
\label{tab}
\end{table}

We here studied the one-dimensional case, for which the effect of LF versus
Brownian search is expected to be most significant. One-dimensional is relevant
for the vertical search of seaborne predators \cite{sims,sims1} as well as random
search along, for instance, natural boundaries such as field-forest boundaries or
the shrubbery growing along streams. Other direct applications include search in
computer algorithms \cite{ilya,ilya1} or the effective one-dimensional search on linear
polymer chains where LFs are effected by jumps to different chain segments at
points where the polymer loops back onto itself \cite{lomholt,igor}.
In a next step it will be of interest to extend these results to two dimensions,
which is the relevant situation for a large number of search and movement processes.
Another important direction of future research is to study the influence of
interdependence on the first-passage properties for processes with infinite variance.
Indeed, when the specific stochastic process is considered in a bounded domain the
analysis of correlations in this process is important \cite{vojta,tobiasg}.
Fractional LFs with long-range dependence have been detected in beat-to-beat heart
rate fluctuations \cite{peng}, in solar flare time series \cite{burnecki}, and they
have been shown to be a model qualitatively mimicking self-organized criticality
signatures in data \cite{watkins1}. Apparently, correlations or spectral power
analysis, strictly speaking, cannot be used for LFs, and alternative measures
of dependence are necessary, see, e.g., the review \cite{agnes}.

In many situations for diffusive processes cognisance of the MFPT is insufficient
to fully characterise the first-passage statistic. This statement was quantified in
terms of the uniformity index statistic in \cite{carlos,thiago}. Instead, it is
important to know the entire PDF of first-passage times, even in finite domains
\cite{aljaz,aljaz1,denis,denis1}. Such notions are indeed relevant for biological
processes, for instance, in scenarios underlying gene regulation, for which the
detailed study reveals a clear dependence on the initial distance, which thus
goes beyond the MFPT \cite{kolesov,otto,prathita}. While we here saw that the
coefficient of variation of the first-passage statistic is below unity, it will
have to be seen, for instance, how this changes to situations of first-arrival
to a partially reactive site. Another feature to be included are many-particle
effects, for instance, flocking behaviour provoking different hunting strategies
\cite{gleb,more?,maria}.

\ack

AP acknowledges funding from the Ministry of Science, Research and Technology of
Iran and Potsdam University in Germany. Computer simulations were performed at
the Shahid Beheshti University (Tehran, Iran) and Potsdam University (Potsdam,
Germany). This research was supported in part by PL-Grid Infrastructure. ACh
and RM acknowledge support from the DFG project 1535/7-1. RM also acknowledges
support from the Foundation for Polish Science (Fundacja na rzecz Nauki Polskiej)
within an Alexander von Humboldt Polish Honorary Research Scholarship.
MM acknowledges support from NCN-DFG Beethoven Grant No.2016/23/G/ST1/04083.

\appendix

\section{Generator and backward Kolmogorov equation for an LF killed upon leaving
the domain}
\label{DX0derivation}

Let $\tau=\min\{t:|x(t)|\geqslant L\}$ be the first-passage time of an LF $x(t)$.
Let us define the corresponding \emph{killed process\/} on $[-L,L]$ as
\begin{equation}
\label{inequality}
\bar{x}(t)=\left\{\begin{array}{lcl}x(t) & & {\rm if\;} t<\tau  \\
\partial & & {\rm if\;} t\geq \tau \\ \end{array} \right..
\end{equation}
Here, $\partial$ is the so-called "cemetery state". It is a domain outside of
the interval $[-L,L]$. Note that the process $\bar{x}(t)$ describes the dynamics
of the LF confined to the interval $[-L,L]$. When the LF leaves the domain,
$\bar{x}(t)$ moves to the cemetery state and stays there forever.

The key property here is that $\bar{x}(t)$ is also a Markov process
\cite{meerschaert}. Therefore one can define its generator $D^{\alpha}_x$ in a
usual way. This generator is equal to the generator of LFs confined to the
interval $[-L,L]$ \cite{meerschaert}. It has the form
\begin{equation}
\label{marcin}
D^{\alpha}_xf(x)=R_{\alpha,\beta}\, _{-L}D_x^{\alpha}f(x)+L_{\alpha,\beta}\,
_xD_L^{\alpha}f(x),
\end{equation}
for appropriately smooth function $f(x)$. Here $_{-L}D_x^{\alpha}$ and $_xD_L^{
\alpha}$ are the fractional derivatives defined in (\ref{eq:lcapu}) and
(\ref{eq:rcapu}), respectively. Moreover, $L_{\alpha,\beta}$ and $R_{\alpha,\beta}$
are the constants defined in equation (\ref{eq:weightcoeffRL}). Here we employ
an important property, namely, that under absorbing boundary conditions the
adjoint operator of the left derivative (\ref{eq:lcapu}) is equal to the right derivative (\ref{eq:rcapu}) and vice versa \cite{podlubny}.

Consequently, it follows from the general theory of Markov processes
\cite{pavliotis} that the
PDF $P_{\alpha,\beta}(x,t|x_0)$ of the killed process starting at $x_0$ satisfies
the backward Kolmogorov equation
\begin{equation}
\frac{\partial P_{\alpha,\beta}(x,t|x_0)}{\partial t}=K_{\alpha}\,D^{\alpha}_{x_0}
P_{\alpha,\beta}(x,t|x_0),
\end{equation}
where $D^{\alpha}_{x_0}$ is given by (\ref{marcin}) with $x$ replaced by $x_0$.
Finally, knowing the generator of $\bar{x}(t)$ and the corresponding backward
Kolmogorov equation one can apply the usual method of finding the mean
first-passage time of the LF described in detail in section \ref{frac-FPT}.

\section{Derivation of MFPT for general $\alpha$-stable process in a finite interval}
\label{appendixMFPT}

Here we compute the MFPT of LFs with stability index $\alpha\in(0,2]$ and skewness
$\beta\in[-1, 1]$ (excluding $\alpha=1$ with $\beta\neq 0$). To determine the
parameters $\mu$ and $\nu$, by substitution of equation (\ref{eq:mfptboundgensolution})
into (\ref{eq:boundmfpt}) we get
\begin{eqnarray}
\nonumber
\frac{K_{\alpha}C_{\alpha,\beta}R_{\alpha,\beta}}{\Gamma(n-\alpha)}\int\limits_{
-L}^{x_0}\frac{\left((L-\zeta)^{\mu}(L+\zeta)^{\nu}\right)^{(n)}}{(
x_0-\zeta)^{\alpha-n+1}}\mathrm{d}\zeta\\
+\frac{(-1)^{n}K_{\alpha}C_{\alpha,\beta}L_{\alpha,\beta}}{\Gamma(n-\alpha)}\int
\limits_{x_0}^{L}\frac{\left((L-\zeta)^{\mu}(L+\zeta)^{\nu}\right)^{(n)}}{
(\zeta-x_0)^{\alpha-n+1}}\mathrm{d}\zeta=-1 .
\end{eqnarray}
Let us first consider the case $n=1$ ($0<\alpha<1$). By taking the first derivative
\begin{eqnarray}
\nonumber
\fl\frac{K_{\alpha}C_{\alpha,\beta}R_{\alpha,\beta}}{\Gamma(1-\alpha)}\int\limits_{
-L}^{x_0}\frac{\nu(L-\zeta)^{\mu}(L+\zeta)^{\nu-1}-\mu(L-\zeta)^{\mu-1}
(L+\zeta)^{\nu}}{(x_0-\zeta)^{\alpha}}\,\mathrm{d}\zeta\\
\fl-\frac{K_{\alpha}C_{\alpha,\beta}L_{\alpha,\beta}}{\Gamma(1-\alpha)}\int\limits_{
x_0}^{L}\frac{\nu(L-\zeta)^{\mu}(L+\zeta)^{\nu-1}-\mu(L-\zeta)^{\mu-1}
(L+\zeta)^{\nu}}{(\zeta-x_0)^{\alpha}}\,\mathrm{d}\zeta=-1 ,
\label{eq:MFPTHypergeo1}
\end{eqnarray}
then, by change of variables $y=(x_0-\zeta)/(x_0+L)$ and $y=(\zeta-x_0)/(L-x_0)$
in the first and second integral on the left hand side, respectively, we have
\begin{eqnarray}
\nonumber
\fl(L+x_0)^{\nu-\alpha}(L-x_0)^{\mu}\\
\times\int\limits_{0}^{1}\frac{\nu(1+\frac{L+x_0}{
L-x_0}y)^{\mu}(1-y)^{\nu-1}-\mu\frac{L+x_0}{L-x_0}(1+\frac{L+x_0}{L-x_0}y)^{\mu-
1}(1-y)^{\nu}}{y^{\alpha}}\mathrm{d}y,
\end{eqnarray}
and
\begin{eqnarray}
\nonumber
\fl(L+x_0)^{\nu}(L-x_0)^{\mu-\alpha}\\
\times\int\limits_0^1\frac{\nu\frac{L-x_0}{L+x_0}(
1-y)^{\mu}(1+\frac{L-x_0}{L+x_0}y)^{\nu-1}-\mu(1-y)^{\mu-1}(1+\frac{L-x_0}{L+x_0}
y)^{\nu}}{y^{\alpha}}\mathrm{d}y.
\end{eqnarray}
Then, defining $z=(L+x_0)/(L-x_0)$ and using the integral representation
of the Gauss hypergeometric function \cite{NNLebedev1972} (see equation (9.1.6)),
\begin{equation}
F(a;b;c;x)=\frac{\Gamma(c)}{\Gamma(b)\Gamma(c-b)}\int\limits_{0}^{1}\frac{
t^{b-1}(1-t)^{c-b-1}}{(1-x t)^{a}}\mathrm{d}t,
\label{eq:hypergeometricfunc}
\end{equation}
where $\mathrm{Re}(c)>\mathrm{Re}(b)>0$ and $|\arg(1-z)|<\pi$, we obtain
\begin{eqnarray}
\nonumber
\fl\frac{K_{\alpha}C_{\alpha,\beta}R_{\alpha,\beta}}{\Gamma(1-\alpha)}(L+
x_0)^{\nu-\alpha}(L-x_0)^{\mu}\Big[\frac{\Gamma(1-\alpha)
\Gamma(1+\nu)}{\Gamma(1+\nu-\alpha)}F(-\mu;1-\alpha;1+\nu-\alpha;-z)\\
\nonumber
\fl-\mu z\frac{\Gamma(1-\alpha)\Gamma(1+\nu)}{\Gamma(2+\nu-\alpha)}F(1-\mu;
1-\alpha;2+\nu-\alpha;-z)\Big]\\
\nonumber
\fl-\frac{K_{\alpha}C_{\alpha,\beta}L_{\alpha,\beta}}{\Gamma(1-\alpha)}(L+
x_0)^{\nu}(L-x_0)^{\mu-\alpha}\Big[\frac{\nu}{z}\frac{
\Gamma(1-\alpha)\Gamma(1+\mu)}{\Gamma(2+\mu-\alpha)}F(1-\nu;1-\alpha;2+\mu
-\alpha;-z^{-1})\\
\fl-\frac{\Gamma(1-\alpha)\Gamma(1+\mu)}{\Gamma(1+\mu-\alpha)}F(-\nu;1-\alpha;1+
\mu-\alpha;-z^{-1})\Big]=-1 .
\label{eq:MFPTHypergeo2}
\end{eqnarray}
Moreover, by applying the relation \cite{NNLebedev1972} (see equation (9.5.9))
\begin{eqnarray}
\nonumber
F(a;b;c;x)=(-x)^{-a}\frac{\Gamma(c)\Gamma(b-a)}{\Gamma(c-a)\Gamma(b)}F(a;1+a-c;
1+a-b;x^{-1})\\
+(-x)^{-b}\frac{\Gamma(c)\Gamma(a-b)}{\Gamma(c-b)\Gamma(a)}F(b;1+b-c;1+b-a;x^{-1}),
\end{eqnarray}
where $|\arg(-x)|<\pi$, $|\arg(1-x)|<\pi$, and $a-b\neq0,\pm1,\pm2,\dots$, one can
write
\begin{eqnarray}
\nonumber
\frac{\Gamma(1-\alpha)\Gamma(1+\nu)}{\Gamma(1+\nu-\alpha)}F(-\mu;1-\alpha;1+\nu
-\alpha;-z)=z^{\mu}\frac{\Gamma(1+\mu-\alpha)\Gamma(1+\nu)}{\Gamma(1+\mu+\nu-
\alpha)}\\
\nonumber
\times F(-\mu;\alpha-\mu-\nu;\alpha-\mu;-z^{-1})+z^{\alpha-1}\frac{\nu\Gamma(1-
\alpha)\Gamma(\alpha-\mu-1)}{\Gamma(-\mu)}\\
\times F(1-\alpha;1-\nu;2+\mu-\alpha;-z^{-1}),
\end{eqnarray}
and
\begin{eqnarray}
\nonumber
\mu z\frac{\Gamma(1-\alpha)\Gamma(1+\nu)}{\Gamma(2+\nu-\alpha)}F(1-\mu;1-\alpha;
2+\nu-\alpha;-z)=\mu z^{\mu}\frac{\Gamma(1+\nu)\Gamma(\mu-\alpha)}{\Gamma(1+\mu+
\nu-\alpha)}\\
\nonumber
\times F(1-\mu;\alpha-\mu-\nu;1+\alpha-\mu;-z^{-1})+\mu z^{\alpha}\frac{\Gamma(1
-\alpha)\Gamma(\alpha-\mu)}{\Gamma(1-\mu)}\\
\times F(1-\alpha;-\nu;1+\mu-\alpha;-z^{-1}).
\end{eqnarray}
By substitution into equation (\ref{eq:MFPTHypergeo2})
\begin{eqnarray}
\nonumber
\fl\frac{K_{\alpha}C_{\alpha,\beta}R_{\alpha,\beta}}{\Gamma(1-\alpha)}(L+x_0)^{\nu-
\alpha}(L-x_0)^{\mu}\Big[\nu z^{\mu}B(\nu,1+\mu-\alpha)F(-\mu;\alpha-\mu-\nu;
\alpha-\mu;-z^{-1})\\
\nonumber
\fl+\nu z^{\alpha-1}B(1-\alpha,\alpha-\mu-1) F(1-\alpha;1-\nu;2+\mu-\alpha;-z^{-1})\\
\nonumber
\fl-\mu z^{\mu}B(1+\nu,\mu-\alpha)F(1-\mu;\alpha-\mu-\nu;1+\alpha-\mu;-z^{-1})-\mu z^{
\alpha}B(1-\alpha,\alpha-\mu)\\
\nonumber
\fl\times F(1-\alpha;-\nu;1+\mu-\alpha;-z^{-1})\Big]\\
\nonumber
\fl-\frac{K_{\alpha}C_{\alpha,\beta}L_{\alpha,\beta}}{\Gamma(1-\alpha)}(L+x_0)^{\nu}
(L-x_0)^{\mu-\alpha}\Big[\frac{\nu}{z}B(1-\alpha,1+\mu) F(1-\nu;1-\alpha;2+\mu-
\alpha;-z^{-1})\\
\fl-\mu B(1-\alpha,\mu)F(-\nu;1-\alpha;1+\mu-\alpha;-z^{-1})\Big] =-1 .
\label{eq:MFPTHypergeo3}
\end{eqnarray}
Here, $B(a,b)=\Gamma(a)\Gamma(b)/\Gamma(a+b)$ is the Beta function and with the
help of the symmetry property of the Gauss hypergeometric function, $F(a;b;c;x)
=F(b;a;c;x)$ \cite{NNLebedev1972} (see equation (9.2.1)), we have
\begin{eqnarray}
\nonumber
\fl F(1-\alpha;1-\nu;2+\mu-\alpha;-z^{-1})=F(1-\nu;1-\alpha;2+\mu-\alpha;-z^{-1})\\
\fl F(1-\alpha;-\nu;1+\mu-\alpha;-z^{-1})=F(-\nu;1-\alpha;1+\mu-\alpha;-z^{-1}).
\end{eqnarray}
By substitution into equation (\ref{eq:MFPTHypergeo3}), we get
\begin{eqnarray}
\nonumber
\fl&\frac{K_{\alpha}C_{\alpha,\beta}R_{\alpha,\beta}}{\Gamma(1-\alpha)}(L+x_0)^{
\mu+\nu-\alpha}\Big[\nu B(\nu,1+\mu-\alpha) F(-\mu;\alpha-\mu-\nu;\alpha-\mu;-z
^{-1})\\
\nonumber
\fl&-\mu B(1+\nu,\mu-\alpha)F(1-\mu;\alpha-\mu-\nu;1+\alpha-\mu;-z^{-1})\Big]\\
\nonumber
\fl&+\frac{K_{\alpha}C_{\alpha,\beta}}{\Gamma(1-\alpha)}(L+x_0)^{\nu-1}(L-x_0)^{1+
\mu-\alpha}\Big[\nu R_{\alpha,\beta}B(1-\alpha,\alpha-\mu-1)-\nu L_{\alpha,\beta}
B(1-\alpha,1+\mu)\Big]\\
\nonumber
\fl&\times F(1-\nu;1-\alpha;2+\mu-\alpha;-z^{-1})-\frac{K_{\alpha}C_{\alpha,\beta}}{
\Gamma(1-\alpha)}(L+x_0)^{\nu}(L-x_0)^{\mu-\alpha}\\
\fl&\times\Big[\mu R_{\alpha,\beta}B(1-\alpha,\alpha-\mu)-\mu L_{\alpha,\beta}B(1-
\alpha,\mu)\Big]F(1-\alpha;-\nu;1+\mu-\alpha;-z^{-1})=-1.
\end{eqnarray}
Then, by rearranging we obtain
\begin{eqnarray}
\nonumber
\fl\frac{K_{\alpha}C_{\alpha,\beta}L_{\alpha,\beta}}{\Gamma(1-\alpha)}(L+
x_0)^{\mu+\nu-\alpha}\Big[\nu B(\nu,1+\mu-\alpha)F(-\mu;\alpha
-\mu-\nu;\alpha-\mu;-z^{-1})\\
\nonumber
\fl-\mu B(1+\nu,\mu-\alpha)F(1-\mu;\alpha-\mu-\nu;1+\alpha-\mu;-z^{-1})\Big]\\
\nonumber
\fl+\frac{K_{\alpha}C_{\alpha,\beta}}{\Gamma(1-\alpha)}(L+x_0)^{\nu}(
L-x_0)^{\mu-\alpha}\Big[\Big(\nu\Big(R_{\alpha,\beta}B(1-\alpha,\alpha
-\mu-1)-L_{\alpha,\beta}B(1-\alpha,1+\mu)\Big)\\
\nonumber
\fl\times z^{-1}F(1-\alpha;1-\nu;2+\mu-\alpha;-z^{-1})-\mu\Big(R_{\alpha,\beta}
B(1-\alpha,\alpha-\mu)-L_{\alpha,\beta}B(1-\alpha,\mu)\Big)\\
\fl\times F(1-\alpha;-\nu;1+\mu-\alpha;-z^{-1})\Big] =-1 .
\label{eq:MFPTHypergeo4}
\end{eqnarray}
The left hand side must be independent of $z$ since $\mu$ and $\nu$ do not depend
on $z$. This requirement leads to the relations below. For the first term on the
left hand side, with the help of $F(a,b=0,c,x)=1$ \cite{NNLebedev1972} (see
section (9.8)), we have
\begin{eqnarray}
\nonumber
&F(-\mu;\alpha-\mu-\nu;\alpha-\mu;-z^{-1})=1\\
&F(1-\mu;\alpha-\mu-\nu;1+\alpha-\mu;-z^{-1})=1 ,
\end{eqnarray}
where
\begin{equation}
b=\alpha-\mu-\nu=0 \to \alpha=\mu+\nu .
\label{eq:bequalzeromunu}
\end{equation}
For the second term, we find
\begin{eqnarray}
\nonumber
&R_{\alpha,\beta}B(1-\alpha,\alpha-\mu-1)-L_{\alpha,\beta}B(1-\alpha,1+\mu)=0\\
&R_{\alpha,\beta}B(1-\alpha,\alpha-\mu)-L_{\alpha,\beta}B(1-\alpha,\mu)=0.
\end{eqnarray}
By definition of the Beta function,
\begin{eqnarray}
\nonumber
&R_{\alpha,\beta}\frac{\Gamma(1-\alpha)\Gamma(\alpha-\mu-1)}{\Gamma(-\mu)}=L_{
\alpha,\beta}\frac{\Gamma(1-\alpha)\Gamma(1+\mu)}{\Gamma(2+\mu-\alpha)}\\
&R_{\alpha,\beta}\frac{\Gamma(1-\alpha)\Gamma(\alpha-\mu)}{\Gamma(1-\mu)}=L_{
\alpha,\beta}\frac{\Gamma(1-\alpha)\Gamma(\mu)}{\Gamma(1+\mu-\alpha)}.
\end{eqnarray}
Using Euler's reflection formula $\Gamma(1-z)\Gamma(z)\sin{(\pi z)}=\pi$, we obtain
\begin{eqnarray}
\nonumber
\frac{R_{\alpha,\beta}}{\sin{(\pi(\alpha-\mu-1))}}=\frac{L_{\alpha,\beta}}{-\sin{
(\pi\mu)}}\\
\frac{R_{\alpha,\beta}}{\sin{(\pi(\alpha-\mu))}}=\frac{L_{\alpha,\beta}}{\sin{
(\pi\mu)}},
\label{eq:weightcoeefrho1}
\end{eqnarray}
which are identical. With the help of the weight coefficients (see equation
(\ref{eq:weightcoeffRL}))
\begin{eqnarray}
\nonumber
L_{\alpha,\beta}&=&-\frac{\sin{(\pi\alpha\rho)}}{\sin{(\pi\alpha)}\cos{(\pi
\alpha(\rho-1/2))}}\\
R_{\alpha,\beta}&=&-\frac{\sin{(\pi\alpha(1-\rho))}}{\sin{(\pi\alpha)}\cos{(\pi\alpha
(\rho-1/2))}},
\label{eq:weightcoeefrho2}
\end{eqnarray}
where $\rho$ is defined in equation (\ref{eq:xiparameter}). Substitution into
equation (\ref{eq:weightcoeefrho1}), we find
\begin{equation}
\frac{\sin{(\pi\alpha(1-\rho))}}{\sin{(\pi(\alpha-\mu))}}=\frac{\sin{(\pi\alpha
\rho)}}{\sin{(\pi\mu)}}.
\end{equation}
Therefore, the parameters $\mu$ and $\nu$ have the following form (see equation
(\ref{eq:bequalzeromunu}))
\begin{equation}
\mu=\alpha\rho, \,\,\, \nu=\alpha-\alpha\rho.
\end{equation}
To determine the normalisation factor, by substituting equation (\ref{eq:muandnu})
into equation (\ref{eq:MFPTHypergeo4}) we obtain
\begin{eqnarray}
\frac{K_{\alpha}C_{\alpha,\beta}R_{\alpha,\beta}}{\Gamma(1-\alpha)}\Big[\nu B(\nu,
1+\mu-\alpha) -\mu B(1+\nu,\mu-\alpha)\Big]=-1.
\end{eqnarray}
Using the Beta function,
\begin{eqnarray}
\frac{K_{\alpha}C_{\alpha,\beta}R_{\alpha,\beta}}{\Gamma(1-\alpha)}\Big[-\alpha
\Gamma(1+\nu)\Gamma(\mu-\alpha) \Big]=-1,
\end{eqnarray}
we get
\begin{eqnarray}
\fl C_{\alpha,\beta}=\frac{\Gamma(1-\alpha)}{\alpha K_{\alpha}R_{\alpha,\beta}
\Gamma(1+\alpha-\alpha \rho)\Gamma(\alpha \rho-\alpha)}
=\frac{1}{\Gamma(1+\alpha)K_{\alpha}}\frac{\sin{(\pi\alpha(\rho-1))}}{R_{\alpha,
\beta}\sin{(\pi\alpha)}},
\end{eqnarray}
where the last equality follows from Euler's reflection formula. Finally by
substitution of $R_{\alpha,\beta}$ (equation (\ref{eq:weightcoeefrho2})), we
get the desire result (\ref{eq:Cnormalfactor}).

\section{Fractional integration of a fractional derivative}
\label{appc}

Here we show the composition rule for the right Riemann-Liouville fractional
integral and the right fractional derivative in the Caputo form of the operator.
The right Riemann-Liouville fractional integral is given by ($p\in\mathrm{Re}>0$)
\cite{Podlubny1999} 
\begin{equation}
_xD_L^{-p}f(x)=\frac{1}{\Gamma(p)}\int\limits_x^L\frac{f(\zeta)}{(\zeta-x)^{1-p}}
\mathrm{d}\zeta,
\end{equation}
and with the right Caputo form of the fractional derivative as ($n-1<q<n$)
\begin{equation}
\label{eq:rcapu1}
_xD_L^q f(x)=\frac{(-1)^n}{\Gamma(n-q)}\int\limits_x^L\frac{f^{(n)}(\zeta)}{
(\zeta-x)^{q-n+1}}\mathrm{d}\zeta,
\end{equation}
we write
\begin{equation}
_xD_L^{-p}\left(_xD_L^qf(x)\right)=\frac{1}{\Gamma(p)}\int\limits_x^L
\frac{_{\zeta}D_L^qf(\zeta)}{(\zeta-x)^{1-p}}\mathrm{d}\zeta.
\end{equation}
Then, with the help of equation (\ref{eq:rcapu1}) we find
\begin{equation}
\fl _xD_L^{-p}\left(_xD_L^qf(x)\right)=\frac{(-1)^n}{\Gamma(p)\Gamma(n-q)}\int
_x^L\frac{1}{(\zeta-x)^{1-p}}\int\limits_{\zeta}^L\frac{f^{(n)}(y)}{(y-\zeta)^{
q-n+1}}\mathrm{d}y\mathrm{d}\zeta.
\end{equation}
Now, we change the integration order,
\begin{equation}
\int\limits_x^L\int\limits_{\zeta}^Lf(x,\zeta,y)\mathrm{d}y\mathrm{d}\zeta=
\int\limits_x^L\int\limits_x^yf(x,\zeta,y)\mathrm{d}\zeta\mathrm{d}y,
\end{equation}
and get
\begin{equation}
\fl _xD_L^{-p}\left(_xD_L^qf(x)\right)=\frac{(-1)^n}{\Gamma(p)\Gamma(n-q)}\int
\limits_x^Lf^{(n)}(y)\int\limits_x^y\frac{1}{(\zeta-x)^{1-p}(y-\zeta)^{q-n+1}}
\mathrm{d}\zeta\mathrm{d}y .
\end{equation}
After change of variable, $\zeta=x+z(y-x)$ in the inner integral, we arrive at
\begin{equation}
\fl _xD_L^{-p}\left(_xD_L^qf(x)\right)=\frac{(-1)^n}{\Gamma(p)\Gamma(n-q)}\int
\limits_x^L\frac{f^{(n)}(y)}{(y-x)^{1-n}}\int\limits_0^1\frac{1}{z^{1-p}(1-z)^{
q-n+1}}\mathrm{d}z\mathrm{d}y.
\end{equation}
Then, with the help of
\begin{equation}
\int\limits_0^1\frac{1}{z^{1-p}(1-z)^{q-n+1}}\mathrm{d}z=\frac{\Gamma(p)\Gamma(
n-q)}{\Gamma(n)},
\end{equation}
we find
\begin{equation}
_xD_L^{-p}\left(_xD_L^qf(x)\right)=\frac{(-1)^n}{\Gamma(n)}\int\limits_x^L
\frac{f^{(n)}(y)}{(y-x)^{1-n}}\mathrm{d}y.
\end{equation}
For our case in section \ref{moment bound one-sided} with $p=q=m\alpha$ and
$f(x_0)=\langle\tau^m\rangle(x_0)$, when $n=1$ ($0<\alpha<1$, $m=1$) this
becomes
\begin{equation}
_{x_0}D_L^{-m\alpha}\left(_{x_0}D_L^{m\alpha}\langle\tau^m\rangle(x_0)\right)=
\langle\tau^m\rangle(x_0)-\langle\tau^m\rangle(L),
\end{equation}
and when $n=2$ ($0<\alpha<1$, $m=2$) after integration by part we get
\begin{equation}
\fl _{x_0}D_L^{-m\alpha}\left(_{x_0}D_L^{m\alpha}\langle\tau^m\rangle(x_0)\right)=
(L-x_0)\frac{\partial\langle\tau^m\rangle(y)}{\partial y}\Big{|}_{y=L}+
\langle\tau^m\rangle(x_0)-\langle\tau^m\rangle(L).
\end{equation}
With a similar procedure for $n\geq3$ it can be deduce that in order to get result
(\ref{eq:Zoiamethodmoment-oneside}), all derivatives of the order $n-1<m\alpha$ of
$\langle\tau^m\rangle(y)$ at $y=L$ should be zero. The fact that $\langle\tau^m
\rangle(y)$ vanishes at $y=L$ is intuitively clear, when the initial point of
the random walker is located right at the absorbing boundary $x_0=L$, it will
be removed immediately. We also note that by differentiating the result
(\ref{eq:moments-oneside-bound}) it is easy to check that the assumption that
all derivatives of $\langle\tau^m\rangle(y)$ vanish at $y=L$ is reasonable.

\section*{References}

\bibliographystyle{iopart-num}

\end{document}